\documentclass{article}
\usepackage{PRIMEarxiv}
\usepackage[utf8]{inputenc} 
\usepackage[T1]{fontenc}    
\usepackage{hyperref}       
\usepackage{url}            
\usepackage{booktabs}       
\usepackage{amsfonts}       
\usepackage{nicefrac}       
\usepackage{microtype}      
\usepackage{lipsum}
\usepackage{fancyhdr}       
\usepackage{graphicx}       
\graphicspath{{media/}}     
\usepackage{subcaption}
\usepackage{multirow} 
\usepackage{amsmath}
\usepackage{diagbox}
\usepackage{siunitx}
\usepackage{comment}
\usepackage{makecell}

\newcolumntype{C}[1]{>{\centering\arraybackslash}m{#1}}
\usepackage{nomencl}
\makenomenclature
\renewcommand{\nomgroup}[1]{%
  \ifstrequal{#1}{S}{\item[\textbf{Symbols}]}{%
  \ifstrequal{#1}{I}{\item[\textbf{Indices}]}{%
  \ifstrequal{#1}{A}{\item[\textbf{Acronyms}]}{}}}}
\title{Periodic Neural Mapping for Unsteady Rotor-Blade Pressure and Aeroelastic Load Prediction

}
\author{
  \textbf{Lionel Salesses}\\
  Cenaero, Gosselies, Belgium \\
  \texttt{lionel.salesses@cenaero.be} \\
  \and
  \textbf{Joachim Dominique}\\
  Cenaero, Gosselies, Belgium \\
  \and
  \textbf{Tariq Benamara} \\
  Cenaero, Gosselies, Belgium \\
  \and
  \textbf{Théo Flament} \\
  Safran Helicopter Engines, Bordes, France \\
  \and
  \textbf{Franck Mastrippolito} \\
  Safran Helicopter Engines, Bordes, France \\
}
\begin{document}
\maketitle
\begin{abstract}
Accurate prediction of unsteady aerodynamic loads remains a major challenge in turbomachinery design. High-fidelity Computational Fluid Dynamics (CFD) simulations are computationally expensive, while aeroelastic Quantities of Interest (QoI) depend sensitively on the temporal evolution of the pressure field. In this work, we introduce a neural-operator-based framework, named periodic Fourier Neural Mapping (p-FNM), for the prediction of unsteady pressure distributions on turbine rotor blades simulated using chorochronic numerical hypothesis. The proposed architecture embeds temporal periodicity directly into the model formulation and learns a continuous mapping from operating conditions and time to pressure fields. Unlike sequential latent-space approaches, the proposed formulation predicts pressure fields independently at any time, thereby preserving temporal continuity and avoiding error accumulation.
The proposed p-FNM is evaluated on a database of unsteady rotor-blade simulations and compared with a state-of-the-art reduced-order baseline based on a variational autoencoder and recurrent neural network, referred to as the Temporal Prediction Model (TPM). Performance is assessed both on pressure-field reconstruction and on the prediction of Generalized Aerodynamic Forces (GAFs), which constitute the primary aeroelastic QoI. Across all considered training datasets, the proposed model consistently outperforms the TPM. On the largest dataset, the p-FNM achieves a pressure-field mean absolute percentage error of 0.46\% and a GAF-magnitude prediction error of 4.42\%, corresponding to relative improvements of approximately 60.7\% and 77.6\% over the TPM baseline, respectively. The minimum weighted phase error reaches 0.060 rad, demonstrating the ability of the proposed framework to accurately preserve the temporal characteristics of the aerodynamic response.
The analysis further shows that GAF prediction is substantially more challenging than pressure-field reconstruction and that temporal coherence plays a critical role in the accurate prediction of spectral aerodynamic quantities. 
These results highlight the potential of neural-operator formulations for learning periodic unsteady aerodynamic manifolds and demonstrate their suitability for reduced-order modeling, aeroelastic analysis, and future machine-learning-assisted turbomachinery design workflows.
\end{abstract}

\nomenclature[S]{$p_s(\mathbf{x},t)$}{Unsteady pressure distribution on the rotor-blade surface}
\nomenclature[S]{$\mathbf{x}$}{Spatial coordinate on the blade surface}
\nomenclature[S]{$x,y$}{Spatial coordinates of the two-dimensional blade-surface representation}
\nomenclature[S]{$t$}{Time variable}
\nomenclature[S]{$T_{BPF}$}{Blade-passing period}
\nomenclature[S]{$\omega_{\mathrm{BPF}}$}{Blade-passing angular frequency}
\nomenclature[S]{$\mathbf{n}(\mathbf{x})$}{Outward unit normal vector to the blade surface}
\nomenclature[S]{$S$}{Blade surface}
\nomenclature[S]{$\boldsymbol{\Phi}_r(\mathbf{x})$}{Structural displacement field associated with mode $r$}
\nomenclature[S]{$F_r(t)$}{Aerodynamic force projected onto structural mode $r$}
\nomenclature[S]{$\hat{F}_{r,k}$}{Complex GAF coefficient associated with structural mode $r$ and harmonic $k$}
\nomenclature[S]{$\rho_{r,k}$}{Magnitude of the GAF coefficient associated with structural mode $r$ and harmonic $k$}
\nomenclature[S]{$\varphi_{r,k}$}{Phase of the GAF coefficient associated with structural mode $r$ and harmonic $k$}
\nomenclature[S]{$\rho$}{Magnitude of the first-harmonic GAF coefficient considered in this study}
\nomenclature[S]{$\varphi$}{Phase of the first-harmonic GAF coefficient considered in this study}
\nomenclature[S]{$\kappa$}{Vector of operating conditions}
\nomenclature[S]{$\alpha$}{Turbine inflow angle [$^{\circ}$]}
\nomenclature[S]{$\beta$}{Blade pitch angle [$^{\circ}$]}
\nomenclature[S]{$d$}{Inter-row spacing}
\nomenclature[S]{$\Pi$}{Stage pressure ratio}
\nomenclature[S]{$\Omega$}{Rotor angular velocity [rad s$^{-1}$]}
\nomenclature[S]{$N_s$}{Number of stator blades}
\nomenclature[S]{$N_r$}{Number of rotor blades}

\nomenclature[S]{$\mathcal{A}$}{Input function space in the operator-learning formulation}
\nomenclature[S]{$\mathcal{U}$}{Output function space in the operator-learning formulation}
\nomenclature[S]{$G$}{Target operator}
\nomenclature[S]{$G_\theta$}{Parametric approximation of the target operator}
\nomenclature[S]{$\theta$}{Trainable model parameters}
\nomenclature[S]{$v_i$}{Intermediate feature representation at FNO layer $i$}
\nomenclature[S]{$C$}{Number of feature channels in an FNO layer}
\nomenclature[S]{$W$}{Spatial width of the discretized pressure field}
\nomenclature[S]{$H$}{Spatial height of the discretized pressure field}
\nomenclature[S]{$\mathcal{T}$}{FNO layer operator}
\nomenclature[S]{$\psi$}{Pointwise nonlinear activation function}
\nomenclature[S]{$\mathcal{K}$}{Nonlocal spectral-convolution operator}
\nomenclature[S]{$\mathcal{F}$}{Fourier transform}
\nomenclature[S]{$\mathcal{F}^{-1}$}{Inverse Fourier transform}
\nomenclature[S]{$R$}{Learned complex-valued spectral convolution weights}
\nomenclature[S]{$\zeta_{\max}$}{Maximum retained Fourier wavenumber in spectral convolutions}
\nomenclature[S]{$L$}{Number of FNO layers}
\nomenclature[S]{$\mathcal{P}$}{Lifting operator}
\nomenclature[S]{$\mathcal{Q}$}{Projection operator}

\nomenclature[S]{$N$}{Number of samples in a batch}
\nomenclature[S]{$T$}{Number of temporal snapshots when used as a tensor dimension}
\nomenclature[S]{$\mathbf{p}^{\mathrm{true}}$}{Reference pressure field}
\nomenclature[S]{$\mathbf{p}^{\mathrm{pred}}$}{Predicted pressure field}
\nomenclature[S]{$p_b^{\mathrm{true}}$}{Reference pressure field associated with sample $b$}
\nomenclature[S]{$p_b^{\mathrm{pred}}$}{Predicted pressure field associated with sample $b$}
\nomenclature[S]{$\bar{p}^{\mathrm{true}}_b$}{Mean reference pressure associated with sample $b$}
\nomenclature[S]{$f_1^{\mathrm{true}}$}{Reference amplitude of the first Fourier mode of the projected-force signal}
\nomenclature[S]{$f_1^{\mathrm{pred}}$}{Predicted amplitude of the first Fourier mode of the projected-force signal}
\nomenclature[S]{$\sigma$}{Standard deviation}
\nomenclature[S]{$E$}{Young's modulus}
\nomenclature[S]{$\nu$}{Poisson's ratio}
\nomenclature[S]{$\rho_{\mathrm{mat}}$}{Material density}
\nomenclature[S]{$R^2$}{Coefficient of determination for pressure prediction}
\nomenclature[S]{$R^2_\rho$}{Coefficient of determination for GAF-magnitude prediction}
\nomenclature[S]{$\Delta\varphi_b$}{Wrapped phase error for sample $b$}


\nomenclature[I]{$b$}{Sample index}
\nomenclature[I]{$t$}{Temporal index}
\nomenclature[I]{$w$}{Spatial index along the width direction}
\nomenclature[I]{$h$}{Spatial index along the height direction}
\nomenclature[I]{$r$}{Structural mode index}
\nomenclature[I]{$k$}{Fourier harmonic index}
\nomenclature[I]{$i$}{FNO-layer index}


\nomenclature[A]{BPF}{Blade Passing Frequency}
\nomenclature[A]{CFD}{Computational Fluid Dynamics}
\nomenclature[A]{CNN}{Convolutional Neural Network}
\nomenclature[A]{DoE}{Design of Experiments}
\nomenclature[A]{FEM}{Finite Element Method}
\nomenclature[A]{FNM}{Fourier Neural Mapping}
\nomenclature[A]{FNO}{Fourier Neural Operator}
\nomenclature[A]{GAF}{Generalized Aerodynamic Force}
\nomenclature[A]{GPU}{Graphics Processing Unit}
\nomenclature[A]{GRU}{Gated Recurrent Unit}
\nomenclature[A]{LCVT}{Latinized Centroidal Voronoi Tessellation}
\nomenclature[A]{LES}{Large-Eddy Simulation}
\nomenclature[A]{LSTM}{Long Short-Term Memory}
\nomenclature[A]{MAE}{Mean absolute error}
\nomenclature[A]{MAPE}{Mean absolute percentage error}
\nomenclature[A]{ML}{Machine Learning}
\nomenclature[A]{MLP}{Multilayer Perceptron}
\nomenclature[A]{MME}{Mean maximum error}
\nomenclature[A]{MSE}{Mean squared error loss}
\nomenclature[A]{NGV}{Nozzle Guide Vane}
\nomenclature[A]{PDE}{Partial Differential Equation}
\nomenclature[A]{p-FNM}{periodic Fourier Neural Mapping}
\nomenclature[A]{POD}{Proper Orthogonal Decomposition}
\nomenclature[A]{QoI}{Quantity of Interest}
\nomenclature[A]{RANS}{Reynolds-Averaged Navier--Stokes}
\nomenclature[A]{RNN}{Recurrent Neural Network}
\nomenclature[A]{ROM}{Reduced-Order Modeling}
\nomenclature[A]{SOAP}{Shampoo with Adam preconditioning}
\nomenclature[A]{t-SNE}{t-distributed Stochastic Neighbor Embedding}
\nomenclature[A]{TPM}{Temporal Prediction Model}
\nomenclature[A]{URANS}{Unsteady Reynolds-Averaged Navier--Stokes}
\nomenclature[A]{V2F}{Vector-to-function mapping}
\nomenclature[A]{VAE}{Variational Autoencoder}
\nomenclature[A]{wMAPE}{Weighted mean absolute percentage error}
\nomenclature[A]{wRMSE$_{\varphi}$}{Weighted root mean square error of the GAF phase}

\printnomenclature
\section{Introduction}
The aerodynamic environment encountered in highly loaded transonic turbine stage turbomachinery is characterized by strong unsteady interactions between stationary and rotating blade rows. As wakes, potential fields and shock structures generated by upstream Nozzle Guide Vanes (NGVs) convect through the rotor passage, they produce highly non-uniform pressure fluctuations on blade surfaces. These interactions give rise to complex spatiotemporal pressure fields involving moving shocks and strong pressure gradients. Beyond their impact on aerodynamic performance, these unsteady loads constitute one of the principal excitation mechanisms responsible for blade vibration and forced-response phenomena.

Assuming weak coupling between the blade structural response and the aerodynamic excitation \cite{farhat2006provably}, such that blade vibrations do not perturb the aerodynamic flow, and under the hypothesis of linear behavior of the blade structure with harmonic motion, the aeroelastic behavior can be characterized by the so-called generalized aerodynamic forces (GAFs). These quantities are obtained by projecting the unsteady pressure distribution onto the structural vibration modes of the blade and provide the coupling between the aerodynamic forcing and the structural response \cite{de2012forced}. Under the weak-coupling assumption, the aeroelastic problem can be treated sequentially, with the aerodynamic field computed first and the structural response subsequently evaluated under the resulting aerodynamic loading. The GAFs are therefore key quantities for assessing blade vibration amplitudes, high-cycle fatigue and mechanical integrity \cite{berthold2024fully}. Because these quantities depend on both the spatial distribution and temporal evolution of the pressure field, their accurate prediction requires a reliable representation of the underlying flow dynamics.


This loading can be assessed through various Computational Fluid Dynamics (CFD) modeling approaches, each offering a different level of fidelity. Unsteady Reynolds-Averaged Navier–Stokes (URANS) methods are commonly preferred over low-order approaches, such as those based on the Euler solvers, due to their ability to capture nonlinear effects of compressibility and viscosity \cite{tucker2013trends, cinnella2004numerical}. Additionally, they are often favored in industry over scale-resolving methods such as Large-Eddy Simulation (LES), which, while more accurate, are considerably more computationally demanding \cite{tucker2011computation}. Nonetheless, the computational cost associated with URANS simulations remains substantial. This becomes particularly restrictive in industrial workflows involving design optimization, uncertainty quantification, sensitivity analysis, or digital-twin applications, where hundreds of flow evaluations may be required \cite{khatouri2022metamodeling}. Reducing the cost of generating unsteady aerodynamic predictions while maintaining sufficient accuracy therefore remains a central challenge.

A common strategy consists in constructing surrogate models \cite{keane2020surrogate}, sometimes directly for scalar quantities of engineering interest. In turbomachinery applications, these quantities depend on the underlying physical problem and may include aerodynamic performance indicators such as stage isentropic efficiency, structural metrics such as maximum von Mises stress \cite{dong2025adaptive}, acoustic measures such as sound power level, or GAFs in the case of forced-response analyses \cite{tran2009multi, ding2026high}, which is the focus of the present work. While such approaches can significantly reduce computational cost, they provide limited insight into the physical mechanisms underlying the predicted response \cite{khatouri2022metamodeling}. Moreover, multiple quantities of interest often originate from the same underlying physical field, which necessitates the development of separate surrogate models despite their shared physical origin. This limitation has motivated increasing interest in field-level surrogate models, which aim to reproduce the complete discretized physical solutions rather than selected derived scalar engineering quantities alone.

Recent advances in machine learning have demonstrated significant potential for accelerating aerodynamic analyses. Most existing approaches focus on either steady flow prediction or direct prediction of integral performance quantities. However, predicting unsteady pressure fields in transonic turbomachinery remains considerably more challenging because the model must simultaneously reconstruct complex spatial structures, including moving shocks, and preserve their temporal evolution. Furthermore, the inherently high dimensionality of pressure fields makes this prediction task particularly demanding, motivating the widespread use of Reduced-Order Modeling (ROM) techniques over the past decades. While accurate pressure reconstruction is important, aeroelastic quantities such as GAFs depend critically on the temporal spectrum of the pressure signal. Small temporal inconsistencies may therefore induce substantial errors in the predicted aerodynamic loads even when pressure reconstruction errors remain moderate.

In \cite{dominique2026reduced}, the authors introduced a reduced-order modeling framework, referred to as the \emph{Temporal Prediction Model} (TPM), for the prediction of unsteady rotor-blade pressure fields. The approach combines a Variational Auto-Encoder (VAE) with a Recurrent Neural Network (RNN), enabling the compression of high-dimensional pressure fields into a latent space and the subsequent modeling of their temporal evolution. As TPM was developed and evaluated on the same database considered in the present work, it serves as the reference baseline, thereby enabling a direct and consistent comparison of predictive performance. 
While TPM accurately reproduces the dominant features of the unsteady pressure field, its sequential latent-space formulation does not explicitly enforce temporal continuity. As a consequence, temporal inconsistencies may arise in the reconstructed pressure dynamics, potentially degrading the prediction of frequency-domain quantities derived from the pressure field, such as the GAFs.

More generally, the decomposition of the problem into separate spatial latent representation and temporal prediction stages introduces modeling assumptions regarding the structure of the latent space and their dynamical trajectory in time. Furthermore, sequential temporal prediction may accumulate errors over long time horizons, potentially degrading phase accuracy and temporal coherence.

Recently, operator-learning methods have emerged as an alternative framework for the approximation of solutions of problem described by sets of partial differential equations. Rather than learning the evolution of reduced latent variables, neural operators seek to approximate mappings between function spaces directly. These methods naturally raise the question of whether preserving the continuous structure of the solution in the modeling framework may lead to improved prediction of derived engineering quantities.

The objective of the present work is therefore twofold. First, we seek to improve the prediction of unsteady pressure fields on turbine rotor blades under varying operating conditions. Second, we aim to preserve prediction accuracy of GAFs derived from these pressure fields. To this end, we introduce a periodic Fourier Neural Mapping (p-FNM) architecture specifically designed for periodic unsteady rotor-pressure prediction. Unlike sequential latent-space approaches \cite{dominique2026reduced}, the proposed framework directly learns the continuous mapping between operating conditions, time, and pressure fields, while explicitly embedding the periodic nature of the simulated physics.

More broadly, this work compares two fundamentally different paradigms for modeling unsteady fluid dynamics. The first relies on discrete-time sequential prediction, where temporal dynamics are learned sequentially using architectures such as RNNs, transformers, or the TPM, often combined with latent-space dimensionality reduction such as the VAE. The second adopts a continuous-time formulation based on neural operators, in which time is treated as a continuous input rather than as a sequence of discrete states. By comparing representative models from these two paradigms, this study assesses their respective abilities to reconstruct unsteady pressure fields and to accurately predict the derived GAFs.

The main contributions of this work are as follows:
\begin{itemize}
    \item A critical analysis of the limitations of sequential latent-space TPM model for the prediction of GAFs derived from unsteady pressure fields.
    \item The introduction of a periodic Fourier Neural Mapping (p-FNM) architecture that combines operator learning with an explicit periodic time representation.
    \item A comprehensive comparison between the proposed framework and the TPM baseline \cite{dominique2026reduced}, including both pressure-field reconstruction metrics and GAF-based metrics.
    \item An analysis of the relationship between temporal coherence, spectral fidelity and GAF prediction accuracy, providing a physical interpretation of the observed differences in performance.
    \item An investigation of direct end-to-end prediction strategies for projected forces and GAF, demonstrating the advantages of pressure-field prediction as an intermediate physical surrogate.
    \item An assessment of the influence of training-dataset size on both pressure and GAF prediction performances.
\end{itemize}
The remainder of this paper is organized as follows. Section~\ref{section:related-work} reviews previous work on machine-learning-based prediction of unsteady pressure fields and aerodynamic loads. Section~\ref{sec:problem-and-dataset} describes the numerical simulation framework, the dataset generation procedure, and the methodology used for the computation of GAFs.
Section~\ref{section:EFNM} presents the proposed p-FNM framework. Numerical results and comparative analyses are then reported in Section~\ref{section:results}. Finally, conclusions and perspectives are provided in Section~\ref{section:conclusion}.

\section{Related Work} \label{section:related-work}
\subsection{Machine Learning for Unsteady Pressure Prediction}
The prediction of unsteady aerodynamic flow fields using machine-learning techniques has emerged as a promising alternative to computationally intensive high-fidelity simulations. Early studies demonstrated that deep-learning-based ROMs can effectively capture the dominant spatio-temporal dynamics of unsteady flows by combining latent-space representations with recurrent forecasting strategies. In particular, Hasegawa \emph{et al.}~\cite{hasegawa2020machine} proposed a reduced-order framework based on a convolutional autoencoder and a Long Short-Term Memory (LSTM) network for predicting unsteady flows around bluff bodies of varying geometries, illustrating the potential of data-driven latent-space models for fluid dynamics applications. Similar ideas have subsequently been applied to aerodynamic pressure prediction, where an autoregressive Convolutional Neural Network (CNN) architecture was used to reconstruct motion-induced unsteady pressure distributions on a wing from CFD-generated data \cite{rozov2021data}. More recently, deep neural networks have been shown capable of accurately reproducing complex transonic unsteady flow fields under previously unseen operating conditions using an autoregressive architecture based on attention U-Net \cite{chen2024deep}. 
Solera \textit{et al.} \cite{solera2024} proposed a nonlinear ROM framework combining a $\beta$-VAE with a transformer network for the prediction of unsteady fluid flows. The $\beta$-VAE is first trained to learn a compact and disentangled latent representation of two-dimensional velocity fields, yielding latent variables that exhibit an interpretable structure similar to Proper Orthogonal Decomposition (POD) modes while providing a more compact representation. A transformer is then trained to model the temporal evolution of the latent variables by processing a history of encoded flow snapshots and autoregressively predicting the latent state at the next time step. The resulting latent sequence is subsequently decoded to reconstruct the flow field. The proposed framework was shown to accurately capture both periodic and chaotic flow dynamics and to outperform alternative latent-space prediction models. A similar approach was proposed in \cite{hemmasian2023reduced}, where a convolutional autoencoder was coupled with a transformer to construct a nonlinear ROM for fluid flows.
In parallel, ensemble-learning strategies have been introduced to improve the robustness of ROMs and mitigate the error accumulation commonly encountered in long-horizon autoregressive predictions \cite{halder2024reduced}. Collectively, these studies highlight the growing potential of machine-learning-based ROMs for the efficient prediction of complex unsteady aerodynamic phenomena.\\
In turbomachinery applications, ROM approaches have been developed to predict unsteady spatio-temporal aerodynamic fields. Qiao \emph{et al.}~\cite{qiao2024reduced} introduced a reduced-order methodology for turbine-cascade aerodynamics based on modal decompositions and machine-learning (ML) regression techniques. More recently, Dominique \emph{et al.}~\cite{dominique2026reduced} proposed a Temporal Prediction Model (TPM) combining a VAE with RNN to predict unsteady pressure fields on turbine rotor blades under various operating conditions. By learning the temporal evolution of latent representations, the TPM achieved accurate pressure-field reconstruction while significantly reducing computational costs relative to URANS simulations. Since the TPM was developed and evaluated on the same database considered in the present work, it constitutes the primary baseline for comparison.
\subsection{Neural Operators for Fluid Dynamics}
Neural operators have recently emerged as a powerful alternative to conventional ROMs for learning solution operators of Partial Differential Equations (PDEs) directly in function spaces. Unlike classical surrogate models that learn mappings between finite-dimensional vectors, neural operators aim to approximate operators acting on continuous functions, thereby enabling improved generalization across parameter spaces and discretizations.

Among these approaches, the Fourier Neural Operator (FNO) has attracted considerable attention due to its ability to efficiently represent long-range interactions through spectral convolutions. The effectiveness of FNO-based architectures for aerodynamic applications was demonstrated in \cite{dai2023fourier}, where boundary-condition treatments were incorporated into the FNO framework for steady airfoil-flow prediction. 
Beyond FNOs, other operator-learning architectures such as Deep Operator Networks (DeepONets) have also been successfully applied to the modeling of unsteady fluid flows, demonstrating the ability of neural operators to learn complex temporal dynamics directly from data \cite{bai2024data}. More recently, deep neural operators have been employed for the prediction of unsteady flow fields at high angles of attack, further highlighting the potential of operator-learning approaches for strongly nonlinear aerodynamic regimes \cite{jin2026prediction}. Neural-operator methodologies have also been extended to turbomachinery applications. In particular, transformer-enhanced neural operators have been proposed for panoramic aerodynamic-performance prediction in turbomachinery cascades, demonstrating the growing interest in operator-learning techniques for complex aerodynamic systems \cite{qineng2025panoramic}.

These developments suggest that neural operators constitute a promising framework for the prediction of parameterized aerodynamic fields. However, their application to periodic unsteady rotor-blade pressure fields and aerodynamic loads remains largely unexplored.

\subsection{Machine Learning for Aerodynamic Load Prediction}
In many industrial applications, the quantities of primary interest are not the flow fields themselves but integrated aerodynamic observables such as forces or moments. Consequently, several studies have investigated ML approaches for the direct prediction of aerodynamic loads.

Machine learning has been successfully applied to aeroelastic load prediction in transonic flow regimes. Recent studies have proposed deep-learning frameworks for the prediction of transonic wing buffet loads induced by structural deformations and vibration modes \cite{zahn2025deep,zahn2023transonic}. These approaches demonstrate that data-driven models can efficiently approximate complex fluid-structure interactions and provide rapid estimates of aeroelastic quantities that are traditionally obtained through expensive coupled simulations.

The feasibility of ML-based aerodynamic-force prediction has also been demonstrated in turbomachinery applications. For example, ML techniques have been employed to predict blade aerodynamic forces in compressors, providing rapid estimates of aerodynamic loading while avoiding costly flow simulations \cite{zhang2021study}. RNN architectures based on Gated Recurrent Units (GRUs) have been shown capable of accurately reproducing the unsteady aerodynamic forces generated by airfoils undergoing large amplitude pitching oscillations, highlighting the ability of sequence-learning models to capture nonlinear aerodynamic memory effects \cite{wu2023unsteady}.

Beyond simulation-based surrogate modeling, ML methods have also been used to infer aerodynamic loads directly from experimental or measurement data. In particular, real-time aerodynamic-force prediction from sparse pressure-sensor measurements has been achieved through hybrid reduced-order and machine-learning approaches, demonstrating the potential of data-driven methods for online monitoring and control applications \cite{duan2024machine}.

Collectively, these studies demonstrate that ML techniques can accurately predict aerodynamic observables such as integrated force coefficients. However, direct load-prediction approaches are generally specialized to a particular QoI and do not provide access to the underlying flow or pressure fields. In contrast, the present work follows a field-based strategy in which unsteady pressure distributions are first predicted and aerodynamic quantities are subsequently derived through physics-based post-processing.

\subsection{Positioning of the Present Work}
The literature reveals two main paradigms for aerodynamic surrogate modeling. The first consists of predicting intermediate flow quantities, such as pressure or velocity fields, and subsequently deriving engineering observables through physics-based post-processing~\cite{dominique2026reduced,qiao2024reduced,hasegawa2020machine}. The second aims at predicting aerodynamic observables directly, including forces, performance coefficients, or aeroelastic loads~\cite{duan2024machine,zahn2025deep,zhang2021study,wu2023unsteady}.

The present work follows the first paradigm and focuses on the prediction of unsteady rotor-blade pressure fields. Building upon the TPM framework introduced in \cite{dominique2026reduced}, we propose a neural-operator-based architecture specifically designed for periodic unsteady flows. Unlike sequential latent-space approaches based on recurrent forecasting, the proposed periodic p-FNM directly learns the mapping between operating conditions, time, and pressure fields. This formulation eliminates error accumulation associated with autoregressive prediction and helps for a better temporal coherency.

A central objective of this work is the accurate prediction of GAFs, which are highly sensitive to temporal coherence and spectral accuracy. We therefore investigate whether a continuous-time neural-operator formulation can improve the preservation of GAF quantity compared to sequential latent-space models. In addition, we analyze the influence of training-dataset size on predictive performance.
\section{Problem Formulation and Dataset} \label{sec:problem-and-dataset}

\subsection{Unsteady Turbine Flow Simulations}

The present study considers the transonic turbine stage developed within the Turbine Aero-Thermal External Flows (TATEF2) project \cite{tatef2}. The configuration consists of a stator with $N_s$ blades followed by a rotor with $N_r$ blades rotating at angular velocity $\Omega$, resulting in a blade-passing period $T_{BPF}=\frac{N_s}{\Omega}$ for the guide vane generated shocks structure. The stage geometry is identical to that employed in \cite{dominique2026reduced}, where a detailed description of the numerical setup and analysis of shocks pattern dynamics can be found.

The aerodynamic flow field is computed using the unsteady RANS formulation of the elsA solver \cite{cambier2011}. The computational domain is discretized with a structured mesh containing approximately $3\times10^6$ cells, including dedicated refinements within the blade boundary layers and rotor tip-clearance region. The mesh quality is ensured through standard industrial criteria, including cell skewness, aspect ratio, expansion ratio, and wall distance. At the inlet, total pressure, total temperature and flow angle are prescribed, while the outlet static pressure is adjusted to impose the desired stage pressure ratio. Chorochronic boundary conditions are applied on both lateral sides, and no-slip conditions are imposed on the solid surfaces. The interface between the fixed stator and the moving rotor domain is defined with so-called chorochronic or phase-lag boundary conditions using an appropriate number of harmonics of the Blade Passing Frequency (BPF) \cite{gerolymos2002}. 

For each operating condition, a steady RANS solution with the k-l Smith turbulence model  is first obtained and subsequently used to initialize a URANS simulation. Once periodic convergence is achieved, the unsteady pressure field is extracted on the rotor blade surface over one complete blade-passing period using $T=192$ uniformly distributed steps in time. The entire process can be performed in average in 810 CPU-h parallelized using 128 CPU on one AMD EPYC 7763 node.

\begin{figure*}[ht!]
\centering
\begin{subfigure}{0.55\textwidth}
    \includegraphics[width=\textwidth,trim=0 0 0 0, clip]{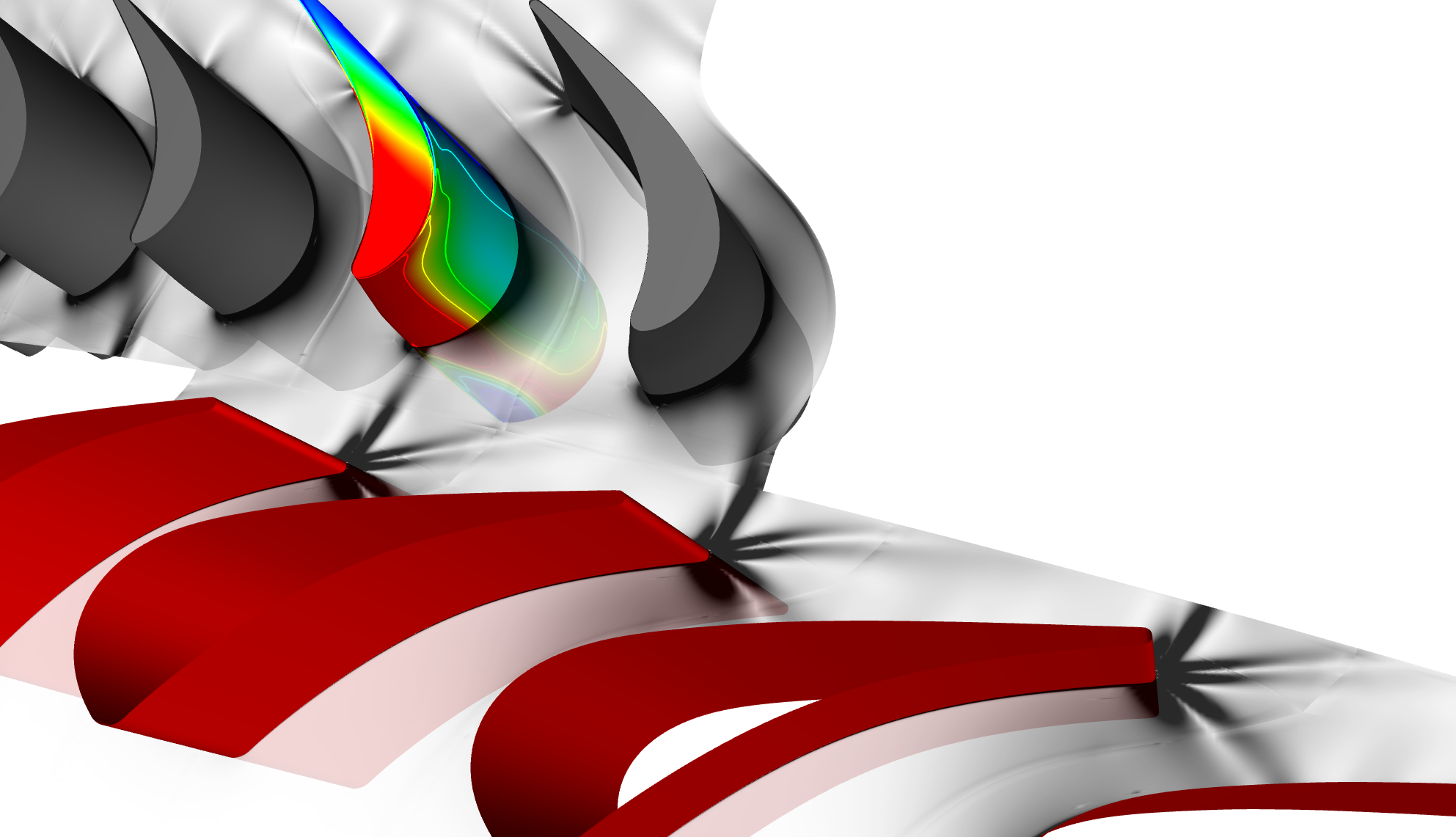}
    \label{fig:ts_schlieren}
    \caption{$t=t_s$}
\end{subfigure}

\begin{subfigure}{0.55\textwidth}
    \includegraphics[width=\textwidth,trim=0 0 0 0, clip]{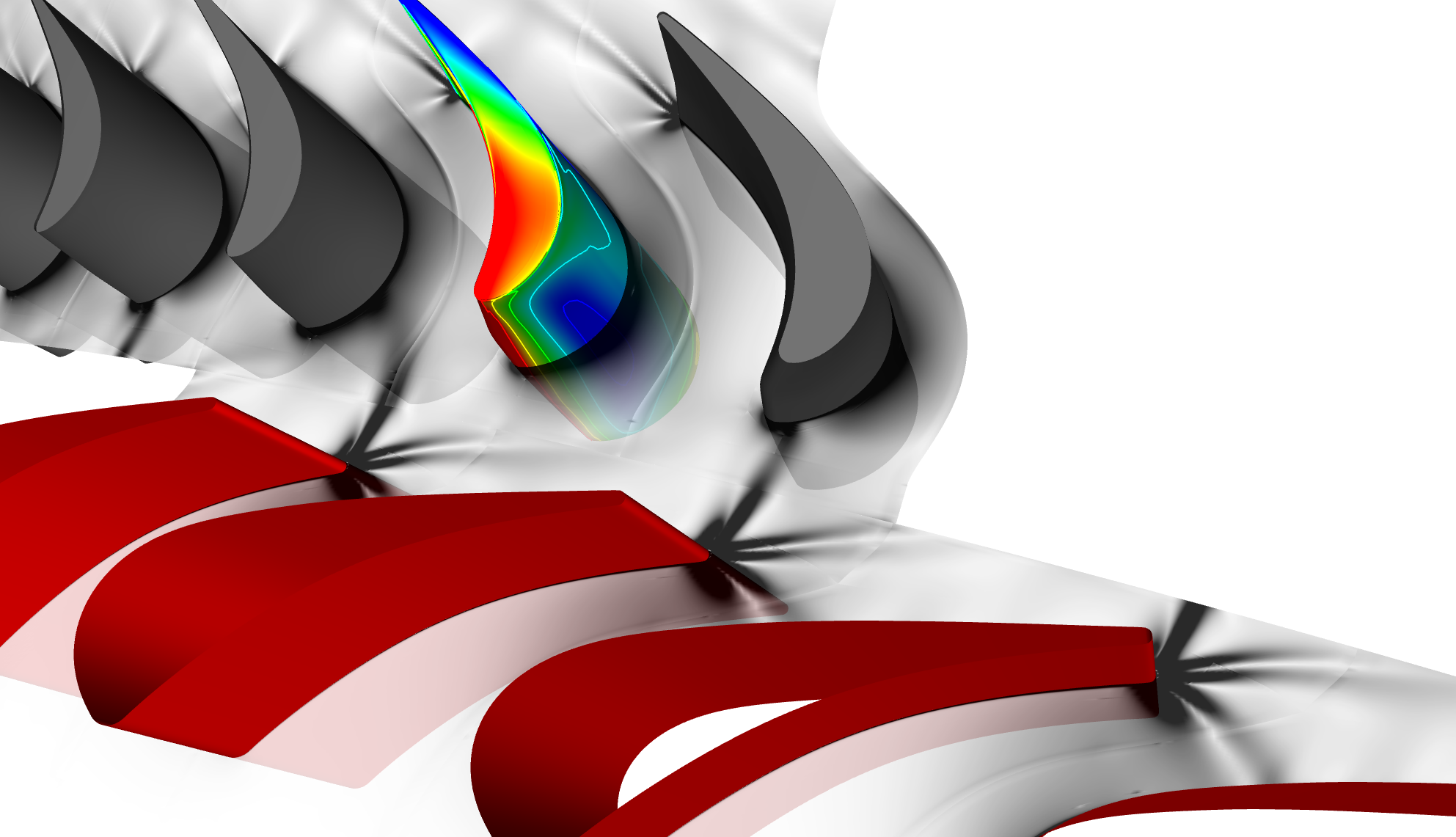}
    \label{fig:ts2_schlieren}
    \caption{$t=t_s+\frac{2}{5}T_{BPF}$}
\end{subfigure}
\begin{subfigure}{0.55\textwidth}
    \includegraphics[width=\textwidth,trim=0 0 0 0, clip]{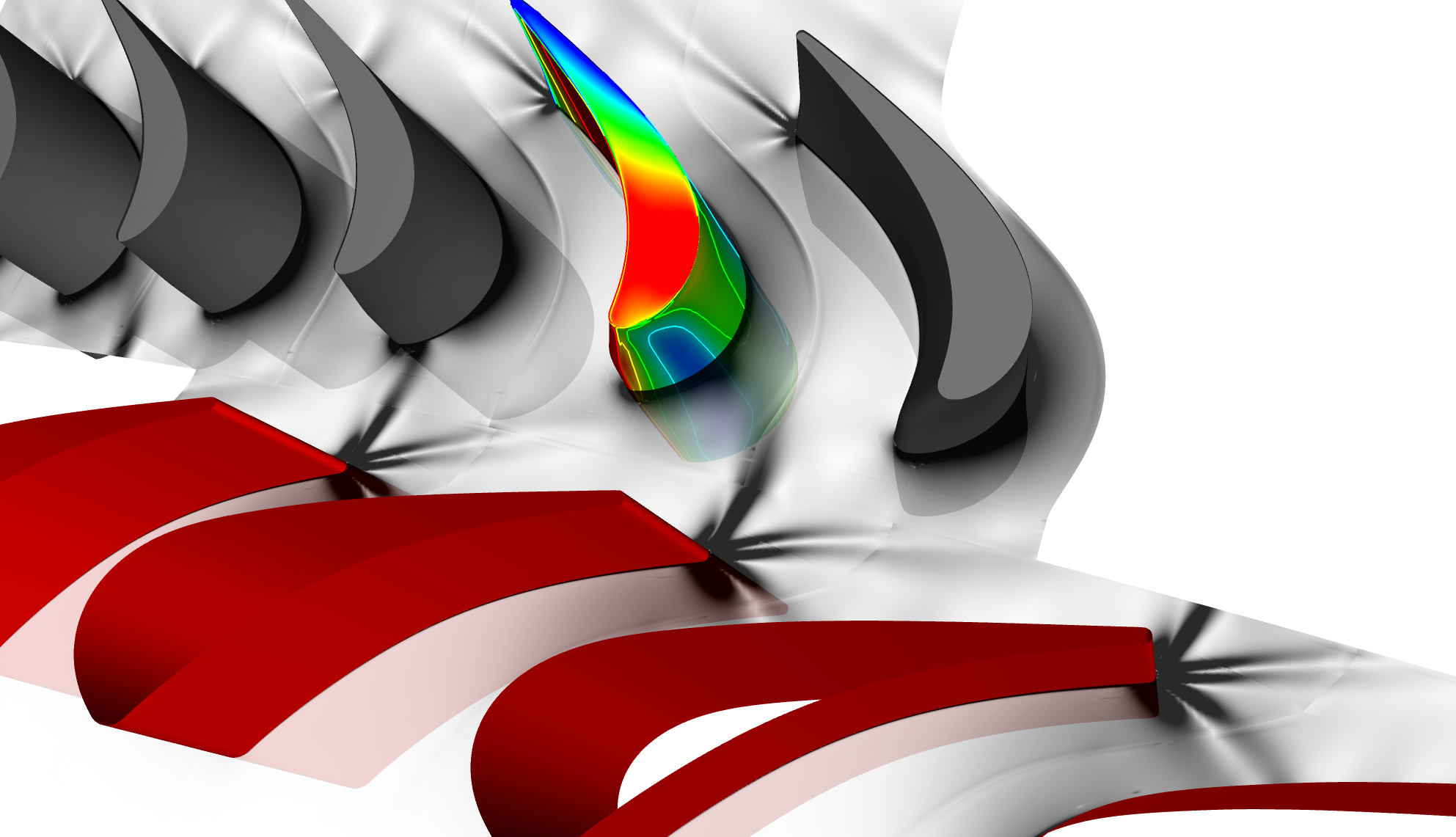}
    \label{fig:ts4_schlieren}
    \caption{$t=t_s+\frac{4}{5}T_{BPF}$}
\end{subfigure}
\caption{Mid-span ($h/H=0.5$) schlieren-like density gradient and static pressure field on the reference blade, showing the correlation between flow structures and aerodynamic loading}
\label{fig:schlieren_pressure_comparison}
\end{figure*}

The resulting flow exhibits features characteristic of highly loaded transonic turbine stages, including stator wakes, moving shocks, and their reflections (Fig. \ref{fig:schlieren_pressure_comparison}). This figure presents the mid-span density gradient, visualized as a schlieren-like image, alongside the corresponding static pressure field on the reference blade. These flow-induced structures generate strongly non-uniform pressure fluctuations on the rotor surface and constitute the primary source of aerodynamic forcing acting on the blade. Because the location and amplitude of these structures evolve continuously throughout the blade-passing period, accurately capturing both their spatial distribution and temporal evolution is essential for preserving the aerodynamic loads and ensuring the accurate prediction of the forced response source terms

\subsection{Generalized Aerodynamic Forces} \label{section:GAF}

The objective of the present work is not only to reconstruct the unsteady pressure field but also to evaluate how reconstruction errors propagate to quantities relevant for forced-response analyses. To this end, the aerodynamic loading is projected onto a structural vibration mode of the blade.

Let $p_s(\mathbf{x},t)$ denote the unsteady pressure distribution on the blade surface $S$ and $\boldsymbol{\Phi}_r(\mathbf{x})$ the complex displacement field associated with structural mode $r$. The modal aerodynamic force is defined as

\begin{equation}
F_r(t)= -
\int_S p_s(\mathbf{x},t) \cdot\mathbf{n}(\mathbf{x}) \cdot \boldsymbol{\Phi}_r(\mathbf{x}) dS,
\end{equation} where $\mathbf{n}$ denotes the outward surface normal vector.

Because the operating flow is periodic with the fundamental BPF, the projected force can be expressed through its Fourier decomposition,
\begin{equation}
    F_r(t)= \sum_k \hat{F}_{r,k}e^{ik \omega_{BPF} t},
\end{equation} where $\omega_{BPF}$ is the blade-passing frequency and $\hat{F}_{r,k}$ are the complex GAF coefficients associated with harmonic $k$.

Each GAF coefficient can be represented either in terms of its real and imaginary components,

\begin{equation} \label{eq:GAF}
    \hat{F}_{r,k} = \Re(\hat{F}_{r,k})+ i\,\Im(\hat{F}_{r,k}),
\end{equation}

or equivalently through its amplitude and phase,

\begin{equation}
    \hat{F}_{r,k} = \rho_{r,k} e^{i\phi_{r,k}},
\end{equation}

where $\rho_{r,k} = |\hat{F}_{r,k}|$ denotes the magnitude of the aerodynamic excitation and $\phi_{r,k}$ its phase relative to the blade-passing cycle. The magnitude quantifies the forcing intensity at a given harmonic, while the phase characterizes its temporal position within the period. Consequently, accurate GAF prediction requires not only correct pressure amplitudes but also preservation of the temporal coherence and spectral content of the underlying pressure field.

For simplicity, the remainder of this work focuses on the first harmonic coefficient, $\hat{F}_{r,1}$, associated with the structural mode $\boldsymbol{\Phi}_r$ considered in this study. This coefficient is hereafter referred to as the GAF coefficient, with its modulus and phase denoted by $\rho$ and $\phi$, respectively. Although the analysis is restricted to this specific coefficient, the proposed methodology is not intrinsically limited to a particular structural mode or harmonic. In particular, the p-FNM predicts the unsteady pressure field independently of the structural model, such that any GAF coefficient can subsequently be obtained by projecting the predicted pressure field onto the corresponding structural mode and extracting the desired Fourier harmonic. The framework can therefore be extended to other structural modes and GAF coefficients without modifying the pressure-prediction model.

The structural deformation field $\boldsymbol{\Phi}_r$ used in the projection is obtained from a finite-element modal analysis performed using the \texttt{SAMCEF} solver. The blade is modeled as a homogeneous steel structure with Young’s modulus $E = 210$ GPa, density $\rho_{\mathrm{mat}} = 7800$ kg/m$^3$, and Poisson’s ratio $\nu = 0.3$. The modal analysis is performed under rotational loading conditions to account for centrifugal forces. However, no additional steady aerodynamic load is applied in the structural computation. For the sake of simplicity the root of the blade is assumed clamped at the hub, while no cyclic symmetry or inter-blade coupling is considered, resulting in an isolated blade configuration. This assumption ensures that the projected aerodynamic forces remain purely real-valued and that no complex coupling between adjacent blades is introduced in the definition of the GAFs. Nevertheless, the methodology proposed in this work is not intrinsically restricted to these simplified mechanical assumptions and can, in principle, be extended to more realistic configurations involving complex structural modes and inter-blade interactions.

In the following sections, both pressure-field errors and GAF-based metrics are used to compare the different surrogate-modeling approaches.

\subsection{Simulation Dataset} \label{section:dataset}

The numerical dataset used to build the surrogate is constructed by sampling its operational parameters. Indeed, prior research \cite{denos2005, paniagua2008, laumert2002_part1, laumert2002_part2} demonstrated the sensitivity of this turbine stage flow characteristics to various operational parameters such as pressure ratios, inflow angles, blade spacing. Thus, this paper maintains constant blades geometry while examining the blade unsteady pressure field across different wind tunnel operating conditions. These operating conditions are characterized by the turbine inflow angle, blade pitch, inter-row spacing, and stage pressure ratio, denoted by $(\alpha, \beta, d, \Pi)\in\mathbb{R}^4$.  

The exploration of this parameter space is carried out using a Design of Experiments (DoE) strategy based on Latinized Centroidal Voronoi Tessellation (LCVT) \cite{romero2006}, implemented through the in-house MINAMO framework \cite{sainvitu2010,beaucaire2019}. This approach ensures a space-filling distribution of sampling points, promoting a uniform coverage of the multidimensional design domain.

Four datasets of increasing size are generated, containing 20, 40, 80, and 160 simulations, respectively, and referred to as \textit{Small} (S), \textit{Medium} (M), \textit{Large} (L), and \textit{Extra Large} (XL), as summarized in Table \ref{tab:datasetssize}. Each dataset is partitioned into training and validation subsets, where the training set is used for model optimization and the validation set is employed for early stopping to mitigate overfitting \cite{zhang2005}. In addition, a separate and fully independent test dataset comprising 40 simulations is generated to provide an unbiased assessment of model performance and to ensure consistent comparisons across different approaches.

To enable the application of machine learning models, the unsteady rotor pressure fields are projected onto a structured two-dimensional grid of size $(W, H)$, yielding an image-like representation of the surface pressure distribution. This mapping is performed without interpolation by unfolding the structured mesh onto a planar representation, after removing the blade tip and introducing a cut at the trailing edge. As a consequence, pressure values located on opposite sides of the trailing edge are treated as independent and are not explicitly correlated within convolutional architectures. While this simplification facilitates the construction of a regular data structure suitable for CNNs, it introduces a modeling assumption that discontinuities at the trailing edge do not significantly affect the learned representations.

The resulting dataset is therefore expressed as a tensor of dimension $(N, T, W, H)$, where $N$ is the number of samples and $T$ denotes the number of time snapshots over one blade-passing period. Although this representation enables efficient processing using deep learning methods, it does not preserve the true geometric distances of the original surface discretization.

\begin{table*}[ht]
    \centering
    \caption{Number of simulations for the different and independent datasets.}
    \begin{tabular}{l|c|c|c|c|c}
    \diagbox{Dataset}{Subset} & Training & Validation & Test & Total & Time per simulation \\
    \hline \hline
    Small & 15 & 5 & & 20 &  \multirow{4}{*}{810 CPU$\cdot$h}\\
    \cline{1-5}
    Medium & 32 & 8 & & 40 & \\
    \cline{1-5}
    Large & 68 & 12 & & 80 & \\
    \cline{1-5}
    Extra Large & 136 & 24 & & 160 &  \\
    \cline{1-5}
    Test & & & 40 & 40 &
    \end{tabular}
    \label{tab:datasetssize}
\end{table*}

\section{Fourier Neural Mapping} \label{section:EFNM}
\subsection{Operator Learning Formulation}
The data used in this work are obtained from Unsteady Reynolds-Averaged Navier-Stokes (URANS) simulations under chorochronic conditions, which enforce phase-shifted periodicity across blade passages. 
Let $p_s(x,y,t;\boldsymbol{\kappa})$ denote the restriction of the three-dimensional pressure field onto the rotor blade surface, where $\boldsymbol{\kappa} \in \mathbb{R}^4$ represents the operating conditions.
Under chorochronic conditions, the solution is periodic, and the parietal pressure can be expressed as a periodic function of time,
\begin{equation}
p_s(x,y,t;\boldsymbol{\kappa}) = p_s(x,y,t+T;\boldsymbol{\kappa}).
\end{equation}
Introducing the phase variable $t \bmod T$, subsequently denoted by $t \in [0,T[$ throughout the remainder of this work, the problem reduces to learning
\begin{equation}
\mathcal{G} : (t, \boldsymbol{\kappa}) \mapsto p_s(\cdot,\cdot, t,\boldsymbol{\kappa}).
\end{equation}
This mapping represents a restriction of the Navier-Stokes evolution operator to a periodic orbit projected onto the blade surface. The learning task is therefore to approximate a nonlinear map from a finite-dimensional parameter space $(t, \boldsymbol{\kappa})$ to a function defined on the blade surface. This perspective naturally motivates the use of operator-learning methods.
\subsection{Neural Operator} 
Classical neural networks are designed to approximate mappings between finite-dimensional Euclidean spaces. In contrast, neural operators are constructed to approximate mappings between infinite-dimensional function spaces. Such approaches have recently demonstrated significant potential in scientific computing \cite{azizzadenesheli2024neural, kovachki2024operator}, particularly in the context of PDEs \cite{wang2021learning, kovachki2023neural, duruisseaux2025fourier}, where the underlying problems are naturally formulated in functional settings. In this framework, the learned operator typically maps input quantities (such as spatially varying coefficients, boundary conditions, or initial conditions) to the corresponding PDE solution.

By operating directly on function spaces, neural operators can mitigate limitations associated with fixed discretizations of the computational domain, thereby enabling discretization-invariant or mesh-independent representations \cite{li2020fourier, kovachki2023neural}.

Formally, let $\mathcal{A}$ and $\mathcal{U}$ denote two function spaces. The objective of operator learning is to approximate a target operator $\mathcal{G} : \mathcal{A} \to \mathcal{U}$ by a parametric map $\mathcal{G}_\theta$, where $\theta$ belongs to a finite-dimensional parameter space. The learning task consists in identifying an optimal parameter set $\theta^\ast$ such that $\mathcal{G}_{\theta^\ast}$ provides an accurate approximation of $\mathcal{G}$ in an appropriate functional norm.

\paragraph{Neural operator motivations.}
In the present context, adopting an operator-learning perspective offers two main advantages.

First, it enables continuous-in-time modeling. Sequential approaches, such as the Temporal Prediction Model (TPM) \cite{dominique2026reduced}, approximate the time-evolution operator through iterative, stepwise predictions. This procedure inherently induces error accumulation and introduces artificial temporal discontinuities due to the discrete-time formulation, as the temporal dynamics are learned through successive time steps. Such discontinuities are particularly detrimental in applications involving moving shocks in the pressure field, where temporal coherence is essential. Moreover, these artifacts can significantly distort the temporal spectrum, thereby affecting derived quantities such as the GAF. In contrast, a neural operator directly approximates the time-evolution operator in a single evaluation, with time (or phase) provided as a continuous input. This formulation enforces a smooth dependency of the predicted pressure field on time and promotes consistent interpolation across the temporal domain. 
A detailed analysis of the temporal coherence achieved by the sequential TPM and the neural-operator-based periodic FNM is presented in Section \ref{section:results:fourier-analysis}.

Second, it facilitates periodicity modeling. Sequential models generally struggle to capture periodic behavior, as periodicity is not inherently encoded in their formulation and must instead be encouraged through additional constraints in the loss function. By contrast, neural operator models can incorporate time as a phase variable, thereby embedding periodic structure directly into the input space. Accurately capturing pressure periodicity is critical, since even small phase errors can introduce discontinuities in the reconstructed signal, leading to spurious spectral components and degradation of quantities such as the GAF.
\subsection{Fourier Neural Operator}
\begin{figure}[htp]
    \centering
    \includegraphics[width=0.50\linewidth]{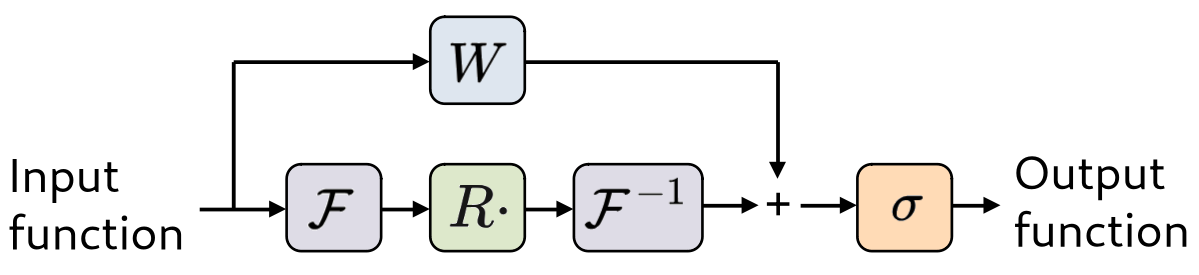}
    \caption{FNO layer architecture}
    \label{fig:fno-layer}
\end{figure}

The Fourier Neural Operator (FNO) is a neural operator architecture \cite{li2020fourier, duruisseaux2025fourier, kovachki2021universal} that parameterizes mappings between function spaces using spectral convolutions. Let $v_i \in \mathbb{R}^{C \times W \times H}$ denote an intermediate field representation at layer $i$, where $C$ is the number of channels and $(W, H)$ are the spatial dimensions. A FNO layer is defined as
\begin{equation}
v_{i+1} = \mathcal{T}(v_i) = \psi \left( \mathcal{W} v_i + \mathcal{K}(v_i) \right),
\end{equation}
where $\mathcal{W}$ is a learned local transformation (typically implemented as a convolution, a pointwise linear transformation or a Multi-Layer Perceptron (MLP) acting along the channel dimension) and $\psi$ is a nonlinear activation function applied pointwise. 
The nonlocal component $\mathcal{K}$ corresponds to a Fourier integral operator (also known as spectral convolution) and is defined as
\begin{equation}
    \mathcal{K}(v) = \mathcal{F}^{-1} \big( R \cdot \mathcal{F}(v) \big),
\end{equation}
where $\mathcal{F}$ denotes the Fourier transform applied over the spatial dimensions. On uniform grids, $\mathcal{F}$ can be efficiently implemented via the fast Fourier transform. The operator $R$ is a learned complex-valued linear mapping acting on a truncated set of Fourier modes, typically restricted to $|\zeta| \leq \zeta_{\max}$. This truncation reduces the computational complexity of the spectral multiplication to $\mathcal{O}(k_{\max})$, as opposed to scaling with the full spatial resolution. Beyond computational considerations, mode truncation introduces a physically motivated inductive bias: for many PDE-governed systems, the dominant large-scale structure of the solution is encoded in low-frequency modes \cite{qin2024toward}, whereas high-frequency modes primarily capture localized fluctuations. Consequently, restricting the spectral convolution to low-frequency components promotes efficient learning of global field structures. \\
However, it is important to note that the truncation of high-frequency modes in the spectral convolution does not prevent the model from capturing fine-scale features. Such components can be recovered through the combined effect of nonlinear activation functions and local transformations $W$, which reintroduce high-frequency content and enrich the representation beyond the truncated spectral support. \\
The overall FNO architecture (see Figure \ref{fig:fno-layer}) is expressed as
\begin{equation}
    \mathcal{G}_\theta = \mathcal{Q} \circ \mathcal{T}_L \circ \cdots \circ \mathcal{T}_1 \circ \mathcal{P},
\end{equation}
where $\lbrace\mathcal{T}_i\rbrace_{i=1}^L$ are FNO layers, and $\mathcal{P}$ and $\mathcal{Q}$ denote lifting and projection operators, respectively. These operators act pointwise to map the input function into a higher-dimensional feature space and subsequently project it back to the target space. In practice, they are implemented as linear transformations or MLPs applied along the channel dimension.

\paragraph{Functional representation.} It is important to note that, in this framework, functions are represented through a truncated set of low-frequency Fourier modes. Given these spectral coefficients, the corresponding field can be reconstructed on a regular grid via the inverse Fourier transform. In the present setting, the computational grid is fixed with resolution $(W, H)$, such that functions are effectively identified with their discrete values on this grid. A key advantage of the FNO formulation, however, lies in its ability to generalize across discretizations. Since the model learns mappings at the level of functions rather than discrete representations, it can be evaluated on grids of varying resolutions while maintaining consistency and coherence in the predicted fields.

\paragraph{Advantages of FNO.}
Overall, the effectiveness of the FNO lies in its capacity to represent nonlinear features across multiple spatial scales. Spectral convolutions act as global operators that capture long-range correlations, while the local transformations within each FNO layer enable the resolution of small-scale structures and high-frequency variations \cite{qin2024toward, liu2024mitigating}. This property is particularly relevant for pressure fields, which typically exhibit a multiscale structure characterized by both smooth regions and sharp gradients, such as shocks.

With respect to the GAF prediction objective, the ability of the Fourier Neural Operator to accurately reconstruct spatial features of the pressure field (e.g., shocks) is also critical for capturing their temporal evolution, thereby enabling a more continuous representation of their motion.
This, in turn, contributes to improved GAF accuracy, as the error in the GAF is directly linked to the temporal spectral error of the predicted pressure field.

From a theoretical standpoint, a universal approximation result has been established for the FNO, demonstrating that it can approximate a broad class of sufficiently regular operators with arbitrary accuracy \cite{kovachki2021universal}.

\paragraph{FNO and finite dimensional input.} The classical FNO framework is designed to approximate mappings between functions defined on continuous function spaces. In the present study, however, the objective is to map a finite-dimensional vector (containing simulation parameters and time) to a pressure field defined over a continuous domain. This corresponds to learning an operator from a finite-dimensional space to an infinite-dimensional function space. A straightforward approach to reconcile this discrepancy is to embed the input parameters into the FNO framework by representing them as constant functions over the spatial domain. However, such a representation is degenerate in the spectral domain, as constant functions correspond to a Dirac delta function concentrated at zero frequency. Given that the standard FNO architecture relies on pointwise functions, local convolutions, and spectral convolutions (implemented as pointwise multiplications in Fourier space), this spectral degeneracy severely restricts the expressive capacity of the model. In particular, it implies that constant inputs are mapped to constant outputs, thereby limiting the model’s ability to generate pressure fields exhibiting rich spatial frequency content. In practice, minor deviations from strict constancy may arise due to numerical artifacts such as padding in convolutional layers together with the effect of nonlinear activation functions, which can introduce perturbations and broaden the resulting frequency spectrum. Nevertheless, these effects are incidental and do not fundamentally resolve the mismatch between the input representation and the desired output complexity. Consequently, the standard FNO formulation is not naturally suited to the problem under consideration. To address this limitation, Section \ref{sec:perFNM} introduces the Fourier Neural Mapping (FNM) \cite{huang2024operator}, an extension of the FNO framework specifically designed to handle mappings from finite-dimensional inputs to continuous function spaces.

\subsection{Periodic Fourier Neural Mapping Architecture} \label{sec:perFNM}
\paragraph{Fourier Neural Mapping.}
The Fourier Neural Mapping, introduced in \cite{huang2024operator}, generalizes the classical FNO framework to accommodate finite-dimensional input spaces, which is particularly suited to the present setting where the inputs consist of a finite-dimensional parameter vector while the output remains a discrete representation of the continuous pressure field. A universal approximation theorem established in \cite{huang2024operator} provides a rigorous theoretical foundation for this architecture.

\begin{figure}[htp]
    \centering
    \includegraphics[width=0.55\linewidth]{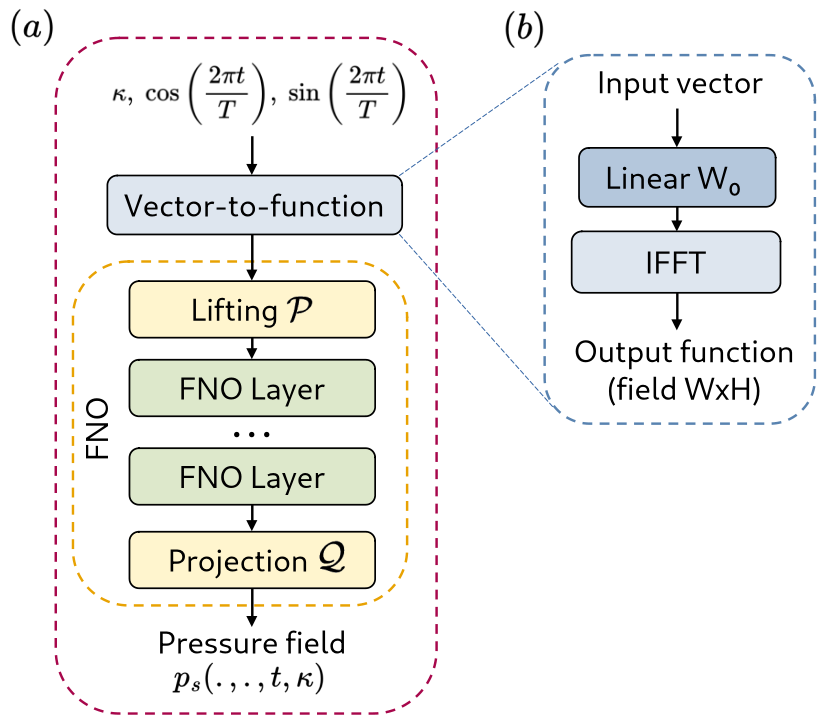}
    \caption{p-FNM architecture (a) and V2F block (b)}
    \label{fig:fnm-arch}
\end{figure}

In practice, the FNM architecture considered here consists of a vector-to-function (V2F) mapping followed by a standard function-to-function FNO (see Figure \ref{fig:fnm-arch}). This construction preserves the key advantages of the FNO (namely, the ability to capture global spatial dependencies and multiscale structures) while adapting it to the parametric setting considered here. The $\text{V2F} : \mathbb{R}^{d+2} \rightarrow L^2(S)$ component, where $S$ denotes the blade surface, applies a parametric transformation (e.g., a linear mapping or a multilayer perceptron) to the input vector, producing a set of coefficients interpreted as truncated low-frequency Fourier modes of a function. The corresponding field is then reconstructed via an inverse Fourier transform. This V2F stage can be viewed as generating an initial approximation of the pressure field conditioned on the input parameters, which is subsequently refined by the FNO layers. The hyperparameters used for the FNM architecture in this study are detailed in Appendix \ref{appendix:fnm-hyperparams}.

\paragraph{Embedded periodicity knowledge.}
To explicitly incorporate periodic structure into the model, the temporal input $t \in [0, T[$ is not provided in its raw form. Instead, a periodic encoding is employed:
\begin{equation}
t \;\mapsto\;
\left( \cos\left(\frac{2\pi t}{T}\right),\; \sin\left(\frac{2\pi t}{T}\right) \right). \label{eq:periodic-emb}
\end{equation}
Providing $t$ directly would require the model to infer periodicity solely from data, in particular by learning the similarity between states at $t=0$ and $t \approx T$, despite the presence of an artificial discontinuity in the input space. In contrast, this encoding maps time onto a continuous representation on the unit circle, thereby eliminating boundary discontinuities and explicitly embedding periodic structure. This facilitates the learning of smooth periodic relationships between the temporal input and the predicted pressure field.
\subsection{Loss \& Training Setup}

All models were trained using the SOAP optimizer \cite{vyas2025soap} with a learning-rate schedule consisting of an initial linear warm-up phase followed by cosine annealing \cite{loshchilov2017sgdr}. Training was conducted on a single NVIDIA A100 GPU. The validation loss was monitored throughout training, and early stopping was employed once no further improvement was observed. The learning rate and batch size were selected to optimize validation performance while satisfying GPU memory constraints.


In this study, the early-stopping patience parameter (defined as the number of consecutive epochs without improvement in validation loss tolerated before terminating training) was fixed to 15 epochs. The SOAP optimizer was selected following comparative experiments with AdamW, as it consistently yielded comparable or improved predictive performance with limited additional computational cost. SOAP relies on a second-order preconditioning strategy that approximates the Hessian of the loss landscape, leading to improved convergence properties and optimization accuracy. In contrast, AdamW is based solely on first-order gradient information.

Model parameters were optimized using the Mean Squared Error (MSE) loss, defined as
\begin{equation}
    \text{MSE}(p^{\mathrm{true}}, p^{\mathrm{pred}}) = \frac{1}{N T W H} \sum_{b=1}^{N} \sum_{t=1}^{T} \sum_{w=1}^{W} \sum_{h=1}^{H} \left| p_{b,t,w,h}^{\mathrm{true}} - p_{b,t,w,h}^{\mathrm{pred}} \right|^2,
\end{equation}
where $N$ denotes the batch size, and $p^{\mathrm{true}}$, $p^{\mathrm{pred}} \in \mathbb{R}^{N \times T \times W \times H}$ represent the ground-truth and predicted pressure fields, respectively.

\section{Experiment and Results} \label{section:results}
All results reported in this section correspond to the mean and variance of the evaluation metrics computed over 30 independent training runs, differing only by the random initialization of the trainable model parameters. This procedure accounts for the variability arising from the stochastic nature of neural-network training. To ensure a fair comparison, all models were designed to have a comparable number of trainable parameters, approximately ($1.15 \times 10^6$). Detailed model sizes are reported in Appendix \ref{appendix:models-size}.
\subsection{Evaluation Metrics} \label{section:metrics}
\paragraph{Metrics on pressure field.}
Model performance on the pressure prediction task was assessed using a comprehensive set of metrics, including the Mean Absolute Error (MAE), Mean Absolute Percentage Error (MAPE), and Mean Maximum Error (MME). They are given by:
\begin{equation}
\begin{aligned}
    &\text{MAE}(p^{\mathrm{true}}, p^{\mathrm{pred}}) = \frac{1}{N T W H} \sum_{b=1}^{N} \sum_{t=1}^{T} \sum_{w=1}^{W} \sum_{h=1}^{H} \left| p_{b,t,w,h}^{\mathrm{true}} - p_{b,t,w,h}^{\mathrm{pred}} \right|, \\
    &\text{MAPE}(p^{\mathrm{true}}, p^{\mathrm{pred}}) = \frac{1}{N T W H} \sum_{b=1}^{N} \sum_{t=1}^{T} \sum_{w=1}^{W} \sum_{h=1}^{H} \frac{\left| p_{b,t,w,h}^{\mathrm{true}} - p_{b,t,w,h}^{\mathrm{pred}} \right|}{\left|p_{b,t,w,h}^{\mathrm{true}}\right|}, \\
    &\text{MME}(p^{\mathrm{true}}, p^{\mathrm{pred}}) = \frac{1}{N} \sum_{b=1}^{N} \; \max_{t, w, h} \left(| p_{b}^{\mathrm{true}} - p_{b}^{\mathrm{pred}} | \right), \;\; \forall (t, w, h) \in [1\mathinner {\ldotp \ldotp} T]\times[1\mathinner {\ldotp \ldotp} W] \times[1\mathinner {\ldotp \ldotp} H].
\end{aligned}
\end{equation}
These metrics provide complementary information regarding the prediction quality. The MAE measures the average absolute prediction error in physical units (kPa), thereby providing a direct interpretation of the typical pressure deviation. The MAPE constitutes a relative counterpart to the MAE and quantifies the prediction error relative to the magnitude of the target pressure values. Finally, the MME characterizes the average worst-case prediction error for each sample, providing an estimate of the upper bound of the absolute pressure error.\\
Additionally, to evaluate the ability of the predictive models to reproduce not only the mean behavior but also the variability of the pressure fields, we consider the coefficient of determination, denoted by $R^2$. The $R^2$ score quantifies the proportion of variance in the target data explained by the model relative to a constant baseline predictor. A value of $R^2 = 1$ corresponds to perfect predictions, whereas squared values in the range $[0, 1[$ indicate partial explanation of the target variance. Conversely, $R^2 < 0$ indicates that the model performs worse than a predictor based solely on the mean of the target data.
The metric is computed independently for each pressure-field sample and subsequently averaged over the batch. This sample-wise formulation prevents the evaluation from being dominated by the most energetic samples, which would otherwise occur if the $R^2$ score were computed globally over the entire dataset. The sample-wise coefficient of determination is defined as
\begin{equation}
    \begin{aligned}
        &R_b^2 = 1 - \frac{\sum\limits_{t=1}^{T} \sum\limits_{w=1}^{W} \sum\limits_{h=1}^{H} \left(p_{b,t,w,h}^{\mathrm{true}} - p_{b,t,w,h}^{\mathrm{pred}}\right)^2}{\sum\limits_{t=1}^{T} \sum\limits_{w=1}^{W} \sum\limits_{h=1}^{H} \left(p_{b,t,w,h}^{\mathrm{true}} - \overline{p}_{b}^{\mathrm{true}}\right)^2},
    \end{aligned}
\end{equation}
where 
\begin{equation}
    \begin{aligned}
        \overline{p}_{b}^{\mathrm{true}} = \frac{1}{T W H} \sum_{t=1}^{T} \sum_{w=1}^{W} \sum_{h=1}^{H} p_{t,w,h}^{\mathrm{true}}, 
    \end{aligned}
\end{equation}
denotes the mean pressure associated with sample $b$. The final reported $R^2$ metric is obtained by averaging over all samples in the batch:
\begin{equation}
    \begin{aligned}
        R^2 = \frac{1}{N} \sum_{b=1}^{N} R_b.
    \end{aligned}
\end{equation}

\paragraph{Metrics on GAF norm.}
The accuracy of the predicted GAF magnitude $\rho$ is additionally evaluated using the Mean Absolute Percentage Error ($\mathrm{MAPE}_\rho$) and the coefficient of determination ($R_\rho^2$), defined as
\begin{equation}
\begin{aligned}
    &\text{MAPE}_\rho(\rho^{\mathrm{true}}, \rho^{\mathrm{pred}}) = \frac{1}{N} \sum_{b=1}^{N} \frac{\left| \rho_{b}^{\mathrm{true}} - \rho_{b}^{\mathrm{pred}} \right|}{\rho_{b}^{\mathrm{true}}},\\
    &R^2_\rho(\rho^{\mathrm{true}}, \rho^{\mathrm{pred}}) = 1 - \frac{\sum\limits_{b=1}^N \left|\rho_{b}^{\mathrm{true}} - \rho_{b}^{\mathrm{pred}} \right|^2}{\sum\limits_{b=1}^N \left|\rho_{b}^{\mathrm{true}} - \overline{\rho}_{b}^{\mathrm{true}} \right|^2}, \;\;\; \text{with} \;\; \overline{\rho}_{b}^{\mathrm{true}} = \frac{1}{N} \sum\limits_{b=1}^N \rho_{b}^{\mathrm{true}}.
\end{aligned}
\end{equation}
\paragraph{Metrics on GAF phase.}
Because phase variables are inherently periodic, conventional arithmetic statistics are not directly suitable for evaluating phase prediction errors. Instead, metrics specifically designed for circular quantities must be employed, as in Directional Statistics (see, e.g., \cite{mardia2009directional, ley2017modern}). To assess the accuracy of the predicted GAF phase $\phi$, we consider the weighted Root Mean Square phase Error ($\mathrm{wRMSE}_\phi$), defined as
\begin{equation}
\begin{aligned}
    &\mathrm{wRMSE}_\phi(\phi^{\mathrm{true}}, \phi^{\mathrm{pred}}) = \sqrt{\frac{\sum\limits_{b=1}^N \rho_b^{\mathrm{true}} \Delta\phi_b^2}{\sum\limits_{b=1}^N \rho_b^{\mathrm{true}}}},\\
    &\text{where} \;\;\Delta\phi_b = \arg\left(e^{i(\phi_b^{\mathrm{true}} - \phi_b^{\mathrm{pred}})}\right) \in [-\pi, \pi], \;\;\forall b\in [1\mathinner {\ldotp \ldotp}N].
\end{aligned}
\end{equation}

$\mathrm{wRMSE}_\phi$, expressed in radians and bounded in $[0,\pi]$, is weighted by the GAF magnitude $\rho^{\mathrm{true}}$ to assign greater importance to high-energy samples, for which phase accuracy is most physically relevant. Conversely, when the GAF magnitude is small, phase errors become less meaningful.
\subsection{Sequential Model Baseline} \label{section:baseline}
\paragraph{Analysis of the TPM architecture.} 
The TPM model introduced in \cite{dominique2026reduced} consists of two main components. First, a VAE is trained on static pressure fields to learn a compact latent representation of the pressure distribution over the rotor blade surface. Subsequently, a RNN is trained to predict temporal sequences of latent vectors conditioned on the operating parameters. These latent representations are then decoded into pressure fields using the pretrained VAE decoder. In this sequential framework, temporal dynamics are modeled recursively through the hidden memory state of the RNN. Consequently, the prediction at a given time step depends on previously predicted latent states. However, due to both the recurrent nature of the model and the use of an intermediate latent-space representation, temporal continuity is not explicitly enforced. This may lead to reduced temporal coherence in the reconstructed pressure fields. The temporal organization of the VAE latent space is further analyzed in Appendix \ref{sec:appendix:vae-latent-analysis}, while the temporal prediction capabilities of the VAE, TPM, and p-FNM models are discussed in Section \ref{section:results:fourier-analysis}.
\paragraph{Training with SOAP.} 

To ensure a fair comparison and to isolate the contribution of the different components introduced in the proposed p-FNM framework, which is trained using the SOAP optimizer, we first compare the original TPM baseline with the same architecture retrained using the SOAP optimizer, hereafter denoted TPM-SOAP. This comparison allows the effect of the optimization strategy to be assessed independently of any architectural modification. The corresponding experiments, detailed in Appendix \ref{appendix:tpm-soap}, show that SOAP consistently improves both pressure-field and GAF predictions while reducing predictive uncertainty for the TPM model and its VAE component across all considered datasets.
Subsequently, in Sections \ref{section:results:pressure} and \ref{section:results:gaf}, the p-FNM is compared against both the original TPM \cite{dominique2026reduced} and TPM-SOAP. This experimental design makes it possible to distinguish the gains attributable to the SOAP optimization from those arising from the proposed continuous-time formulation and the p-FNM architecture itself.

\subsection{p-FNM Pressure Prediction Results}\label{section:results:pressure}
\begin{figure}[!htp]
    \centering    
    \begin{subfigure}{0.40\textwidth}
        \centering
        \includegraphics[width=\linewidth]{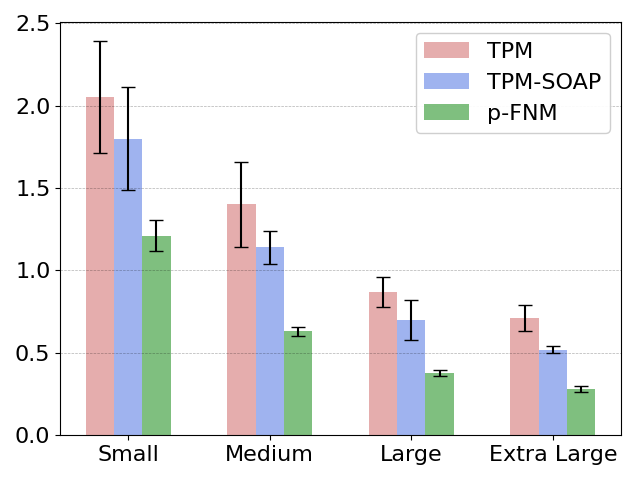}
        \caption{Mean Absolute Error [kPa] of the predicted pressure. Lower is better.}
        \label{fig:press-mae}
    \end{subfigure}
    \hspace{7mm}
    \begin{subfigure}{0.40\textwidth}
        \centering
        \includegraphics[width=\linewidth]{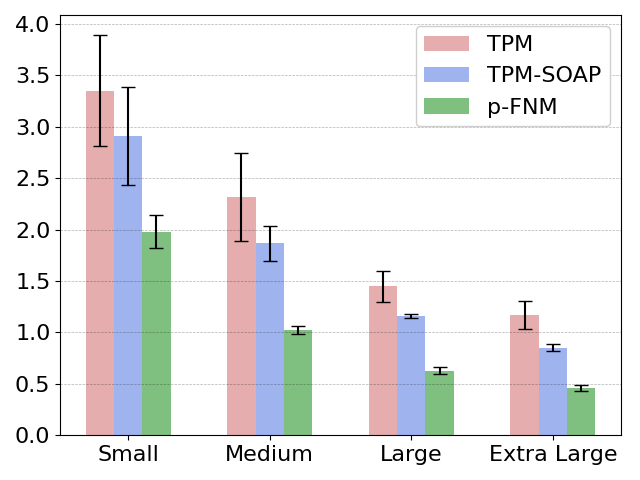}
        \caption{Mean Absolute Percentage Error [\%] of the predicted pressure. Lower is better.}
        \label{fig:press-mape}
    \end{subfigure}
    \par\medskip
    \begin{subfigure}{0.40\textwidth}
        \centering
        \includegraphics[width=\linewidth]{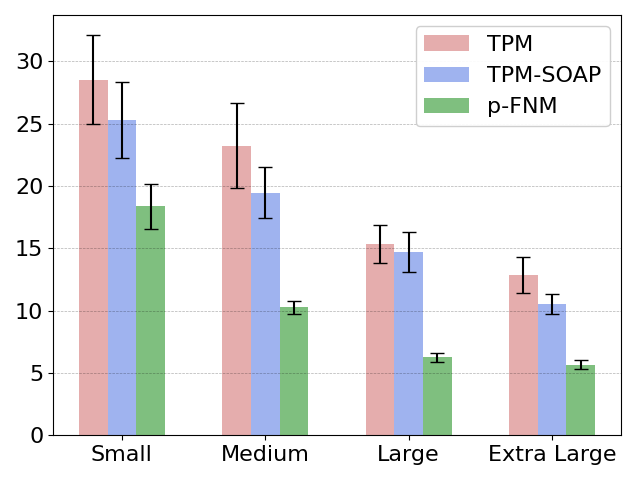}
        \caption{Mean Max Error [kPa] of the predicted pressure. Lower is better.}
        \label{fig:press-mme}
    \end{subfigure}
    \hspace{7mm}
    \begin{subfigure}{0.40\textwidth}
        \centering
        \includegraphics[width=\linewidth]{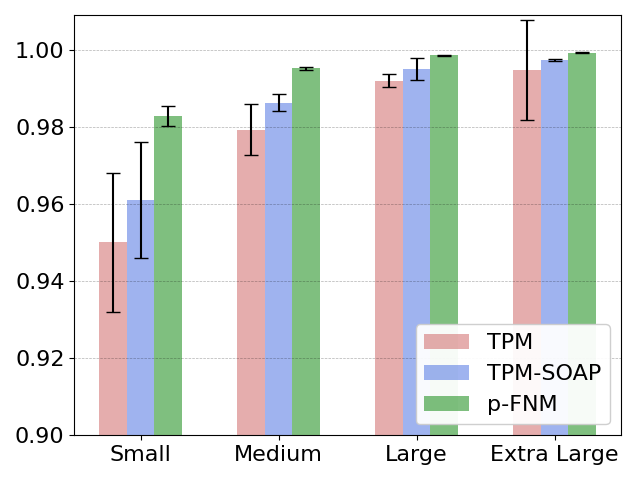}
        \caption{$R^2$ score of the predicted pressure. Higher is better.}
        \label{fig:press-r2}
    \end{subfigure}
    \caption{Comparison of the TPM, TPM-SOAP and p-FNM models with respect to the pressure prediction metrics across the different training datasets. The metrics are evaluated on the test dataset. The black error bars represent the metric uncertainty quantified by the $\pm 1.96\sigma$ confidence interval computed over 30 independent training runs.}
    \label{fig:press-metrics}
\end{figure}

The performances of the p-FNM model compared with the TPM and TPM-SOAP models are reported in Figures \ref{fig:press-metrics}. Detailed quantitative results are reported in Tables \ref{tab:vae-vs-vae-soap-all} and \ref{tab:fnm-metrics-detailed} of Appendices \ref{appendix:tpm-soap} and \ref{appendix:detail-res}, respectively. Figures \ref{fig:press-mae}, \ref{fig:press-mape}, \ref{fig:press-mme}, and \ref{fig:press-r2} present the MAE, MAPE, MME, and $R^2$ metrics respectively, for the TPM, TPM-SOAP, and p-FNM models across the different training datasets.
The results indicate that both predictive accuracy and uncertainty consistently improve with the training dataset size. As discussed in Section \ref{section:baseline}, the p-FNM systematically outperforms the TPM-SOAP, which already improves upon the original TPM baseline introduced in \cite{dominique2026reduced}.
The MAE and MAPE results demonstrate that the p-FNM achieves highly accurate mean pressure predictions, with relative errors ranging from $1.98\%$ for the smallest dataset to $0.46\%$ for the largest one. The MME results further highlight the robustness of the p-FNM approach by quantifying the average worst-case pointwise pressure prediction error. Finally, the high $R^2$ scores demonstrate that the p-FNM accurately reproduces not only the mean pressure distribution but also its variability, with values exceeding $0.98$ even for the smallest training dataset.
\paragraph{Model errors and shock-wave reconstruction.}
Tables \ref{fig:pressure-pred-sim0-t14} and \ref{fig:pressure-pred-sim1-t107} in Appendix \ref{appendix:pressure-preds} show that the largest prediction errors of the machine-learning models are primarily concentrated in the vicinity of the pressure shock waves. Due to the strong pressure gradients associated with these structures, even small shifts in localization of the predicted shock can generate large and highly sharp pressure discrepancies, thereby contributing significantly to the increase of the MME metric.
The predominance of errors near shock regions can also be attributed to the intrinsic difficulty of modeling moving discontinuities, particularly in low-data regimes such as the small dataset configuration. In addition, shock waves correspond to sharp discontinuities in the pressure profile and therefore contain substantial high-frequency components in the spatial spectrum. Such high-frequency structures are generally more challenging to represent with neural networks trained with a MSE loss.
For the VAE-based approach, the stochastic nature of the latent representation additionally induces a smoothing effect in the reconstructed pressure fields, which further limits the accurate representation of sharp discontinuities. More specifically, for a given latent variable $z$ and target field $x$, the minimizer of the mean squared error reconstruction loss in a classical VAE framework is given by the conditional expectation $\mathbb{E}[x \mid z]$. Consequently, the decoder tends to produce the average of all plausible reconstructions associated with the latent variable $z$, leading to blurred predictions and reduced sharpness of localized structures such as shock waves. Since the TPM relies on the pretrained VAE decoder for pressure-field reconstruction, its predictive accuracy is inherently constrained by the reconstruction capabilities of the decoder itself. 
Neural-operator-based approaches also exhibit an intrinsic bias towards low-frequency components \cite{qin2024toward}, nevertheless, the p-FNM provides significantly more accurate shock-wave reconstruction than the TPM baseline. In particular, the quality of the reconstructed shock structures improves substantially as the amount of training data increases.
Despite these improvements, the prediction of strong pressure gradients remains challenging because of the limited representation of spatial high-frequencies. Additional strategies could be employed to further enhance the prediction of strong pressure gradients by mitigating spectral bias effects. Examples include residual \cite{qin2024toward} or multiscale correction techniques \cite{YOU2026114530} specifically designed to improve the representation of high-frequency components at the expense of an increased model complexity.
\subsection{Temporal Fourier Analysis}\label{section:results:fourier-analysis}
The GAF is computed from the projected-force signal (see Section \ref{section:GAF}), more specifically through the extraction of the first harmonic of its temporal Fourier transform. Consequently, the analysis focuses on both the projected-force signal and its first temporal Fourier mode.
Errors in the GAF prediction are primarily driven by temporal inaccuracies rather than purely spatial reconstruction errors. Indeed, spatial errors in the pressure field contribute to the projected-force error only through their scalar product with the mechanical mode. However, if the spatial reconstruction error was constant in time, it would not affect the first Fourier mode of the projected-force signal and therefore would not contribute to the GAF error. Consequently, temporal fluctuations in the spatial prediction error have a substantially greater impact on GAF accuracy than the magnitude of the spatial error alone.\\
\paragraph{Projected-force analysis.} We first present in Table \ref{fig:projF-all} the predicted projected-force signals for two randomly selected simulations from the test set. The reference projected-forces are compared with the predictions obtained using the VAE, TPM, and p-FNM models trained on the different datasets.
The TPM predictions remain significantly less accurate than those produced by the p-FNM and the standalone VAE reconstructions. Despite being trained only on the spatial reconstruction of static pressure fields and therefore lacking any explicit temporal modeling, the VAE nevertheless produces relatively coherent temporal force evolutions due to its satisfactory pressure-field reconstruction capability (see Appendix \ref{sec:appendix:vae-latent-analysis}). However, the projected-force predictions generated by the VAE exhibit noticeable temporal oscillations. These fluctuations originate from the stochastic latent-space sampling mechanism, which introduces noise into the reconstructed pressure fields that is not entirely attenuated by the spatial integration used to compute the projected force.
While the TPM employs the pretrained VAE decoder to reconstruct pressure fields from the latent representations, its projected-force predictions are significantly degraded compared with those of the standalone VAE. 
Overall, the p-FNM consistently outperforms the TPM architecture for projected-force prediction. Even for the smallest training datasets, the p-FNM successfully captures the global evolution of the projected-forces, whereas the TPM fails to reproduce its main characteristics. Increasing the size of the training dataset further improves the p-FNM predictions. In contrast, the prediction quality of both the VAE and TPM appears to reach a performance plateau, with only marginal improvements observed when moving from the large to the extra-large dataset.
Finally, because temporal periodicity is explicitly embedded within the p-FNM architecture, the predicted projected-force signals naturally preserve the expected periodic behavior.

\begin{figure}[!htp]
    \centering    
    \begin{subfigure}{0.40\textwidth}
        \centering
        \includegraphics[width=\linewidth]{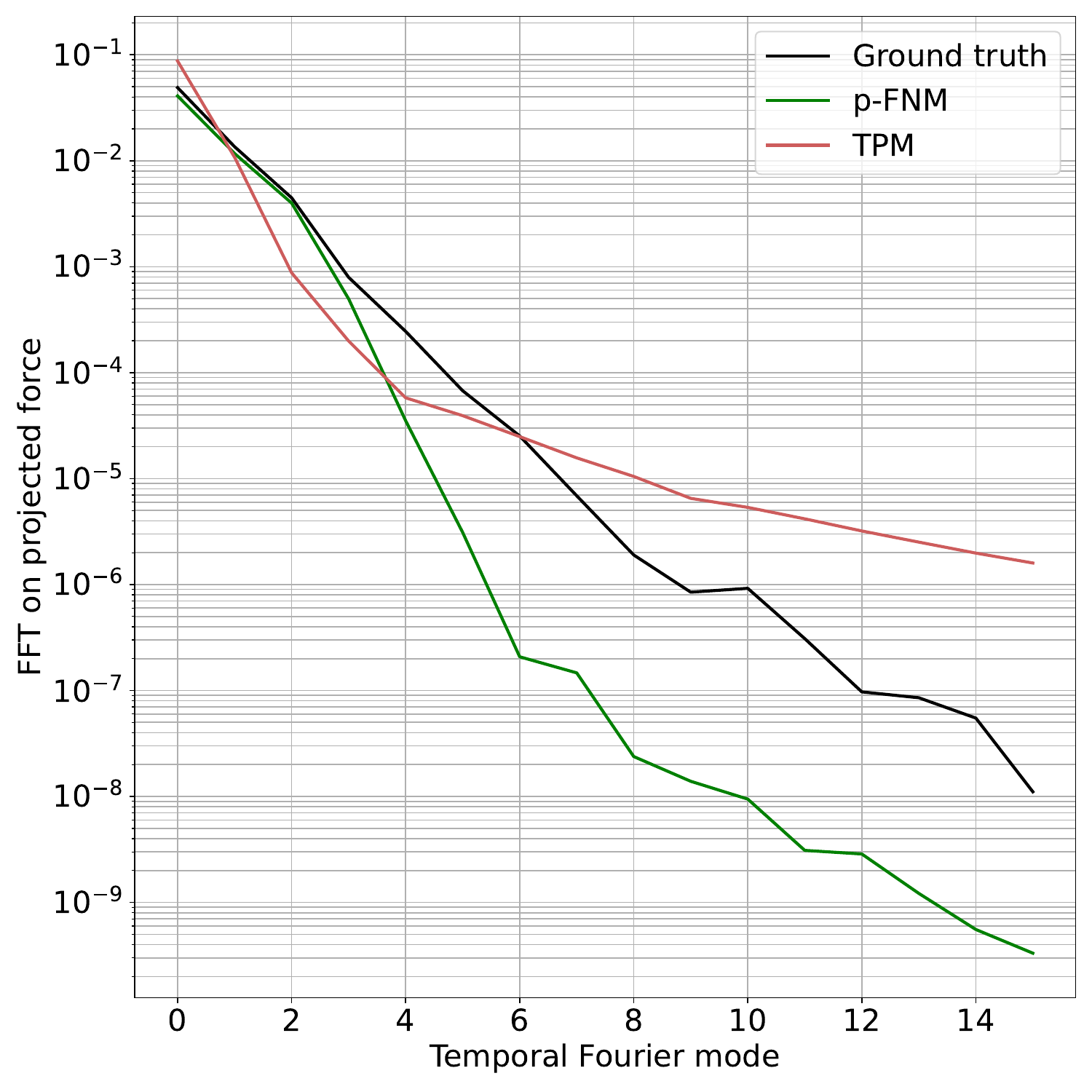}
        \caption{Training dataset: small.}
        \label{fig:projF-spectrum-S}
    \end{subfigure}
    \hspace{7mm}
    \begin{subfigure}{0.40\textwidth}
        \centering
        \includegraphics[width=\linewidth]{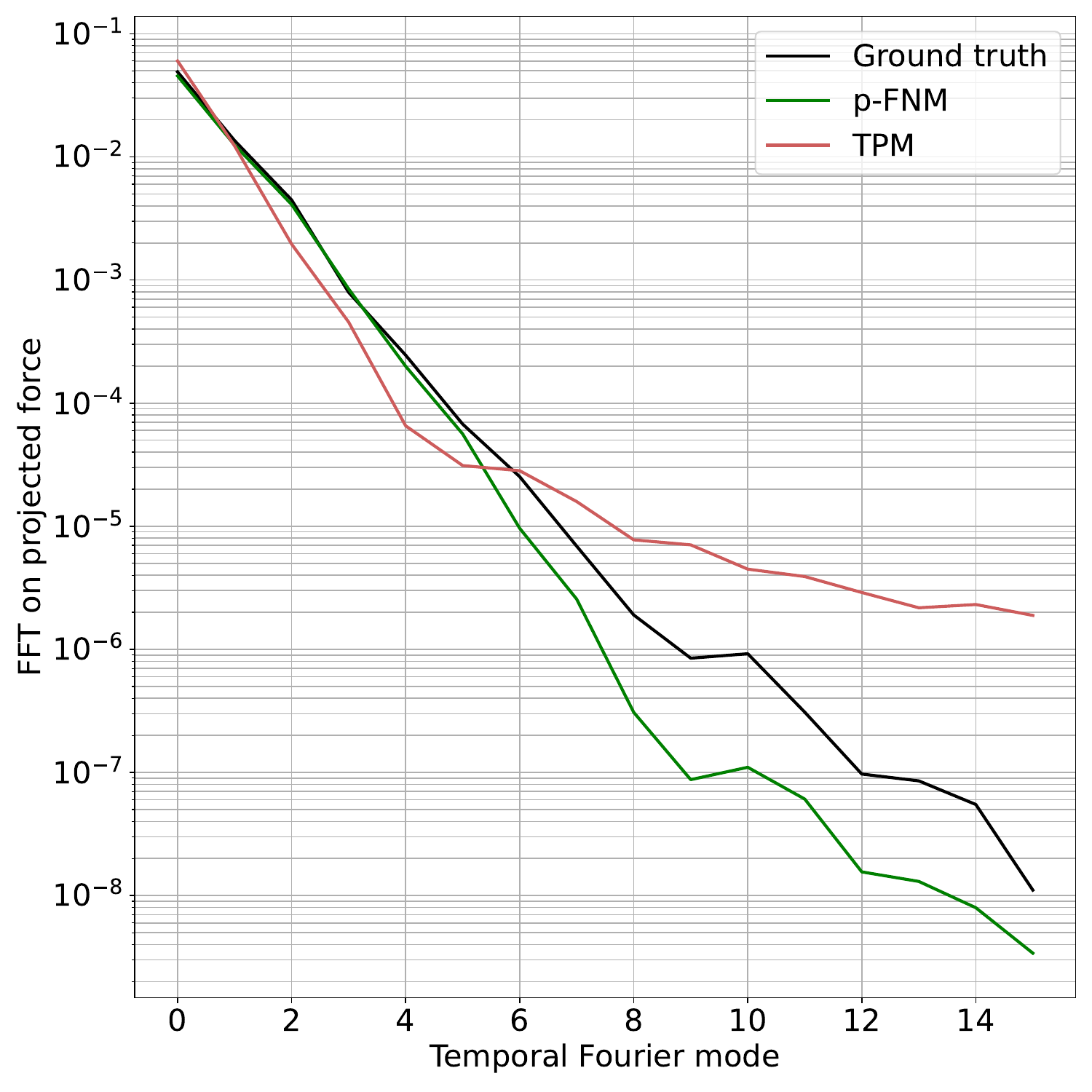}
        \caption{Training dataset: medium.}
        \label{fig:projF-spectrum-M}
    \end{subfigure}
    \par\medskip
    \begin{subfigure}{0.40\textwidth}
        \centering
        \includegraphics[width=\linewidth]{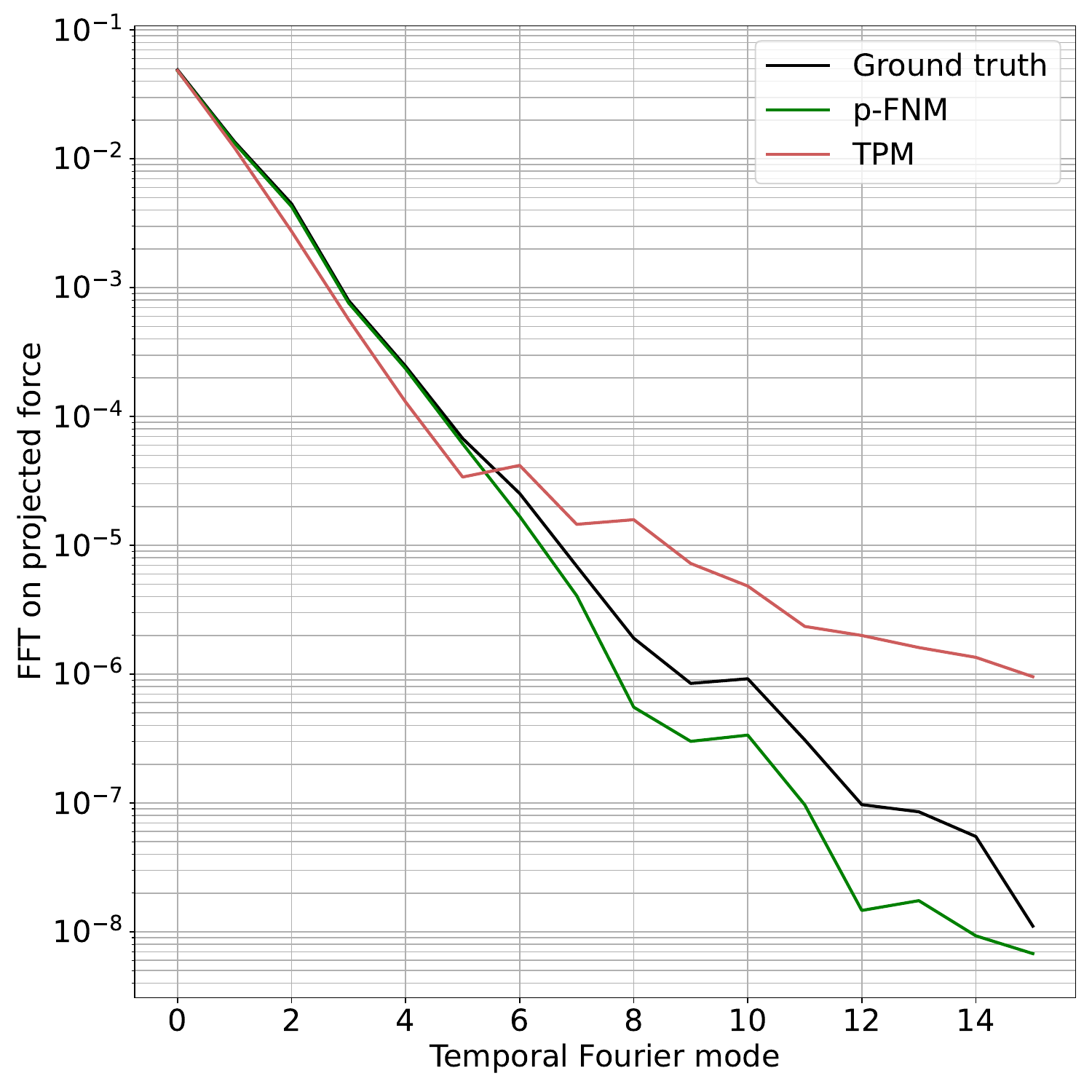}
        \caption{Training dataset: large.}
        \label{fig:projF-spectrum-L}
    \end{subfigure}
    \hspace{7mm}
    \begin{subfigure}{0.40\textwidth}
        \centering
        \includegraphics[width=\linewidth]{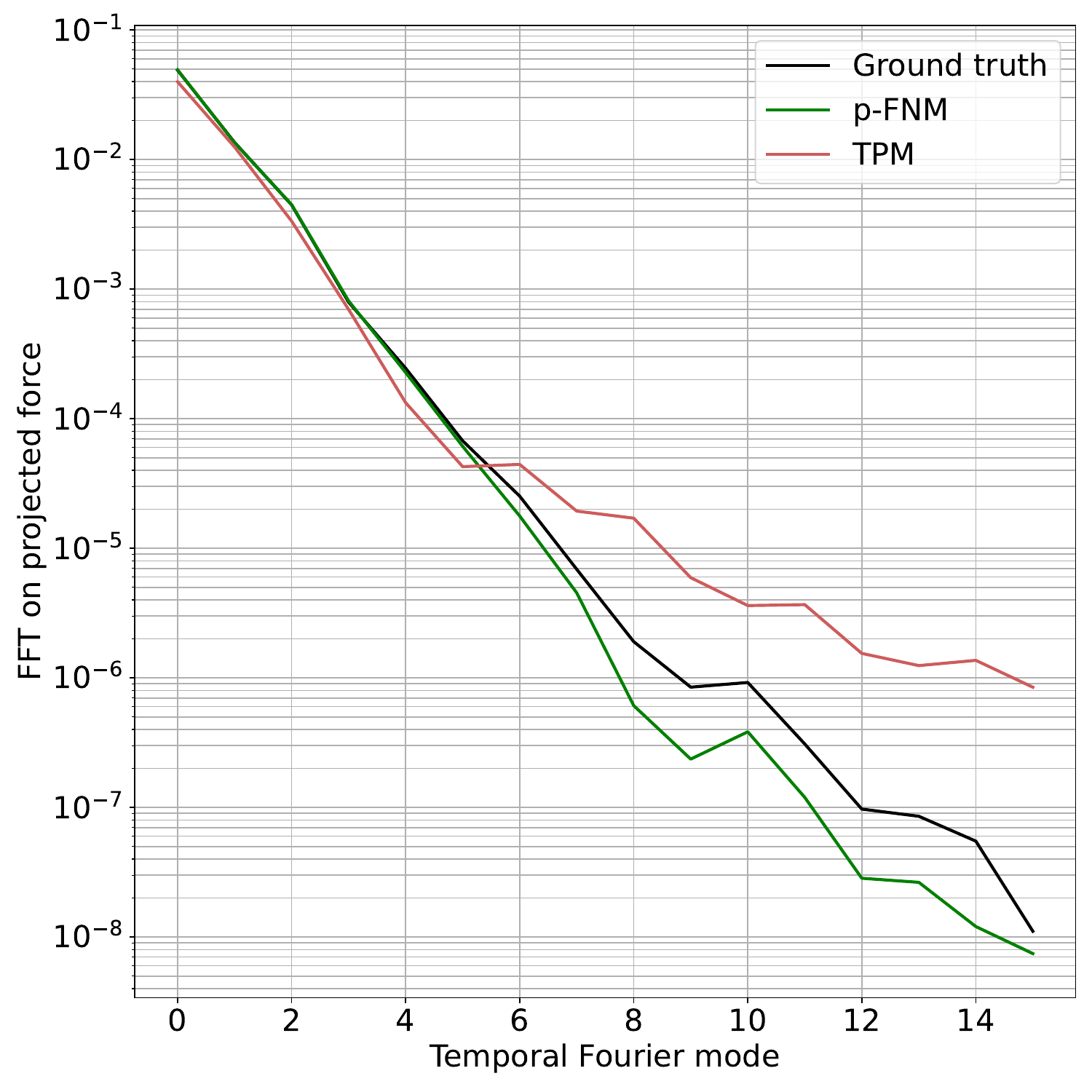}
        \caption{Training dataset: extra-large.}
        \label{fig:projF-spectrum-XL}
    \end{subfigure}
    \caption{Comparison of the TPM, and p-FNM models with respect to the temporal spectrum of the projected-force signal evaluated on the test dataset. All the Fourier harmonics of the blade-passing frequency used in the chorochronic boundary conditions are displayed.} 
    \label{fig:projF-spectrum}
\end{figure}

\begin{figure}[htbp]
  \centering
  \includegraphics[width=0.45\linewidth]{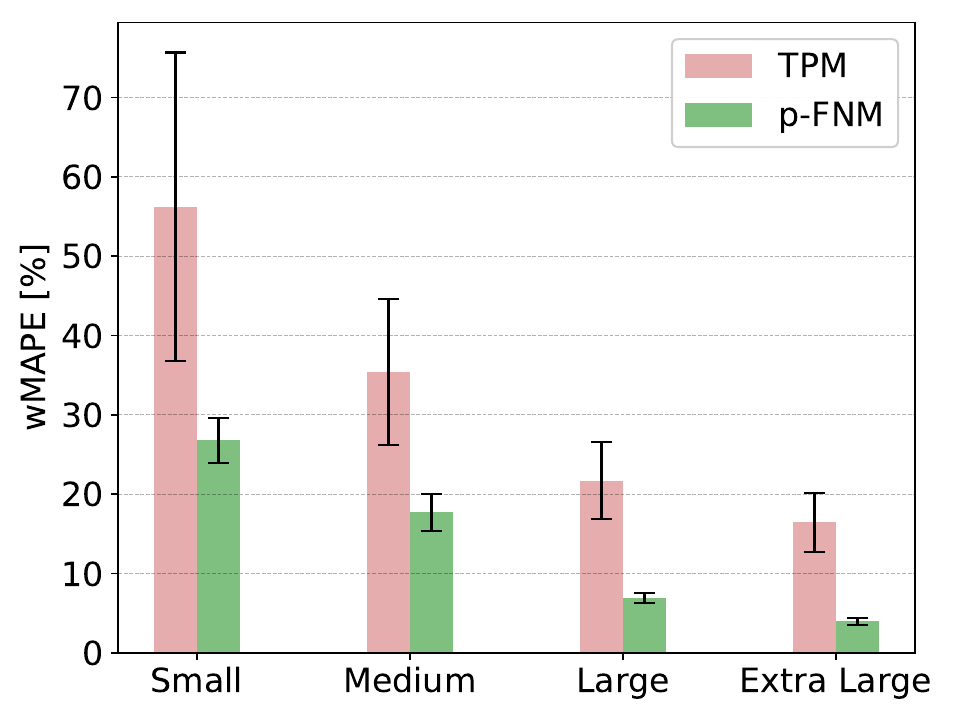}
  \caption{Weighted Mean Absolute Percentage Error (wMAPE) associated with the prediction of the first Fourier mode of the projected-force for the TPM and p-FNM models across the different training datasets. The wMAPE is evaluated on the test dataset. The black error bars represent the metric uncertainty quantified by the $\pm 1.96\sigma$ confidence interval computed over 30 independent training runs.}
  \label{fig:projF-spectrum:mode1}
\end{figure}

\begin{table}[!htp]
    \centering
    \caption{Comparison between predicted and reference projected-force signals for the VAE, TPM, and p-FNM models across the different training datasets for two representative test-cases from the test dataset. The black curve denotes the reference solution, whereas the orange, red, and green curves correspond to the VAE, TPM, and p-FNM predictions, respectively. For visual reference, each grid cell corresponds to 0.0005 units of projected force along the vertical axis.}
    \vspace*{4mm}
    \begin{tabular}{c|c|c}
        & Sim \#1 & Sim \#2 \\
        \hline
        \rotatebox[origin=c]{90}{Small} &
        \raisebox{-0.5\height}{\includegraphics[width=0.30\linewidth]{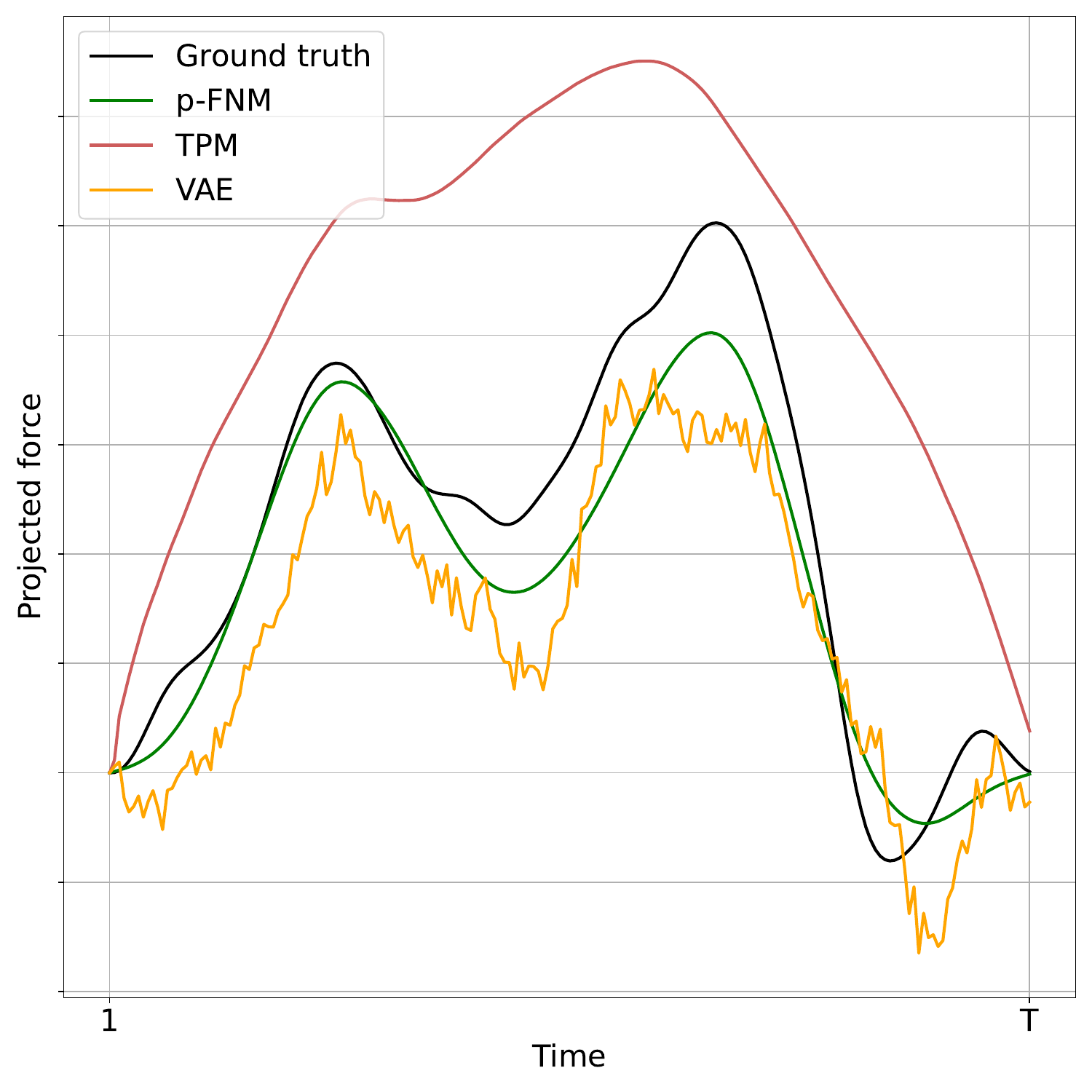}} &
        \raisebox{-0.5\height}{\includegraphics[width=0.30\linewidth]{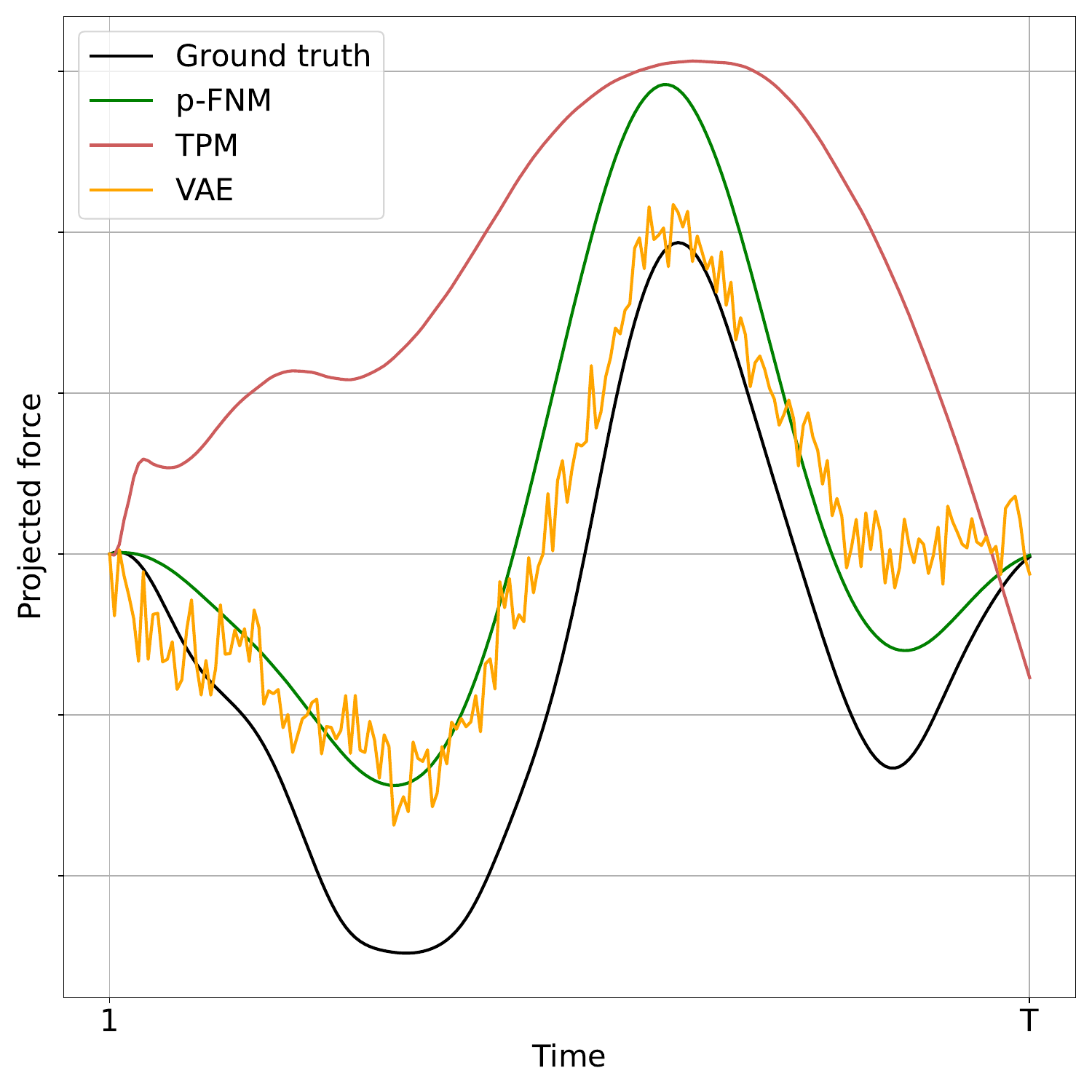}} \\
        \hline
        \rotatebox[origin=c]{90}{Medium} &
        \raisebox{-0.5\height}{\includegraphics[width=0.30\linewidth]{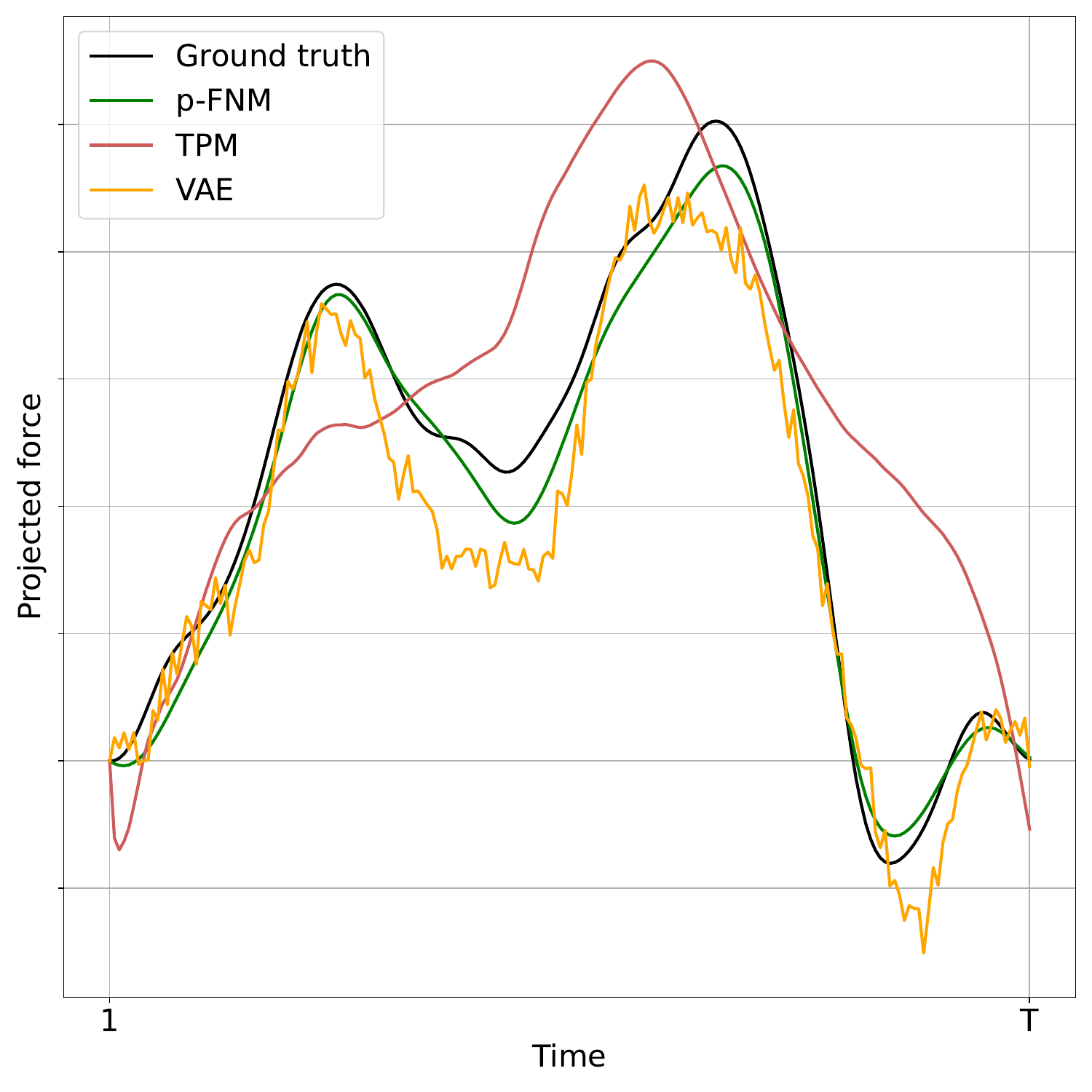}} &
        \raisebox{-0.5\height}{\includegraphics[width=0.30\linewidth]{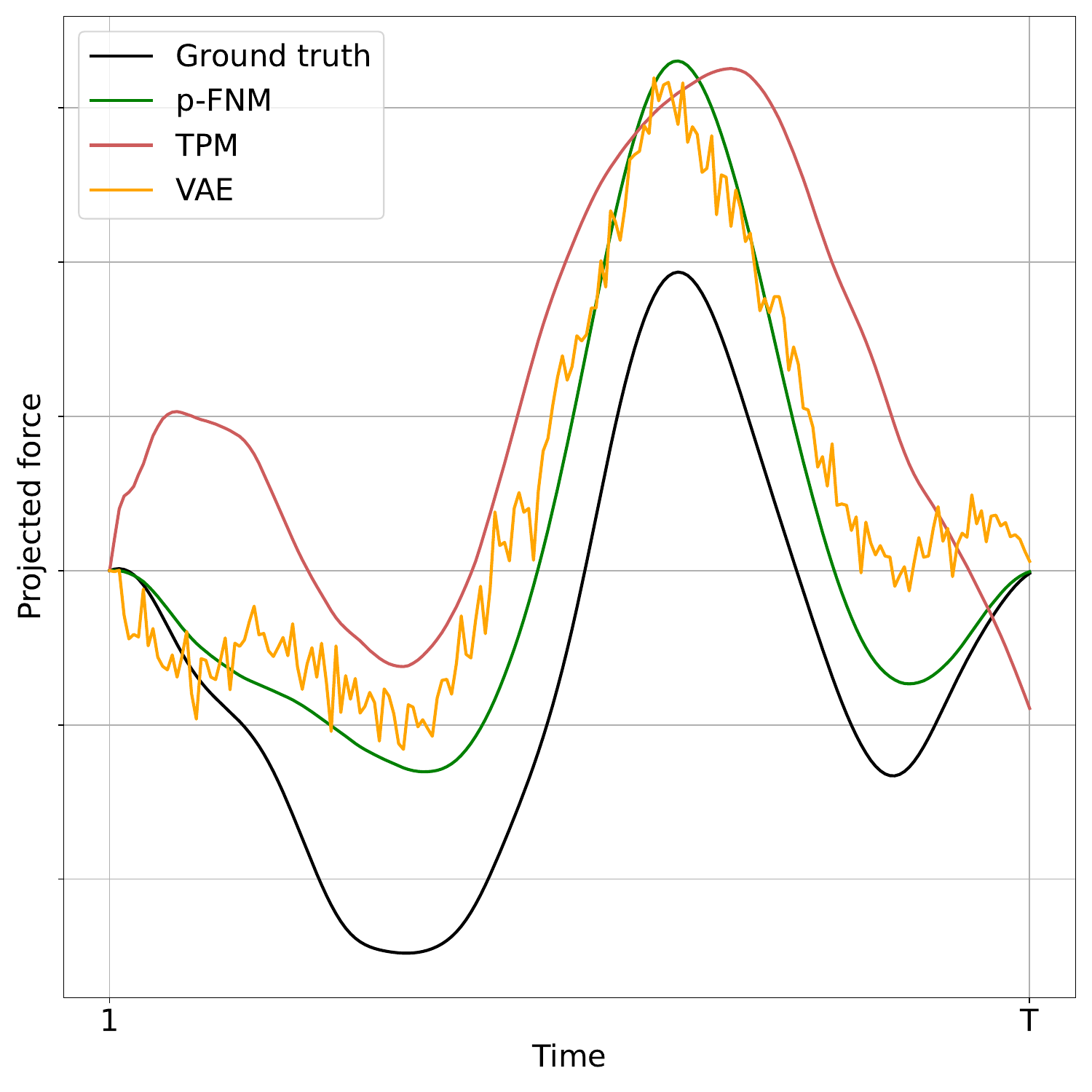}} \\
        \hline
        \rotatebox[origin=c]{90}{Large} &
        \raisebox{-0.5\height}{\includegraphics[width=0.30\linewidth]{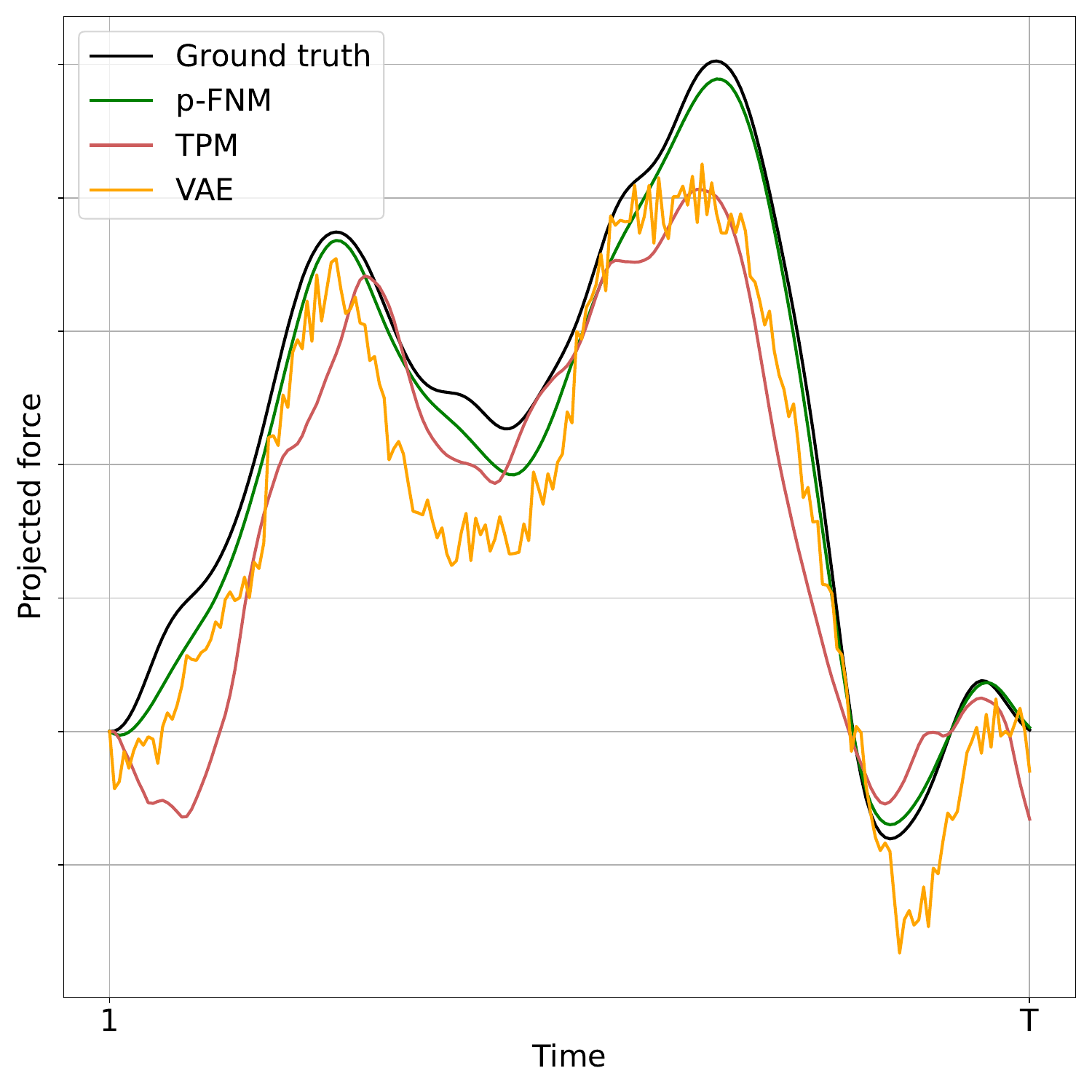}} &
        \raisebox{-0.5\height}{\includegraphics[width=0.30\linewidth]{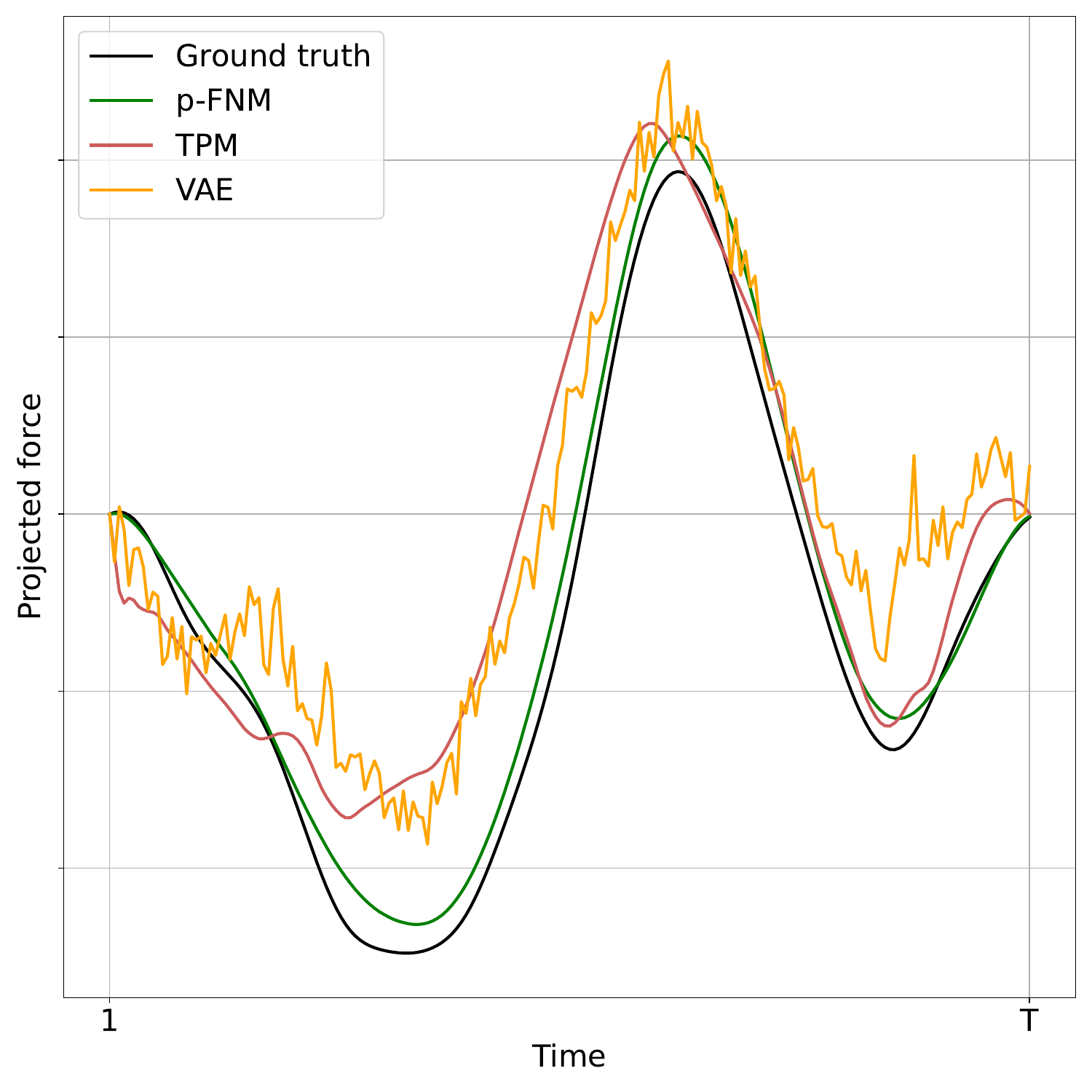}} \\
        \hline
        \rotatebox[origin=c]{90}{Extra Large} &
        \raisebox{-0.5\height}{\includegraphics[width=0.30\linewidth]{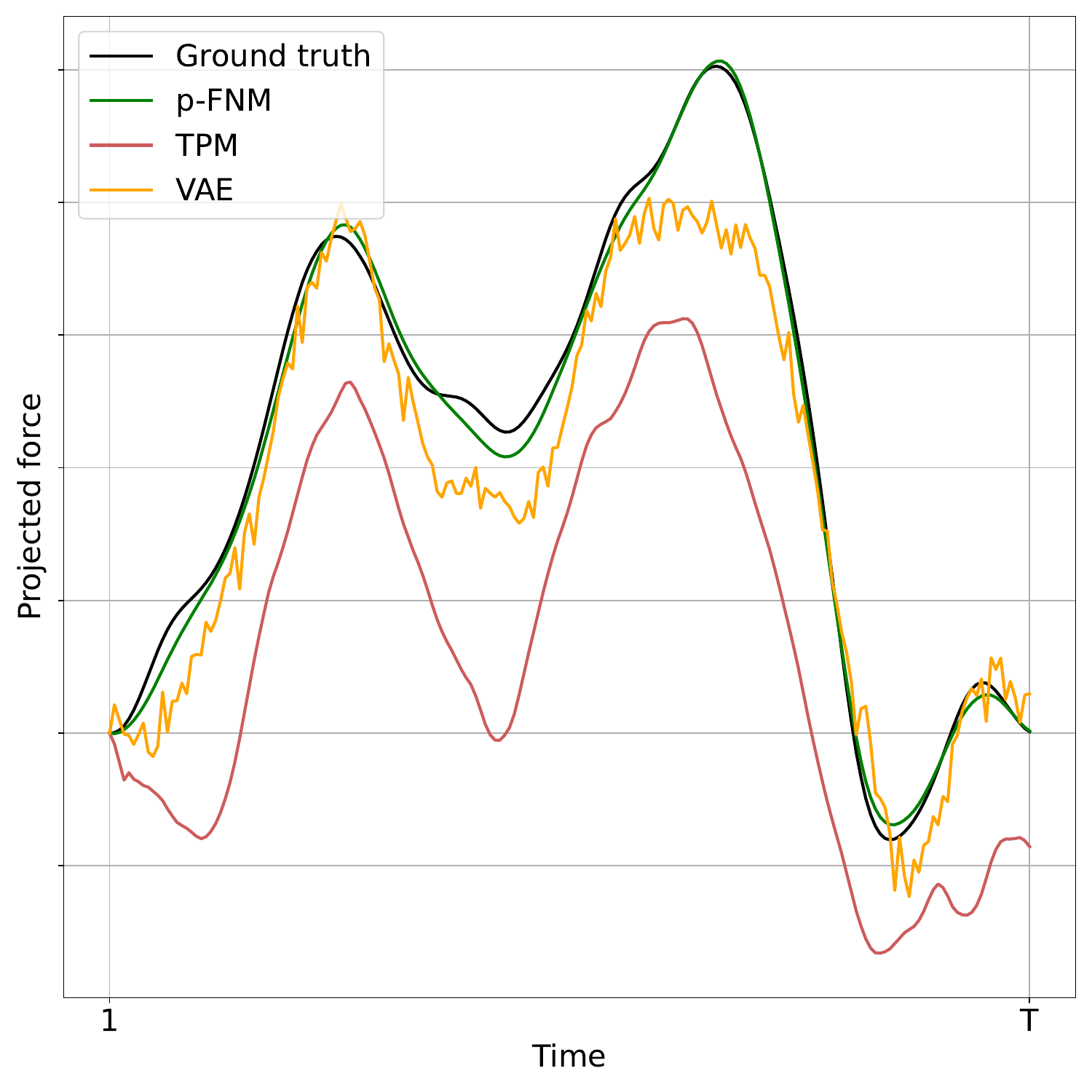}} &
        \raisebox{-0.5\height}{\includegraphics[width=0.30\linewidth]{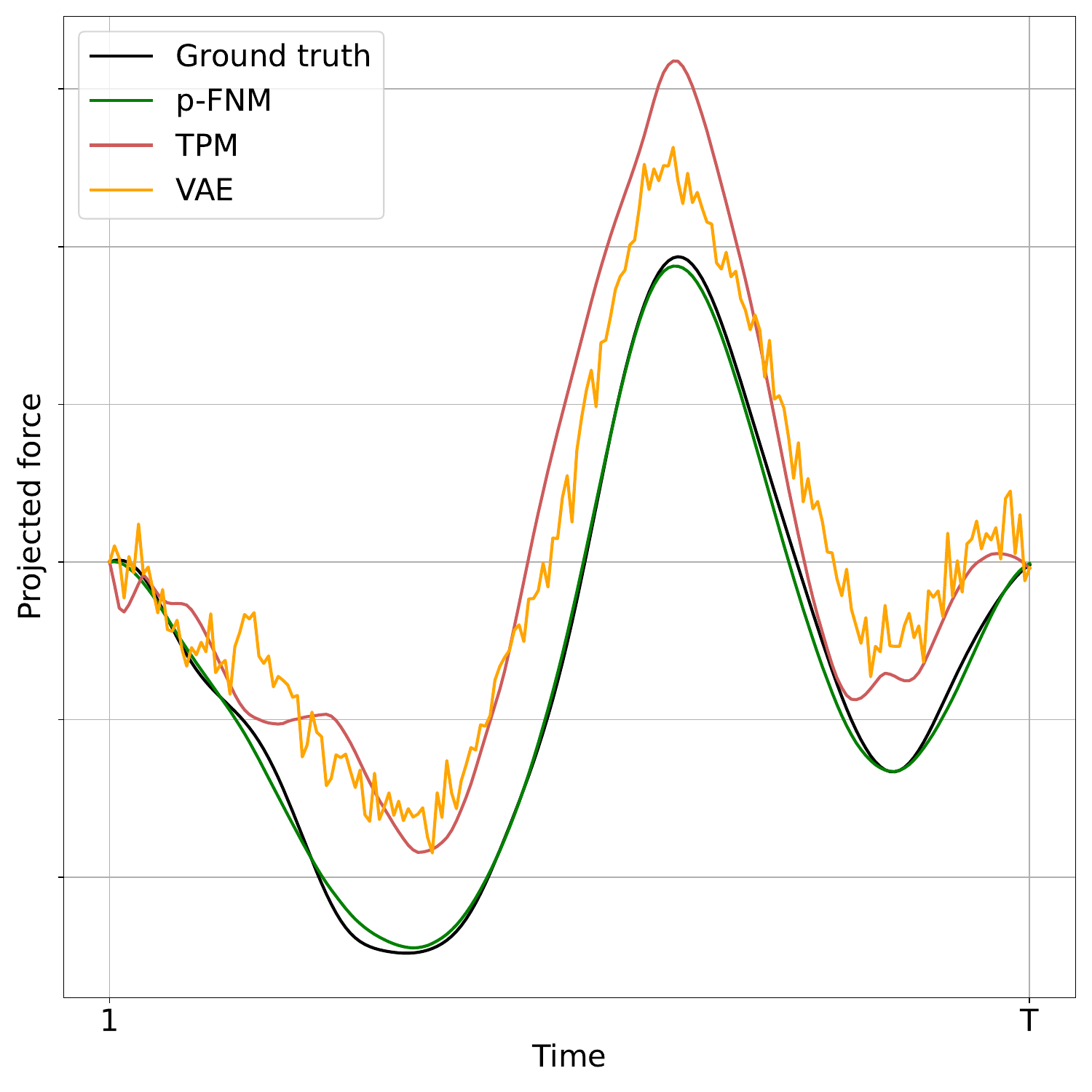}} \\
    \end{tabular}
    \label{fig:projF-all}
\end{table}
\paragraph{Fourier analysis of the projected-force.} In Figure \ref{fig:projF-spectrum} we plotted the temporal spectrum of the projected-force signal for the TPM and p-FNM models trained on the different datasets. The spectrum curves are obtained by taking the average of the temporal spectrum on all the training runs and all the test set simulations. One can see that the p-FNM outperforms the TPM on the low frequencies, closely matching the reference average spectrum. 
For the highest frequencies the p-FNM average spectrum is better than the TPM except for the small dataset, but when adding mode data to the training set the p-FNM average spectrum closes the gap to the reference.
\paragraph{First Fourier mode of the projected-force.} The weighted Mean Absolute Percentage Error (wMAPE) associated with the prediction of the first Fourier mode of the projected-force signal by the TPM and p-FNM models is reported in Figure \ref{fig:projF-spectrum:mode1}. The wMAPE is defined as
\begin{equation*}
    \mathrm{wMAPE}(f_1^{\mathrm{true}}, f_1^{\mathrm{pred}}) = \frac{\sum\limits_{b=1}^{N} \left| f_1^{\mathrm{true}} - f_1^{\mathrm{pred}} \right|}{\sum\limits_{b=1}^{N} \left| f_1^{\mathrm{true}} \right|},
\end{equation*}
where $f_1^{\mathrm{true}}$ and $f_1^{\mathrm{pred}}$ denote the amplitudes of the first Fourier mode of the reference and predicted projected-force signals, respectively. The wMAPE metric is here preferred over the conventional MAPE because the amplitude of the first Fourier mode can become very small for certain samples, leading to artificially large or diverging relative errors in the standard MAPE formulation due to a near-zero denominator. Figure \ref{fig:projF-spectrum:mode1} clearly shows that the p-FNM consistently outperforms the TPM model in predicting the first Fourier mode of the projected-force signal. This improved spectral reconstruction capability directly explains the superior performance of the p-FNM for GAF prediction.
\subsection{Aerodynamic Forces Results}\label{section:results:gaf}
\begin{figure}[!htp]
    \centering
    \begin{subfigure}{0.40\textwidth}
        \centering
        \includegraphics[width=\linewidth]{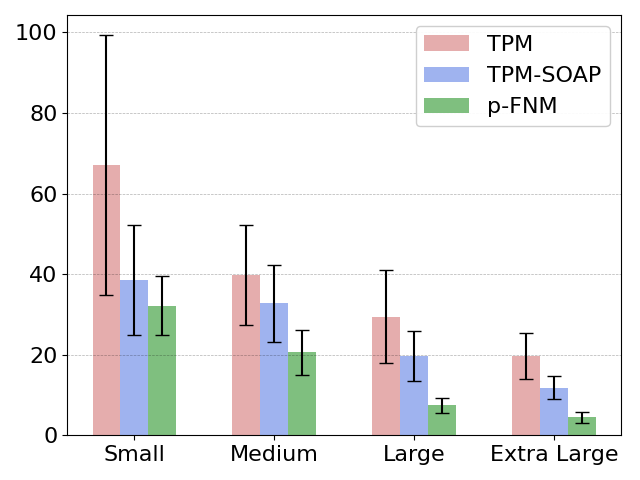}
        \caption{$\mathrm{MAPE}_\rho$: Mean Absolute Percentage Error [\%] of the predicted GAF norm. Lower is better.}
        \label{fig:gaf-rho-mape}
    \end{subfigure}
    \begin{subfigure}{0.40\textwidth}
        \centering
        \includegraphics[width=\linewidth]{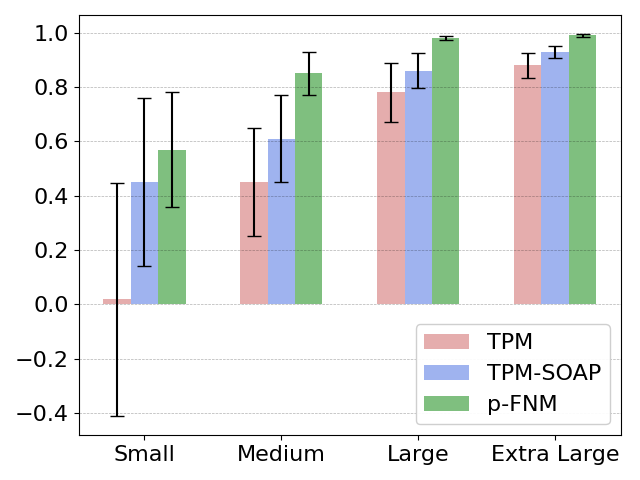}
        \caption{$R^2_\rho$: The $R^2$ score of the predicted GAF norm. Higher is better.}
        \label{fig:gaf-rho-r2}
    \end{subfigure}
    \begin{subfigure}{0.40\textwidth}
        \centering
        \includegraphics[width=\linewidth]{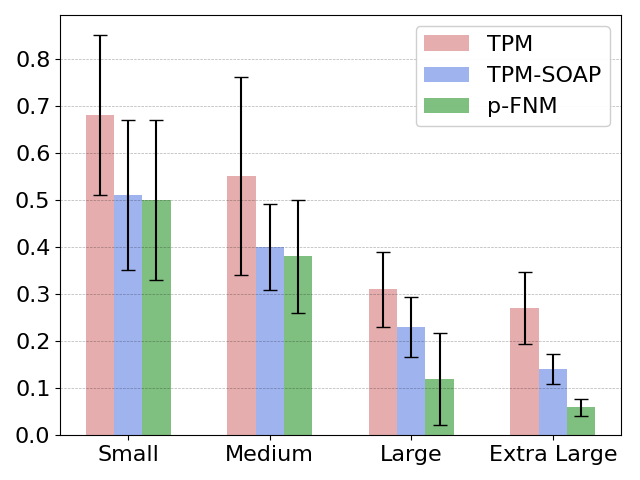}
        \caption{$\mathrm{wRMSE}_\phi$: Weighted RMSE [radians] of the prediction GAF phase. Lower is better.}
        \label{fig:gaf-phi-rmse}
    \end{subfigure}
    \caption{Comparison of the TPM, TPM-SOAP and p-FNM models with respect to the GAF prediction metrics across the different training datasets. The metrics are evaluated on the test dataset. The black error bars represent the metric uncertainty quantified by the $\pm 1.96\sigma$ confidence interval computed over 30 independent training runs.}
    \label{fig:gaf-metrics}
\end{figure}

\begin{figure}[!htp]
    \centering    
    \begin{subfigure}{0.40\textwidth}
        \centering
        \includegraphics[width=\linewidth]{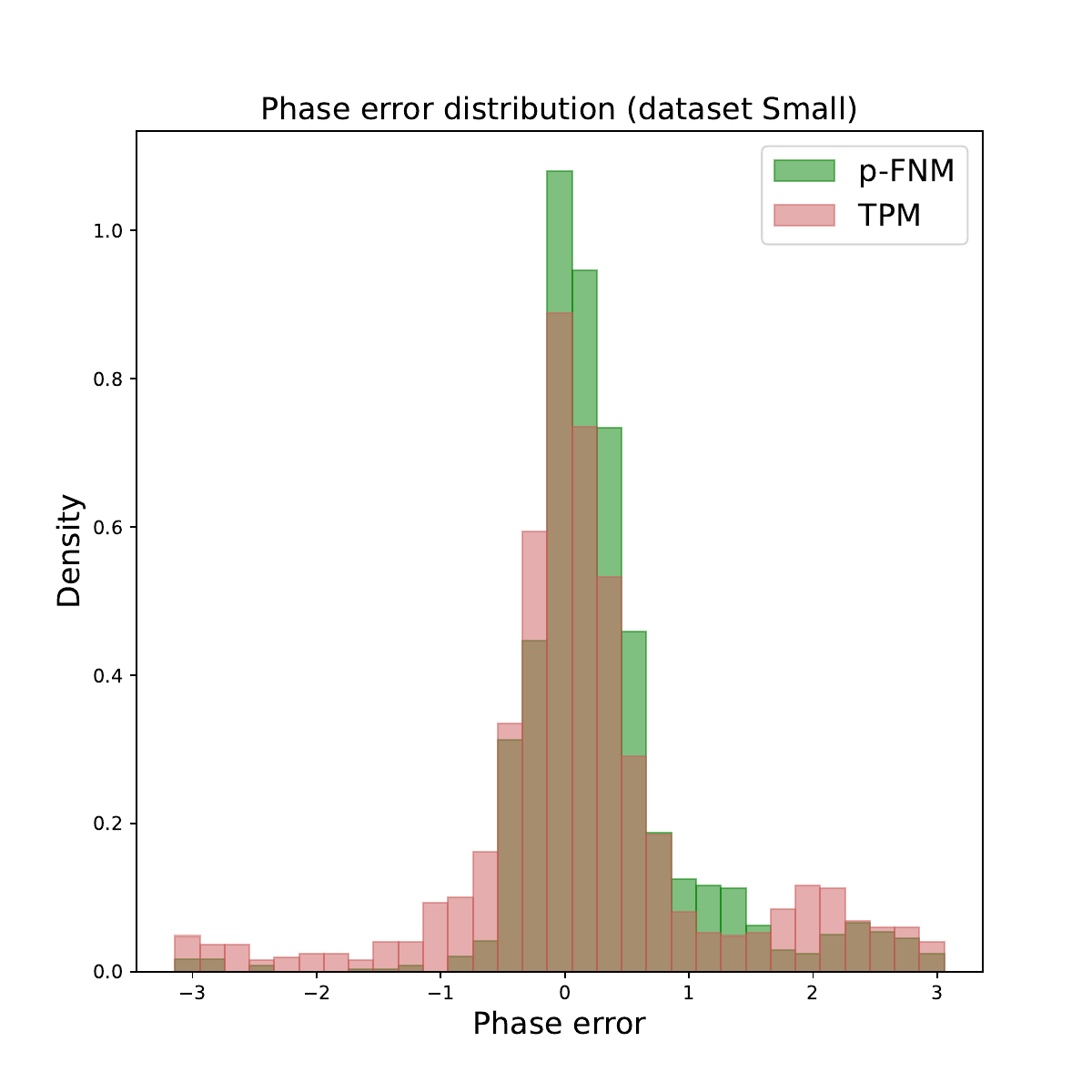}
        \caption{Training dataset small (S)}
        \label{fig:gaf-phase-err:s}
    \end{subfigure}
    \begin{subfigure}{0.40\textwidth}
        \centering
        \includegraphics[width=\linewidth]{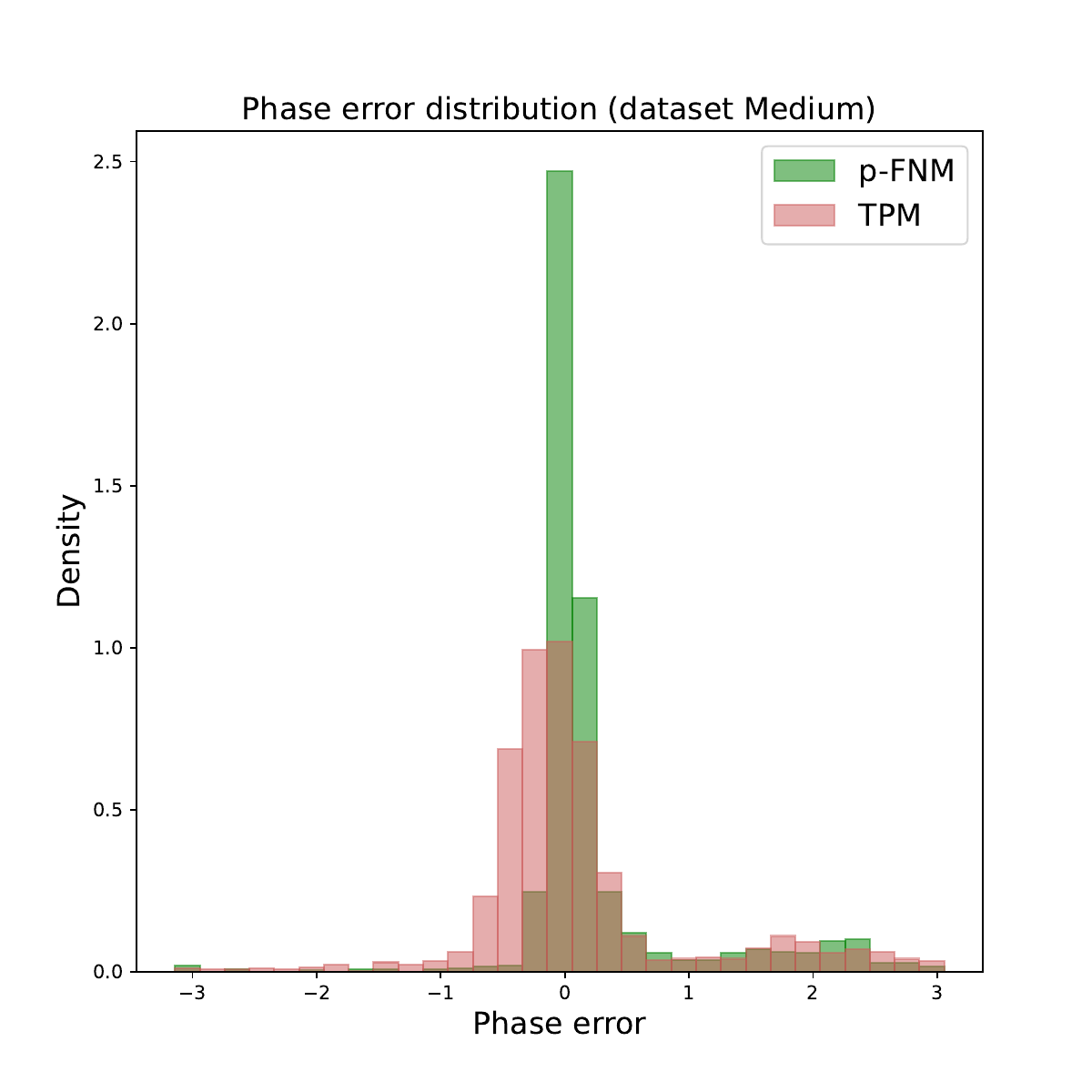}
        \caption{Training dataset medium (M)}
        \label{fig:gaf-phase-err:m}
    \end{subfigure}
    \begin{subfigure}{0.40\textwidth}
        \centering
        \includegraphics[width=\linewidth]{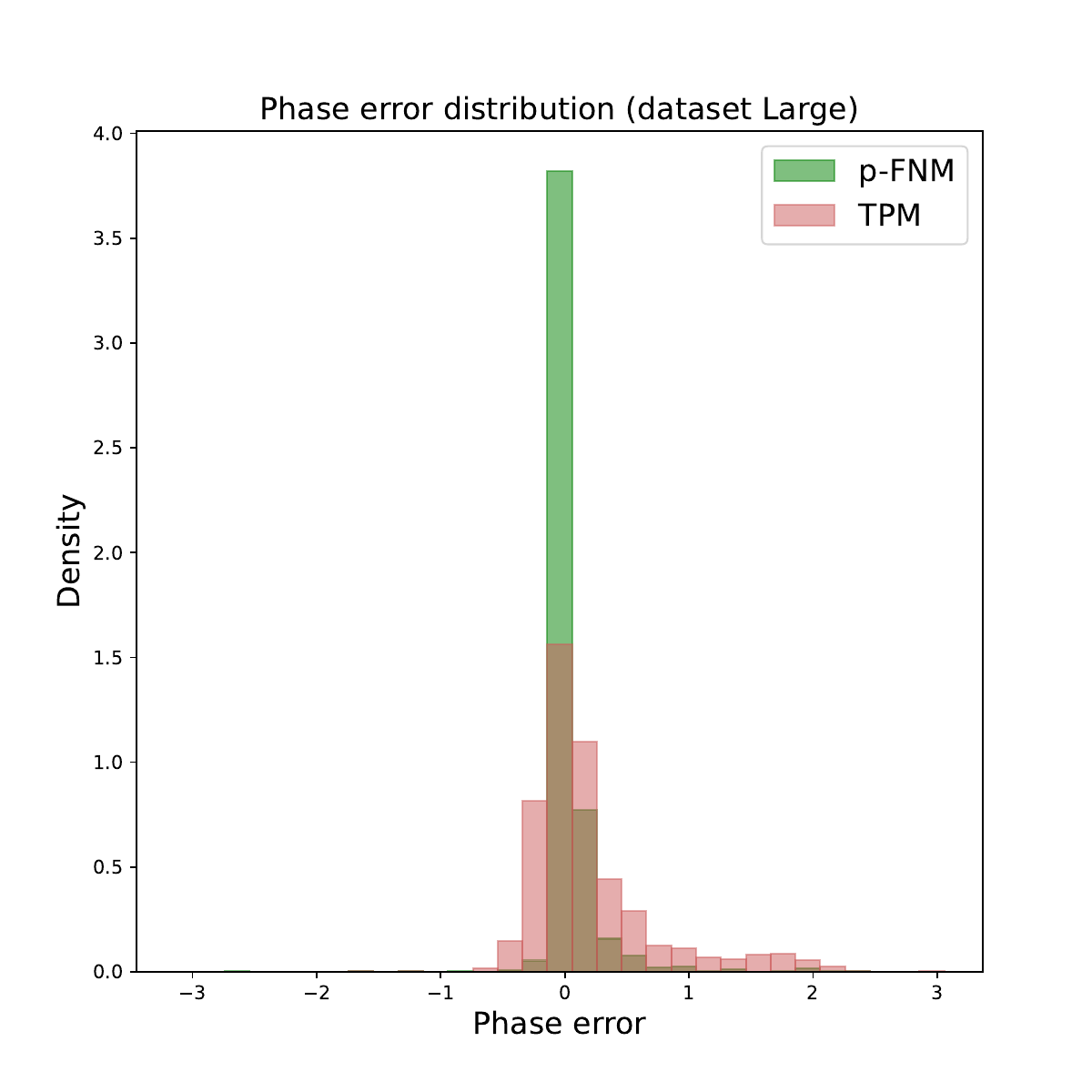}
        \caption{Training dataset large (L)}
        \label{fig:gaf-phase-err:l}
    \end{subfigure}
    \begin{subfigure}{0.40\textwidth}
        \centering
        \includegraphics[width=\linewidth]{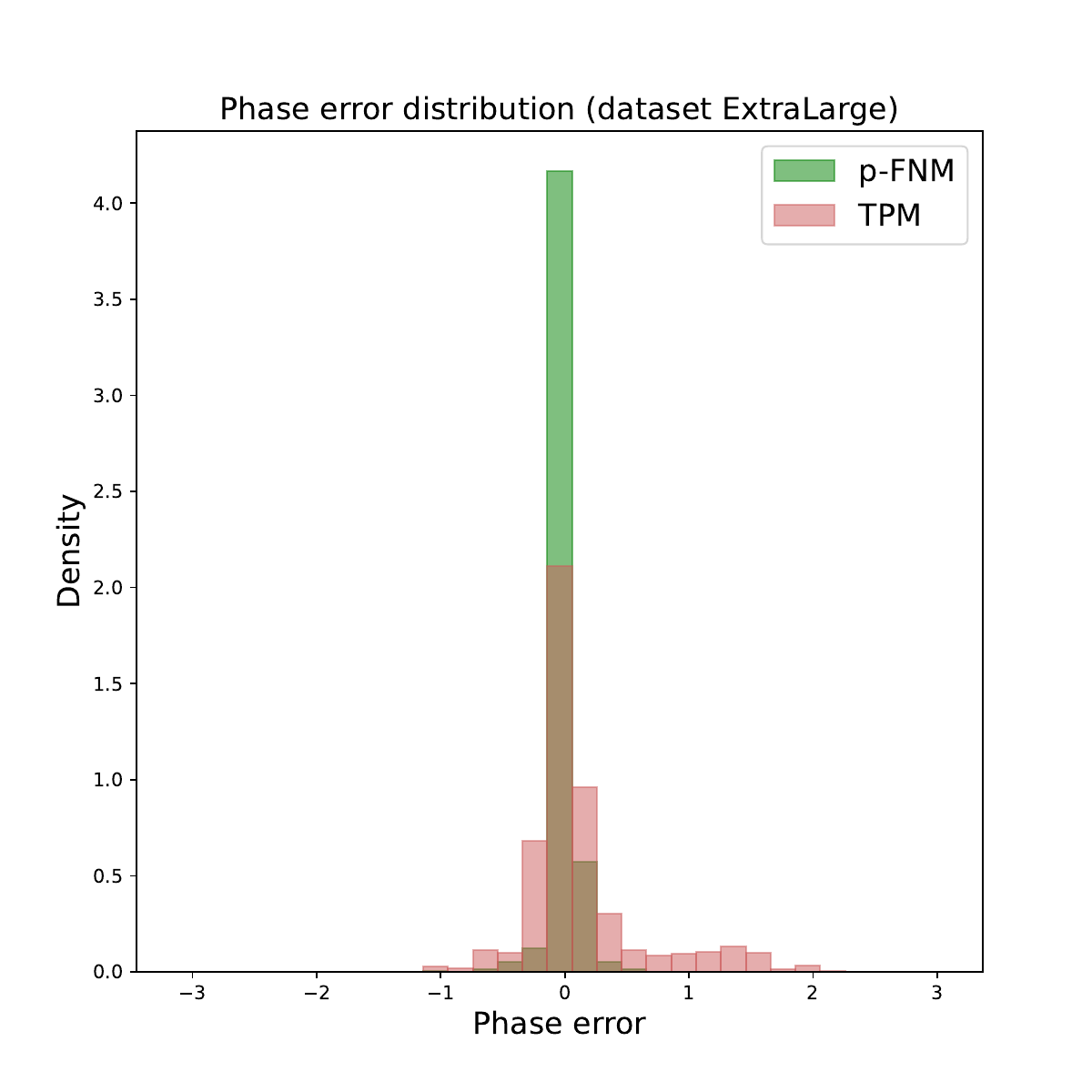}
        \caption{Training dataset extra large (XL)}
        \label{fig:gaf-phase-err:xl}
    \end{subfigure}
    \caption{Comparison of the GAF phase prediction error distributions (in radians) for the TPM and p-FNM models across the different training datasets. The models are evaluated on the test dataset. The phase error is defined within the interval $[-\pi, \pi]$. Optimal performance corresponds to error distributions centered around zero with minimal dispersion.}
    \label{fig:gaf-phase-err-dist}
\end{figure}

The metrics associated with the prediction of the GAF magnitude and phase are reported in Figures \ref{fig:gaf-rho-mape}, \ref{fig:gaf-rho-r2}, and \ref{fig:gaf-phi-rmse}, comparing the predictive accuracy and uncertainty of the TPM, TPM-SOAP, and the p-FNM models across the different training datasets. Detailed quantitative results are provided in Tables \ref{tab:vae-vs-vae-soap-all} and \ref{tab:fnm-metrics-detailed} of Appendices \ref{appendix:tpm-soap} and \ref{appendix:detail-res}, respectively.
\paragraph{GAF magnitude prediction.} 
The predicted versus reference GAF magnitudes for the different training datasets and for both the TPM and p-FNM models are presented in Table \ref{fig:gaf-norm-err}. For both architectures, the prediction accuracy improves with the training dataset size, as evidenced by the progressive concentration of the predicted samples around the identity line. In addition, the p-FNM systematically exhibits reduced dispersion compared with the TPM baseline, indicating improved robustness and predictive consistency. Prediction uncertainty across the 30 independent training runs is also represented through error bars. The uncertainty decreases with increasing dataset size for both models, while remaining consistently lower for the p-FNM. \\
Beyond this qualitative analysis, the metrics introduced in Section \ref{section:metrics} provide a quantitative assessment of GAF magnitude prediction performances.
Figure \ref{fig:gaf-rho-mape} reports the MAPE on the predicted GAF magnitude. Both the prediction error and associated uncertainty decrease consistently as the amount of training data increases. Nevertheless, the p-FNM systematically outperforms both the TPM and TPM-SOAP. The TPM baseline exhibits particularly large relative errors, reaching $67.11\%$ for the smallest dataset and remaining above $10\%$ even for the extra-large dataset. While the TPM-SOAP model improves these results it still fails to reduce the relative error below $10\%$. In contrast, the p-FNM achieves substantially lower errors, decreasing from values above $10\%$ for the small and medium datasets to $4.42\%$ for the extra-large dataset.\\
The ability of the models to reproduce the variability of the GAF magnitude is evaluated through the $R^2$ metric, reported in Figure \ref{fig:gaf-rho-r2}. For the smallest dataset, the TPM baseline achieves an $R^2$ score of only $0.018$, indicating performance comparable to a constant mean predictor. As the dataset size increases, the TPM progressively improves and reaches an $R^2$ score of $0.88$ for the largest datasets, demonstrating an ability to capture GAF magnitude variability. The TPM-SOAP further increases the $R^2$ score while reducing prediction variability across training runs. However, the p-FNM consistently outperforms both TPM variants, demonstrating a superior ability to reproduce GAF magnitude fluctuations with reduced uncertainty. Although performance remains moderate for the small ($R^2 = 0.57$) and medium ($R^2 = 0.85$) datasets, the p-FNM achieves excellent predictive accuracy for the large and extra-large datasets, reaching a maximum $R^2$ score of $0.99$.
\paragraph{GAF phase prediction.} 
The distributions of the GAF phase prediction error are shown in Figures \ref{fig:gaf-phase-err:s}, \ref{fig:gaf-phase-err:m}, \ref{fig:gaf-phase-err:l}, and \ref{fig:gaf-phase-err:xl}. For the smallest dataset, the p-FNM already exhibits a slight advantage over the TPM, with phase errors more concentrated around zero. This advantage becomes increasingly pronounced as the training dataset size grows. In particular, the phase error distributions associated with the p-FNM progressively collapse to zero, whereas the TPM distributions remain significantly more dispersed. \\
Quantitative results for $\mathrm{wRMSE}_\phi$ are reported in Figure \ref{fig:gaf-phi-rmse}. The p-FNM consistently outperforms both TPM variants across all training datasets. For the small and medium datasets, the performance gain relative to TPM-SOAP remains moderate. However, the gap between the two architectures becomes substantially larger for the large and extra-large datasets. Overall, the phase prediction error decreases monotonically with increasing dataset size, reaching a minimum value of $0.06\;\mathrm{rad}$ for the p-FNM on the extra-large dataset, corresponding to less than $1\%$ of the unit-circle circumference.
\paragraph{Synthesis of the GAF prediction results.}
Overall, these results demonstrate that GAF prediction constitutes a substantially more challenging task than pressure-field reconstruction. Nevertheless, the p-FNM consistently outperforms both the TPM variants across all metrics associated with GAF magnitude and phase prediction. Furthermore, both predictive accuracy and uncertainty improve systematically with the amount of training data.
The experiments also indicate that achieving satisfactory performance on GAF-related metrics requires significantly larger datasets than those needed for accurate pressure-field prediction. Despite this increased difficulty, the architecture of the p-FNM enables substantially improved GAF prediction compared with the TPM baseline under an equivalent training-data budget.
\begin{table}[!htp]
    \centering
    \caption{Comparison between predicted and reference GAF norm values for the TPM and p-FNM models across the different training datasets. The models are evaluated on the test dataset. The black dashed line indicates perfect agreement, while the green and red dashed lines correspond to relative errors of $\pm 5\%$ and $\pm 10\%$, respectively. The markers represent the mean predicted GAF values, and the gray error bars denote the associated prediction uncertainty quantified by the $\pm 1.96\sigma$ confidence interval. Each grid cell corresponds to 0.0005 GAF units along both axes.}
    \vspace*{4mm}
    \begin{tabular}{c|c|c}
        & TPM & p-FNM \\
        \hline
        \rotatebox[origin=c]{90}{Small} &
        \raisebox{-0.5\height}{\includegraphics[width=0.30\linewidth]{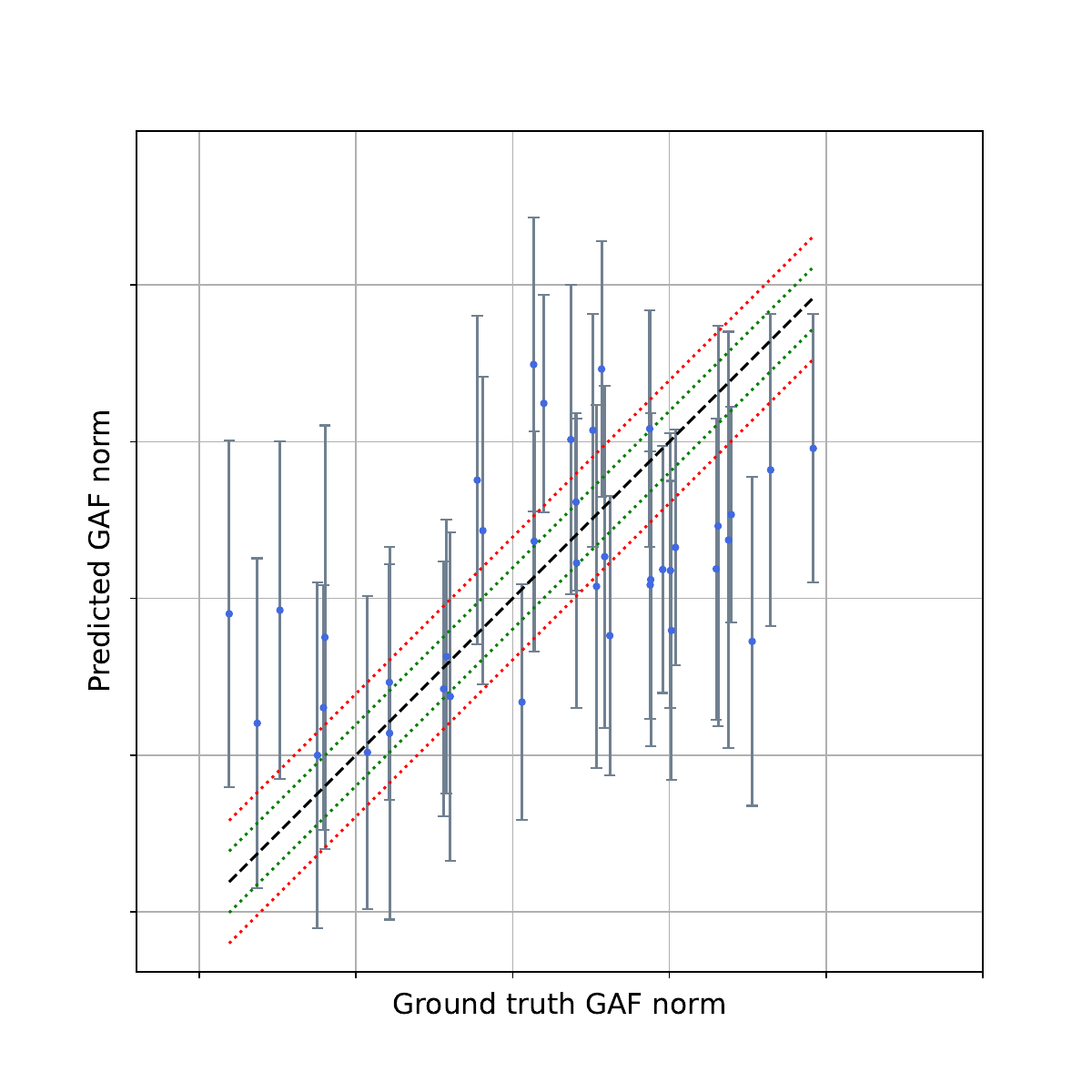}} &
        \raisebox{-0.5\height}{\includegraphics[width=0.30\linewidth]{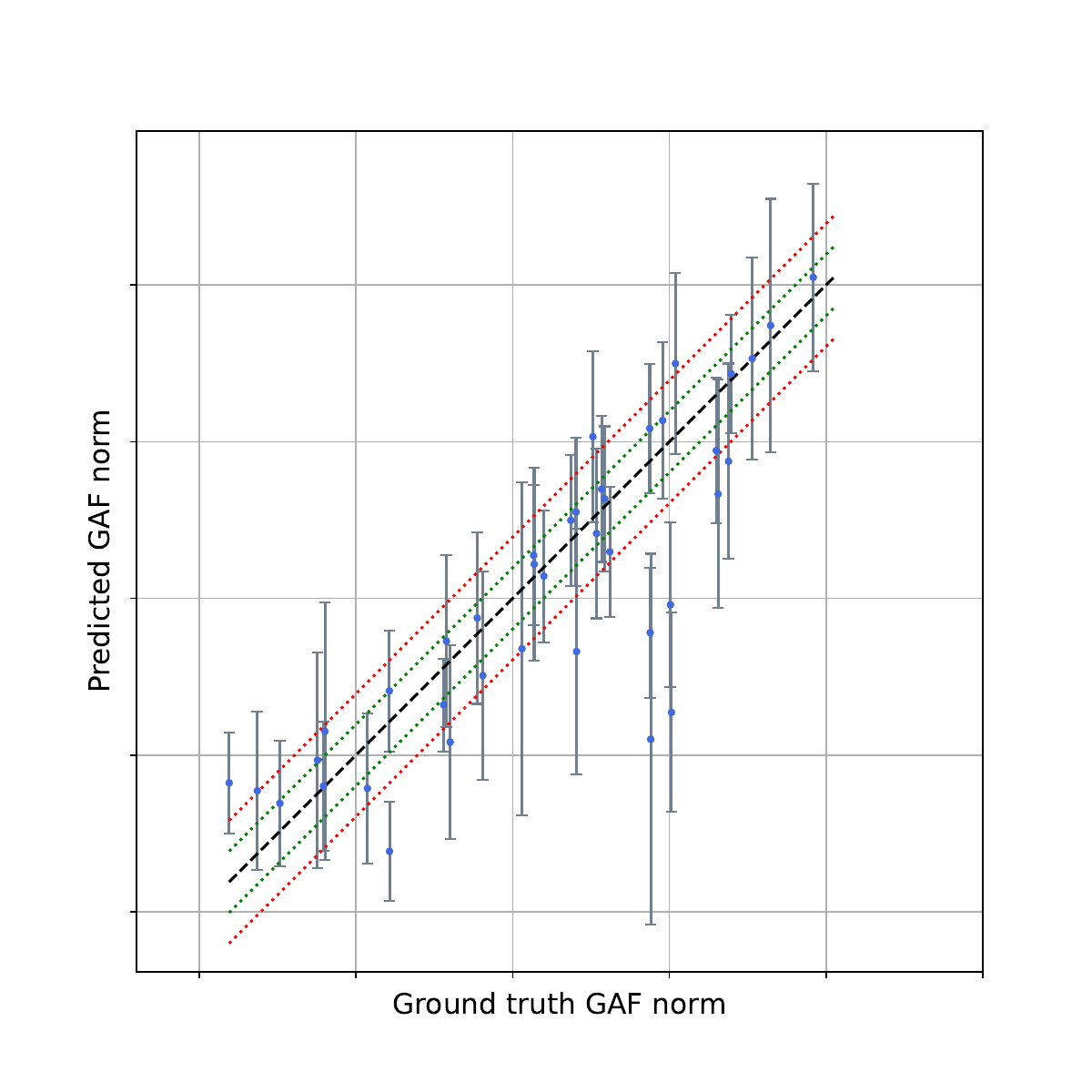}} \\
        \hline
        \rotatebox[origin=c]{90}{Medium} &
        \raisebox{-0.5\height}{\includegraphics[width=0.30\linewidth]{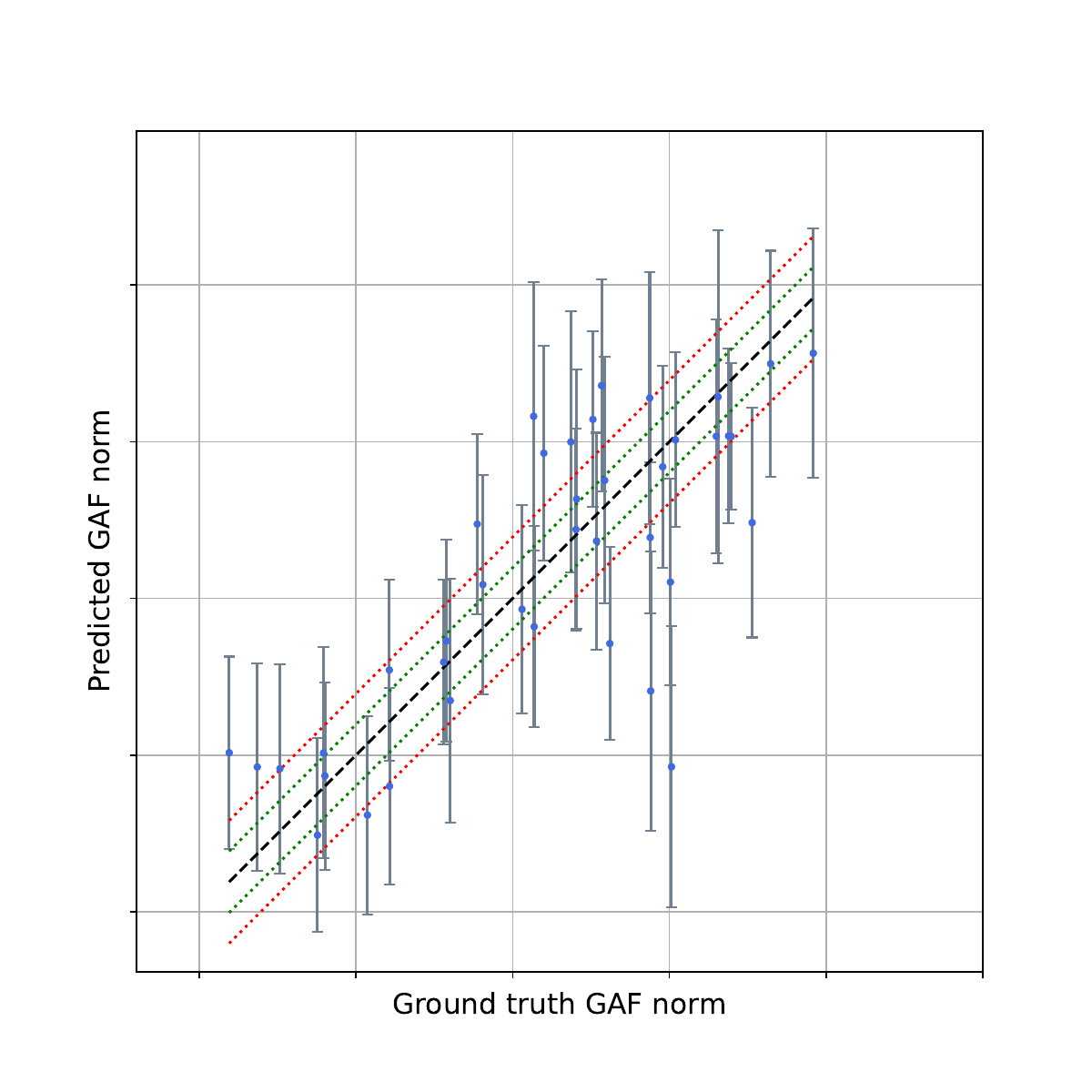}} &
        \raisebox{-0.5\height}{\includegraphics[width=0.30\linewidth]{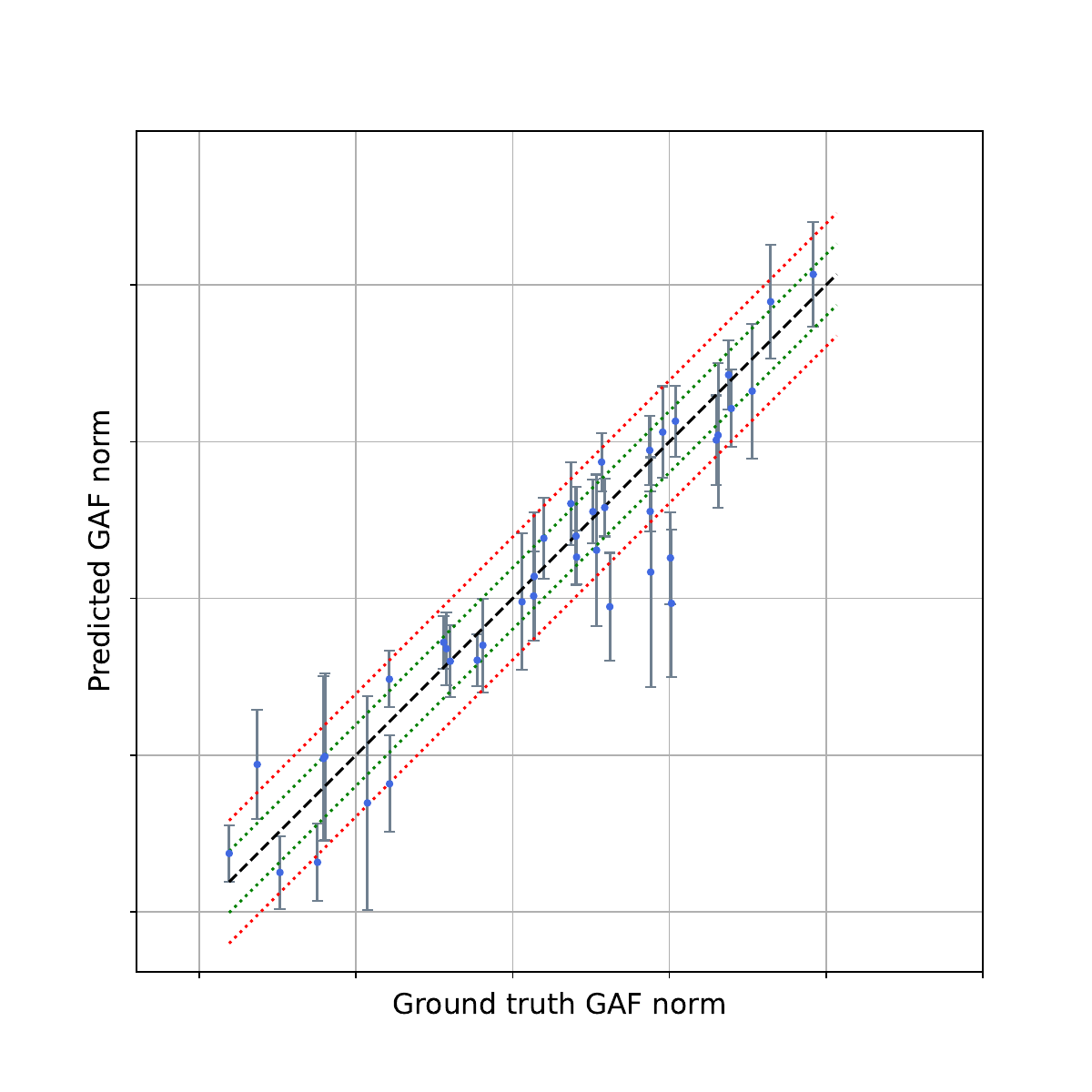}} \\
        \hline
        \rotatebox[origin=c]{90}{Large} &
        \raisebox{-0.5\height}{\includegraphics[width=0.30\linewidth]{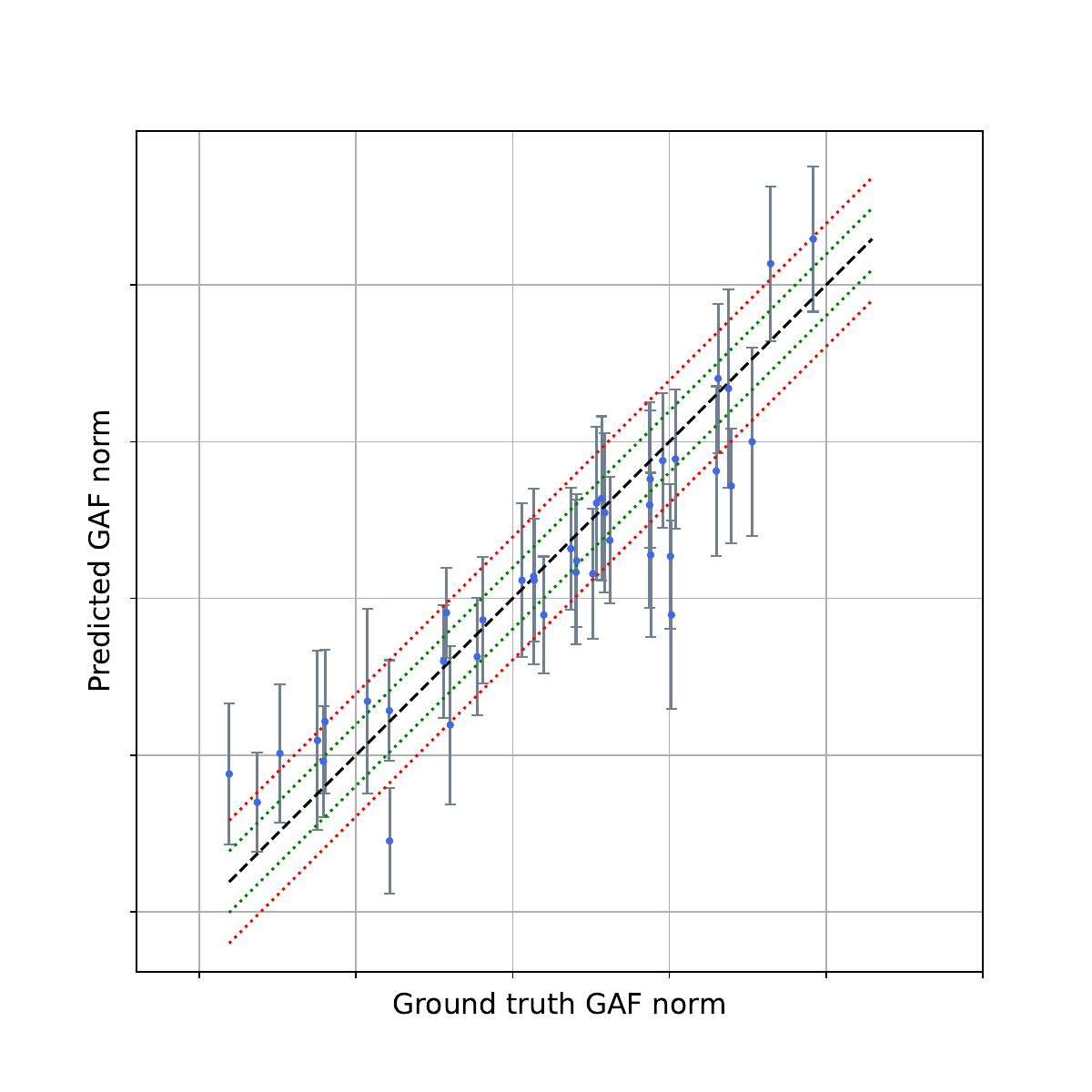}} &
        \raisebox{-0.5\height}{\includegraphics[width=0.30\linewidth]{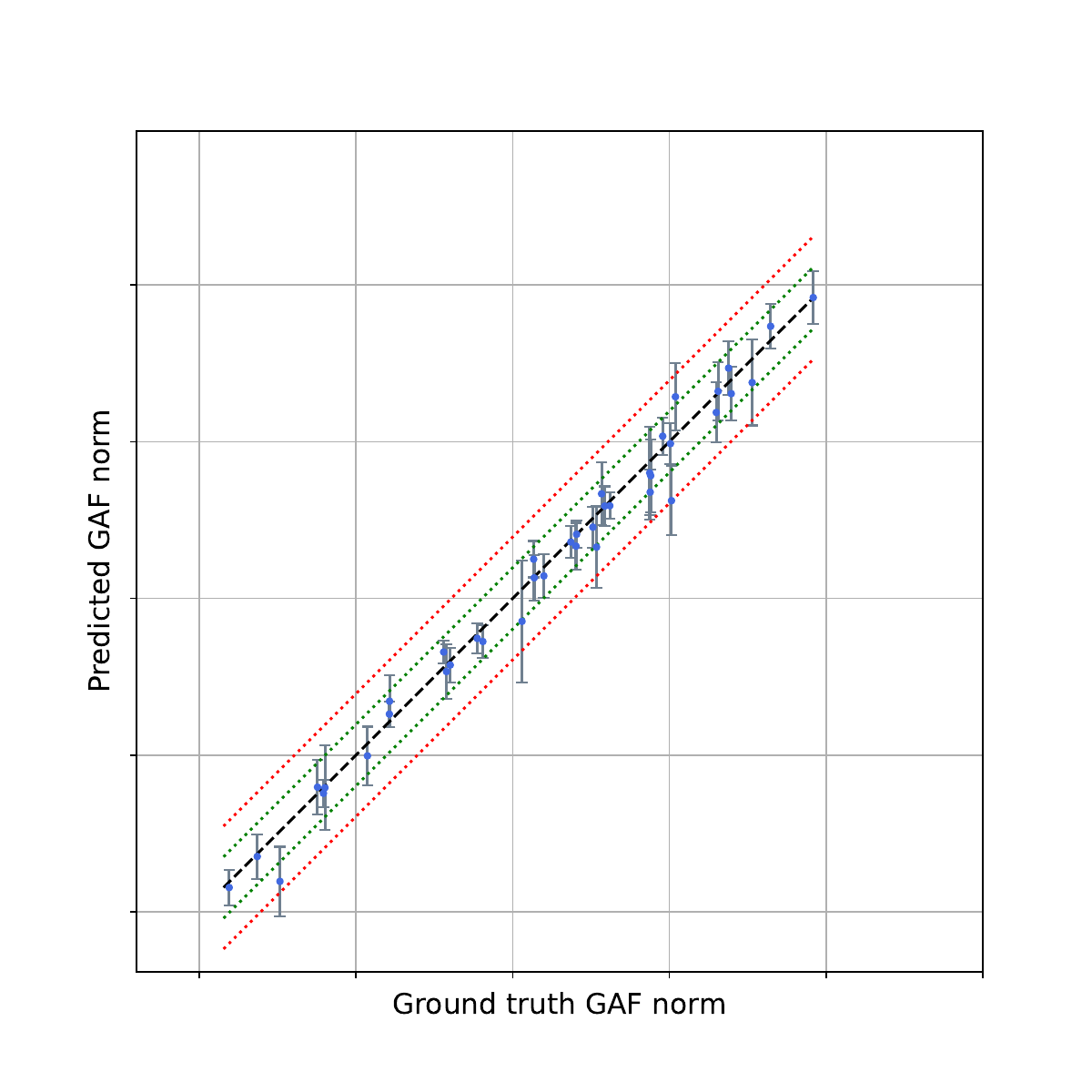}} \\
        \hline
        \rotatebox[origin=c]{90}{Extra Large} &
        \raisebox{-0.5\height}{\includegraphics[width=0.30\linewidth]{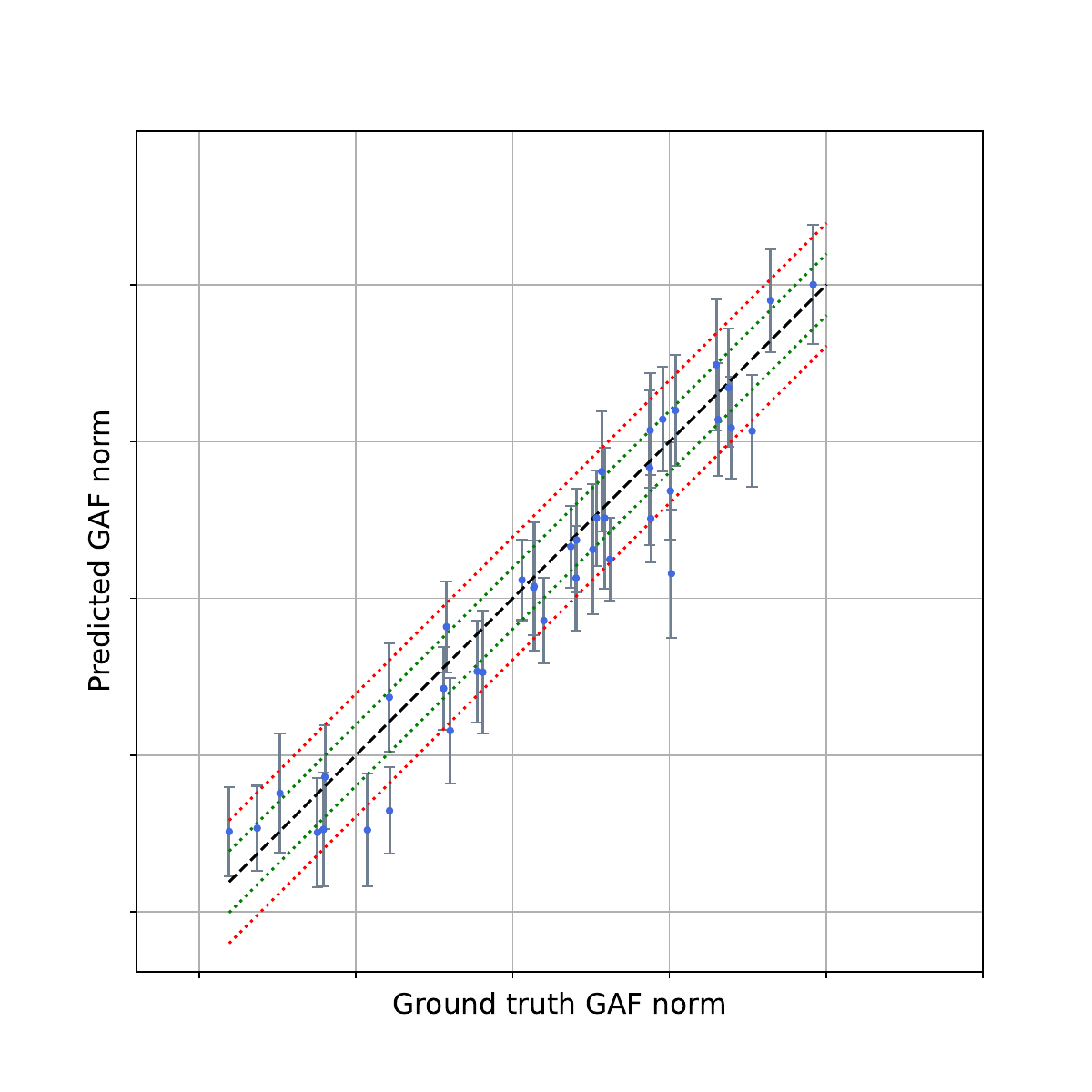}} &
        \raisebox{-0.5\height}{\includegraphics[width=0.30\linewidth]{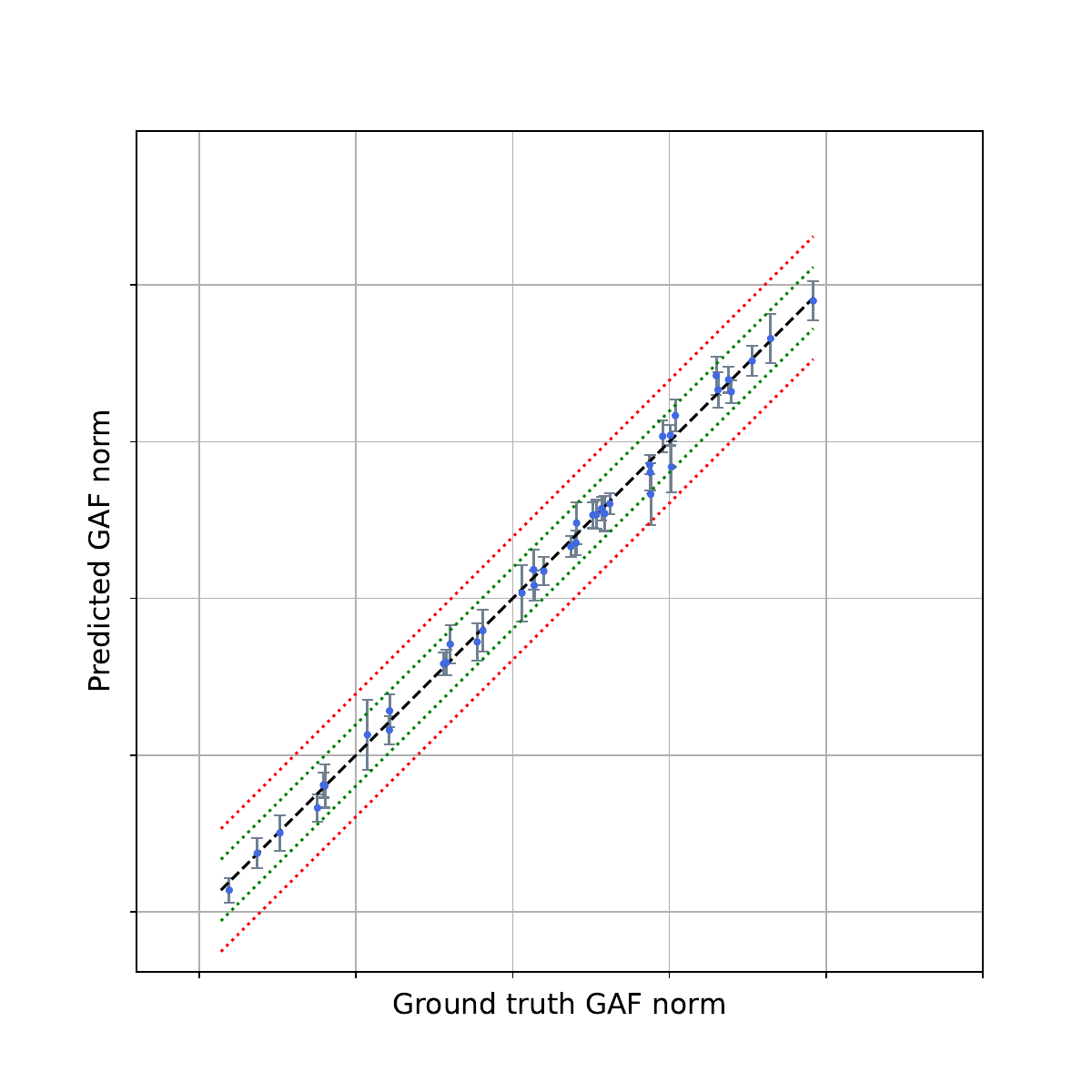}}
         
    \end{tabular}
    \label{fig:gaf-norm-err}
\end{table}

\subsection{Training Time Comparison}\label{section:training-time-comparison}
\begin{figure}[htbp]
  \centering
  \includegraphics[width=0.45\linewidth]{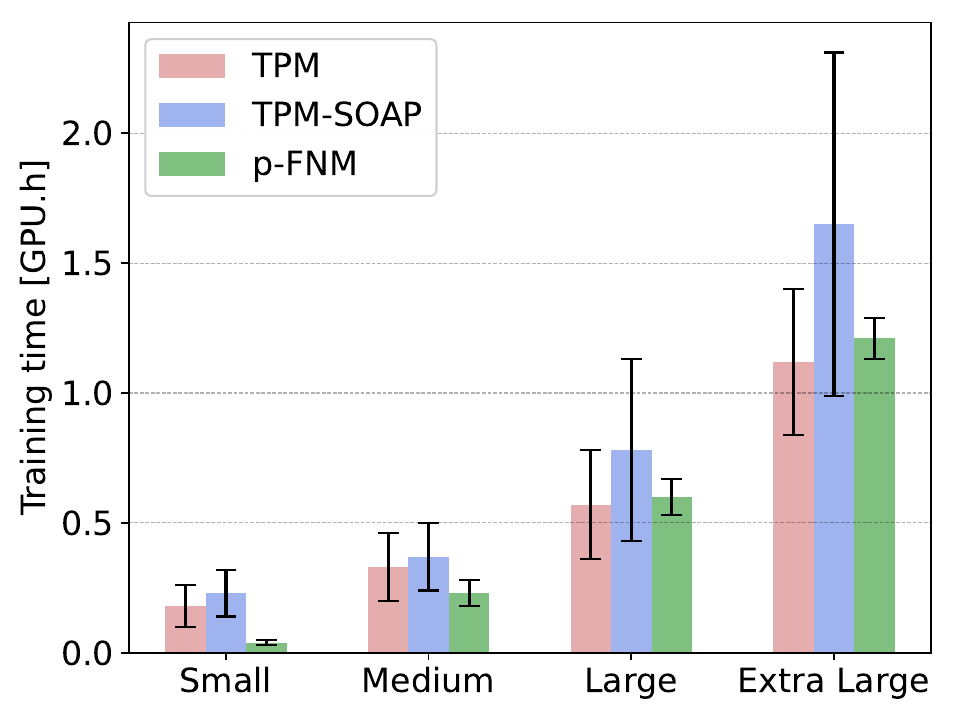}
  \caption{Comparison of the TPM, TPM-SOAP, and p-FNM models in terms of training time (GPU$\cdot$h) across the different training datasets. The black error bars indicate the training-time uncertainty quantified by the $\pm1.96\sigma$ confidence interval computed over all independent training runs.}
  \label{fig:training-time-comp}
\end{figure}

Figure \ref{fig:training-time-comp} presents the average training times and their associated uncertainties for the TPM, TPM-SOAP, and the p-FNM model. For all architectures, both the mean training time and its variability increase with the size of the training dataset, reflecting the additional computational cost associated with processing larger amounts of simulation data.
The results further indicate that the p-FNM model is computationally more efficient than both the TPM and TPM-SOAP models for the small and medium datasets. For the large and extra-large datasets, the training cost of the p-FNM remains lower than that of the TPM-SOAP, while becoming slightly higher than that of the original TPM baseline.
\subsection{Direct GAF Prediction} \label{sec:direct-gaf-pred}
\begin{figure}[!htp]
    \centering
    \begin{subfigure}{0.40\textwidth}
        \centering
        \includegraphics[width=\linewidth]{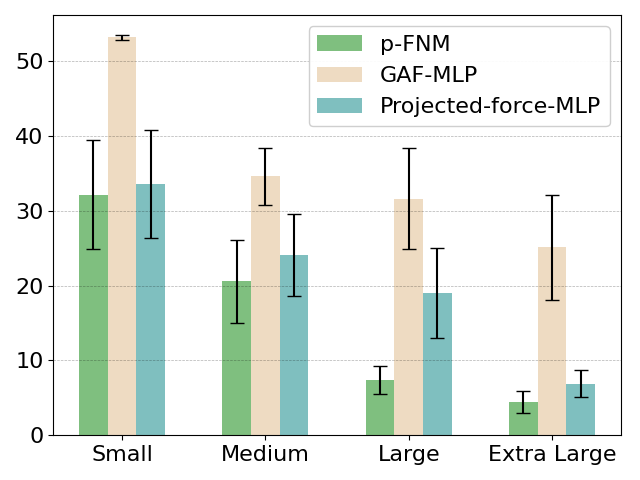}
        \caption{$\mathrm{MAPE}_\rho$ (\%). Lower is better.}
        \label{fig:e2e-mlp-comp:gaf-mape}
    \end{subfigure}
    \begin{subfigure}{0.40\textwidth}
        \centering
        \includegraphics[width=\linewidth]{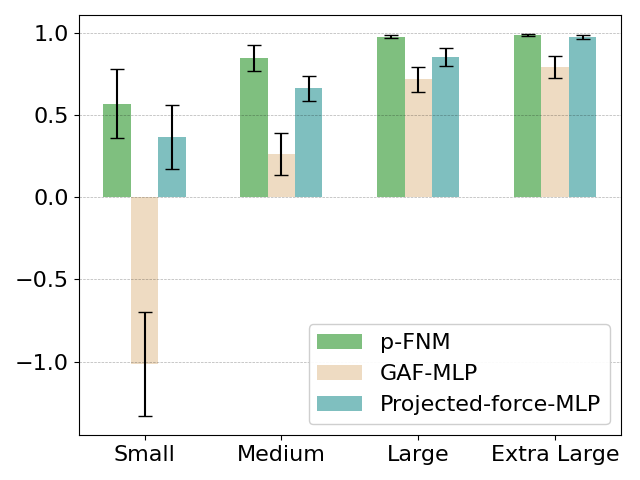}
        \caption{$R^2_\rho$. Higher is better.}
        \label{fig:e2e-mlp-comp:gaf-rho-r2}
    \end{subfigure}
    \begin{subfigure}{0.40\textwidth}
        \centering
        \includegraphics[width=\linewidth]{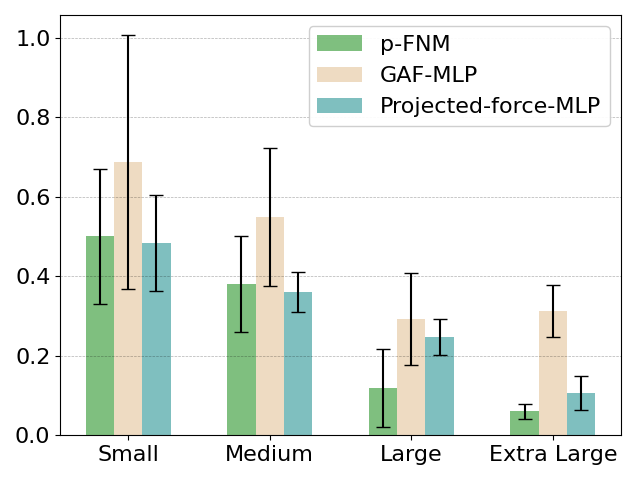}
        \caption{$\mathrm{wRMSE}_\phi$. Lower is better.}
        \label{fig:e2e-mlp-comp:gaf-phi-rmse}
    \end{subfigure}
    \caption{Comparison of the p-FNM, GAF-MLP and projected-force-MLP models on GAF metrics across the different training datasets. The metrics are evaluated on the test dataset. The black error bars indicate the training-time uncertainty quantified by the $\pm1.96\sigma$ confidence interval computed over all independent training runs.}
    \label{fig:e2e-mlp-results}
\end{figure}

In the present study, the projected force and the GAF are obtained analytically from the unsteady pressure fields predicted by the ML models. An important question is whether this pressure-field surrogate constitutes the most effective strategy, or improved performance could be achieved by directly predicting either the projected-force signal or the GAF using dedicated machine-learning models.\\
The question of predicting an intermediate physical field rather than directly targeting an observable quantity has also been investigated by \cite{huang2024operator}. Through both theoretical arguments and numerical experiments involving advection-diffusion and airfoil-flow problems, the authors showed that end-to-end approaches generally require substantially larger datasets to achieve a level of accuracy comparable to that obtained when first learning an intermediate physical field and subsequently deriving the quantity of interest.\\
To investigate this question, two MLP architectures were tested. The first, referred to as projected-force-MLP, predicts the projected-force signal from the operating conditions and the time variable. The second, denoted GAF-MLP, directly predicts the GAF from the operating conditions. For the projected-force-MLP, the temporal input is encoded using the same periodic embedding as the p-FNM (Equation \ref{eq:periodic-emb}), thereby explicitly incorporating the periodic nature of the dynamics. The GAF-MLP predicts the GAF in polar form. To avoid discontinuities associated with angular wrapping, the model predicts the cosine and sine of the phase rather than the phase angle itself.
For a fair comparison, both MLP models were trained using the SOAP optimizer. Their hyperparameters were tuned to maximize predictive performance while maintaining a model size comparable to that of the p-FNM, namely approximately 1.15 million trainable parameters.
The resulting comparison is reported in Figure \ref{fig:e2e-mlp-results}, where the models are evaluated using the GAF metrics across all training datasets. The GAF-MLP exhibits substantially lower performance than both the p-FNM and the projected-force-MLP for all metrics and dataset sizes. Although its performance improves as additional training data become available, this improvement remains limited and does not close the gap with the competing approaches. This behavior is likely explained by the reduced amount of supervision available to the model. In the GAF-MLP formulation, each simulation contributes only a single training sample, whereas the p-FNM and projected-force-MLP exploit every time step of every simulation, resulting in a substantially larger and richer training set.
The projected-force-MLP demonstrates more competitive performance and benefits consistently from increased dataset size. Nevertheless, its accuracy generally remains below that of the p-FNM. The only exception is the GAF phase prediction on the small and medium datasets, where both approaches achieve comparable results. As the amount of training data increases, however, the p-FNM exhibits larger gains in both predictive accuracy and robustness, leading to a widening performance gap.
These results support the choice of predicting the pressure field as an intermediate physical quantity from which the GAF is subsequently derived. This strategy yields substantially better results than direct GAF prediction and generally outperforms direct projected-force prediction.
Beyond predictive performance, two additional considerations motivate the pressure-field formulation. First, both the projected-force-MLP and GAF-MLP are intrinsically tied to the selected mechanical model and its underlying assumptions. Any modification of the mechanical model would therefore require retraining the corresponding surrogate. In contrast, the pressure-field surrogate remains independent of the downstream mechanical analysis, providing greater flexibility. Second, pressure-field predictions enable the computation of a wide range of derived quantities beyond the GAF, while also offering direct access to the underlying flow physics. From an engineering and design perspective, the ability to visualize and analyze the unsteady pressure distribution constitutes a significant advantage.
\subsection{About a Physics-Based Loss for the p-FNM} \label{sec:result:physics-loss}

Preliminary experiments were conducted to investigate the incorporation of a projected-force-based loss term alongside the pressure reconstruction loss, with the objective of improving GAF prediction accuracy. However, this multi-objective formulation consistently yielded inferior performance compared with training based solely on the pressure-field loss. In addition, the inclusion of a projected-force loss introduces a dependence on the selected mechanical mode and the associated modeling assumptions, thereby reducing the flexibility of the framework and requiring retraining whenever these assumptions are modified. For these reasons, the p-FNM presented in this work is trained exclusively using the pressure reconstruction loss.

Nevertheless, the integration of physics-informed objectives remains a promising research direction. The simultaneous optimization of pressure reconstruction and physics-based losses may introduce gradient conflicts between competing objectives, potentially deteriorating convergence, as discussed by \cite{Liu2024ConFIG}. Such issues could be alleviated through dedicated multi-objective optimization strategies specifically designed to resolve conflicting gradients. Furthermore, physics-informed loss terms have been shown to improve the prediction of shock structures and other localized flow features \cite{ghoreishi2026physics, mizuno2026physics}, suggesting potential benefits for the accurate reconstruction of strong pressure gradients.

An alternative approach would consist of preserving a general-purpose unsteady pressure predictor while training a secondary surrogate model that maps the predicted pressure fields to derived quantities such as the projected force or the GAF. Such a correction model could compensate for pressure-prediction errors in a manner specifically tailored to the target aeroelastic quantities, while retaining the flexibility and interpretability associated with pressure-field predictions.

\newpage   
\section{Conclusion} \label{section:conclusion}
This work introduced a neural-operator framework, named periodic Fourier Neural Mapping (p-FNM), for the prediction of unsteady pressure fields on turbine rotor blades and the subsequent estimation of GAFs under chorochronic hypothesis. The proposed approach combines a Fourier Neural Operator (FNO) architecture with an explicit periodic embedding of time, enabling the direct learning of a continuous mapping from operating conditions and phase to pressure fields.

The proposed framework was assessed on a database of high-fidelity unsteady simulations and compared with the Temporal Prediction Model (TPM) introduced in \cite{dominique2026reduced}, as well as with a SOAP-optimized variant of this baseline (TPM-SOAP). For a comparable number of trainable parameters, the p-FNM consistently achieved superior predictive performances across all considered training datasets. Improvements were observed not only for pressure-field reconstruction metrics but also, and more importantly, for GAF magnitude and phase predictions, while simultaneously reducing predictive uncertainty.

The analysis highlighted that GAF prediction constitutes a substantially more demanding task than pressure reconstruction alone. Although pressure errors are integrated spatially when computing aerodynamic loads, temporal inconsistencies in the predicted pressure fields strongly affect the spectral content of the projected-force signal and therefore the resulting GAF. Through pressure-field visualizations, Fourier analyses, and comparisons with end-to-end load-prediction models, we showed that preserving temporal continuity is a key requirement for accurate GAF predictions. The continuous time-wise formulation of the p-FNM model proved particularly effective in this respect and outperformed the sequential latent-space paradigm adopted by the TPM.

The influence of training dataset size was also investigated. For both pressure and GAF prediction, predictive accuracy and robustness monotonically improves as additional simulation data become available. 

Several directions for future research can be identified. First, the integration of physics-informed objectives could improve the prediction of localized flow structures such as shock waves and further enhance aerodynamic load accuracy. Second, uncertainty-aware and multi-fidelity formulations could be investigated to improve data efficiency in low-data regimes. Finally, extending the proposed framework toward design optimization appears particularly promising, as the p-FNM provides rapid access to both unsteady pressure fields and derived aerodynamic quantities that could constitutes a potential building block for future machine-learning-assisted turbomachinery design workflows.

Overall, the results demonstrate that neural-operator formulations specifically designed to exploit the periodic nature of unsteady turbomachinery flows provide a robust and accurate alternative to sequential reduced-order models. The proposed periodic FNM significantly improves the prediction of both pressure fields and GAF while maintaining a computational cost that is negligible compared with that of high-fidelity numerical simulations.

\section*{Acknowledgments}
The presented work has been performed in the framework of the HE-ART project, part of the Clean Aviation Joint Undertaking, and funded by the European Union’s Horizon 2020 research and innovation programme under grant agreement N° 101102013. It was also supported by the "ARIAC by DigitalWallonia4.ai" research project (grant agreement No 2010235– TRAIL institute) and benefited from computational resources made available on the Tier-1 supercomputer of the Fédération Wallonie-Bruxelles, infrastructure funded by the Walloon Region under grant agreement N° 1910847.
\section*{AI Disclosure Statement}
Generative AI tools, such as ChatGPT, were used for language editing and improving the clarity of the manuscript. All intellectual content, research design, and data analysis were conducted solely by the authors.
\bibliographystyle{unsrt}  
\bibliography{references}  
\appendix
\clearpage
\section{Model Complexity} \label{appendix:models-size}
\begin{table}[ht]
    \centering
    \caption{Number of trainable parameters for the deep learning architectures considered in this study. The TPM parameter count is reported both for the recurrent prediction module alone and for the complete model, including the VAE component.}
    \begin{tabular}{l|c}
        Model & Parameters count\\
        \hline
        VAE & 1,083,649 \\
        \hline
        TPM (RNN module only) & 75,104 \\
        \hline
        TPM (full architecture) & 1,158,753 \\
        \hline
        p-FNM & 1,156,094 \\
        \hline
        GAF-MLP & 1,169,427 \\
        \hline
        Projected-force-MLP & 1,199,425 \\
    \end{tabular}
    \label{tab:models-size}
\end{table}
\section{Updated TPM Baseline with SOAP Optimizer} \label{appendix:tpm-soap}
The TPM baseline introduced in \cite{dominique2026reduced} was originally trained using AdamW. To ensure a fair comparison and to assess the impact of the optimizer, both the TPM model and its VAE component were retrained using the SOAP optimizer. The corresponding results are reported in Table \ref{tab:vae-vs-vae-soap-all}. For both optimization schemes, hyperparameters were tuned to maximize predictive performance. The only differences between the original TPM and its SOAP-optimized variant (TPM-SOAP) concern the optimizer and its associated training hyperparameters, namely the learning-rate schedule and learning rate. All other aspects, including the model architecture, parameter count, and remaining training hyperparameters, were kept unchanged.
The experiments indicate that the use of SOAP consistently improves pressure prediction metrics while reducing predictive uncertainty for both the VAE and TPM models. Similar trends are observed for GAF magnitude prediction, where both accuracy and uncertainty are enhanced. Regarding GAF phase prediction, SOAP generally yields comparable or improved accuracy. Although a slight increase in uncertainty is observed for the standalone VAE on the small and medium datasets, uncertainty is consistently reduced for the TPM model across all datasets. Training times reported in Table \ref{tab:vae-vs-vae-soap-time} indicate that the use of SOAP increases the computational cost of training for both the VAE and TPM models. Depending on the dataset, the increase in training time ranges from $12\%$ to $47\%$ and is accompanied by comparable or larger variability across independent training runs. Overall, training the TPM baseline with SOAP yields improved predictive performance and reduced uncertainty for both pressure-field reconstruction and GAF estimation across all considered datasets, at the expense of increased training time.

\begin{table}[ht]
    \centering
    \caption{Comparison of the VAE and TPM models \cite{dominique2026reduced} with their variants trained using the SOAP optimizer. Metrics are reported as the mean $\pm$ 1.96 std for the test dataset across 30 training runs.}
    \begin{tabular}{c|l|c|c|c|c}
        Dataset & Metric & VAE & VAE with SOAP & TPM & TPM-SOAP \\
        \hline 
        \multirow[c]{7}{*}{\centering\rotatebox{90}{Small}} & Pressure MAE [kPa] & 0.97 \scriptsize{$\pm$ 0.08} & 0.87 \scriptsize{$\pm$ 0.03} & 2.05 \scriptsize{$\pm$ 0.34}  & 1.80 \scriptsize{$\pm$ 0.31}\\
        & Pressure MAPE [\%] & 1.59 \scriptsize{$\pm$ 0.14} & 1.42 \scriptsize{$\pm$ 0.047} & 3.35 \scriptsize{$\pm$ 0.54}  & 2.91 \scriptsize{$\pm$ 0.48}\\
        & Pressure MME [kPa] & 18.74 \scriptsize{$\pm$ 1.63} & 16.98 \scriptsize{$\pm$ 1.10} & 28.53 \scriptsize{$\pm$ 3.58} & 25.30 \scriptsize{$\pm$ 3.04}\\
        & Pressure $R^2$ & 0.9904 \scriptsize{$\pm$ 0.0014} & 0.9914 \scriptsize{$\pm$ 0.0013} & 0.9501 \scriptsize{$\pm$ 0.018} &  0.9610 \scriptsize{$\pm$ 0.015} \\
        & GAF norm $\mathrm{MAPE}_\rho$ [\%] & 27.29 \scriptsize{$\pm$ 11.28} & 25.56 \scriptsize{$\pm$ 6.59} & 67.11 \scriptsize{$\pm$ 32.24} & 38.52 \scriptsize{$\pm$ 13.75}\\
        & GAF norm $R^2_\rho$ & 0.70 \scriptsize{$\pm$ 0.16} & 0.71 \scriptsize{$\pm$ 0.089} & 0.018 \scriptsize{$\pm$ 0.43} & 0.45 \scriptsize{$\pm$ 0.31} \\
        & GAF phase $\mathrm{wRMSE}_\phi$ [rad] & 0.53 \scriptsize{$\pm$ 0.11} & 0.53 \scriptsize{$\pm$ 0.14} & 0.68 \scriptsize{$\pm$ 0.17} & 0.51 \scriptsize{$\pm$ 0.16} \\
        \hline  
        \multirow[c]{7}{*}{\centering\rotatebox{90}{Medium}} & Pressure MAE [kPa] & 0.83 \scriptsize{$\pm$ 0.060} & 0.73 \scriptsize{$\pm$ 0.014} & 1.40 \scriptsize{$\pm$ 0.26} & 1.14 \scriptsize{$\pm$ 0.10} \\
        & Pressure MAPE [\%] & 1.37 \scriptsize{$\pm$ 0.090} & 1.20 \scriptsize{$\pm$ 0.023} & 2.32 \scriptsize{$\pm$ 0.43} & 1.87 \scriptsize{$\pm$ 0.17} \\
        & Pressure MME [kPa] & 16.38 \scriptsize{$\pm$ 1.29} & 15.14 \scriptsize{$\pm$ 1.27} & 23.24 \scriptsize{$\pm$ 3.44} & 19.47 \scriptsize{$\pm$ 2.05} \\
        & Pressure $R^2$ & 0.9932 \scriptsize{$\pm$ 9.1e-4} & 0.9947 \scriptsize{$\pm$ 1.7e-4} & 0.9793 \scriptsize{$\pm$ 0.0067} & 0.9862 \scriptsize{$\pm$ 0.0022} \\
        & GAF norm $\mathrm{MAPE}_\rho$ [\%] & 26.24 \scriptsize{$\pm$ 7.07} & 25.73 \scriptsize{$\pm$ 5.87} & 39.84 \scriptsize{$\pm$ 12.46} & 32.77 \scriptsize{$\pm$ 9.51} \\
        & GAF norm $R^2_\rho$ & 0.76 \scriptsize{$\pm$ 0.083} & 0.80 \scriptsize{$\pm$ 0.065} & 0.45 \scriptsize{$\pm$ 0.20} & 0.61 \scriptsize{$\pm$ 0.16} \\
        & GAF phase $\mathrm{wRMSE}_\phi$ [rad] & 0.53 \scriptsize{$\pm$ 0.069} & 0.50 \scriptsize{$\pm$ 0.096} & 0.55 \scriptsize{$\pm$ 0.21} & 0.40 \scriptsize{$\pm$ 0.092} \\
        \hline 
        \multirow[c]{7}{*}{\centering\rotatebox{90}{Large}} & Pressure MAE [kPa] & 0.74 \scriptsize{$\pm$ 0.041} & 0.66 \scriptsize{$\pm$ 0.013} & 0.87 \scriptsize{$\pm$ 0.092} & 0.70 \scriptsize{$\pm$ 0.12} \\
        &  Pressure MAPE [\%] & 1.22 \scriptsize{$\pm$ 0.073} & 1.08 \scriptsize{$\pm$ 0.022} & 1.45 \scriptsize{$\pm$ 0.15} & 1.16 \scriptsize{$\pm$ 0.020} \\
        & Pressure MME [kPa] & 15.04 \scriptsize{$\pm$ 1.08} & 14.19 \scriptsize{$\pm$ 1.08} & 15.34 \scriptsize{$\pm$ 1.53} & 14.69 \scriptsize{$\pm$ 1.62} \\
        & Pressure $R^2$ & 0.9948 \scriptsize{$\pm$ 6.1e-4} & 0.9959 \scriptsize{$\pm$ 1.72e-4} & 0.9920 \scriptsize{$\pm$ 0.0017} & 0.9950 \scriptsize{$\pm$ 0.0029} \\
        & GAF norm $\mathrm{MAPE}_\rho$ [\%] & 17.79 \scriptsize{$\pm$ 6.51} & 12.70 \scriptsize{$\pm$ 3.99} & 29.44 \scriptsize{$\pm$ 11.51} & 19.79 \scriptsize{$\pm$ 6.20}\\
        & GAF norm $R^2_\rho$ & 0.88 \scriptsize{$\pm$ 0.056} & 0.92 \scriptsize{$\pm$ 0.033} & 0.78 \scriptsize{$\pm$ 0.11} & 0.86 \scriptsize{$\pm$ 0.064} \\
        & GAF phase $\mathrm{wRMSE}_\phi$ [rad] & 0.25 \scriptsize{$\pm$ 0.085} & 0.20 \scriptsize{$\pm$ 0.033} & 0.31 \scriptsize{$\pm$ 0.079} & 0.23 \scriptsize{$\pm$ 0.063} \\
        \hline 
        \multirow[c]{7}{*}{\centering\rotatebox{90}{Extra Large}} & Pressure MAE [kPa] & 0.69 \scriptsize{$\pm$ 0.033} & 0.62 \scriptsize{$\pm$ 0.014} & 0.71 \scriptsize{$\pm$ 0.080} & 0.52 \scriptsize{$\pm$ 0.020} \\
        & Pressure MAPE [\%] & 1.12 \scriptsize{$\pm$ 0.054} & 1.02 \scriptsize{$\pm$ 0.024} & 1.17 \scriptsize{$\pm$ 0.14} & 0.85 \scriptsize{$\pm$ 0.033} \\
        & Pressure MME [kPa] & 14.18 \scriptsize{$\pm$ 1.17} & 13.53 \scriptsize{$\pm$ 0.89} & 12.86 \scriptsize{$\pm$ 1.41} & 10.51 \scriptsize{$\pm$ 0.79} \\
        & Pressure $R^2$ & 0.9955 \scriptsize{$\pm$ 4.27e-4} & 0.9963 \scriptsize{$\pm$ 1.83e-4} & 0.9947 \scriptsize{$\pm$ 0.013} & 0.9973 \scriptsize{$\pm$ 2.23e-4}\\
        & GAF norm $\mathrm{MAPE}_\rho$ [\%] & 13.73 \scriptsize{$\pm$ 4.81} & 10.78 \scriptsize{$\pm$ 2.59} & 19.73 \scriptsize{$\pm$ 5.72} & 11.84 \scriptsize{$\pm$ 2.93}\\
        & GAF norm $R^2_\rho$ & 0.92 \scriptsize{$\pm$ 0.037} & 0.95 \scriptsize{$\pm$ 0.019} & 0.88 \scriptsize{$\pm$ 0.045} &  0.93 \scriptsize{$\pm$ 0.023} \\
         & GAF phase $\mathrm{wRMSE}_\phi$ [rad] & 0.22 \scriptsize{$\pm$ 0.094} & 0.17 \scriptsize{$\pm$ 0.035} & 0.27 \scriptsize{$\pm$ 0.077} & 0.14 \scriptsize{$\pm$ 0.032} \\
    \end{tabular}
    \label{tab:vae-vs-vae-soap-all}
\end{table}
\begin{table}[ht]
    \centering
    \caption{Comparison of models training time [GPU.h] for the VAE and TPM models \cite{dominique2026reduced} with their variants trained using the SOAP optimizer. Metrics are reported as the mean $\pm$ 1.96 std for the test set across 30 training runs.}
    \begin{tabular}{l|c|c|c|c|c|c}
        Dataset & VAE & VAE with SOAP & TPM alone & \makecell{TPM alone\\(SOAP variant)} & TPM total & \makecell{TPM-SOAP\\total}\\
        \hline 
        Small & 0.16 \scriptsize{$\pm$ 0.08} & 0.20 \scriptsize{$\pm$ 0.08} & 0.02 \scriptsize{$\pm$ 0.003} & 0.03 \scriptsize{$\pm$ 0.01} & 0.18 \scriptsize{$\pm$ 0.08} & 0.23 \scriptsize{$\pm$ 0.09} \\
        \hline  
        Medium & 0.29 \scriptsize{$\pm$ 0.12} & 0.32 \scriptsize{$\pm$ 0.11} & 0.04 \scriptsize{$\pm$ 0.01} & 0.05 \scriptsize{$\pm$ 0.02} & 0.33 \scriptsize{$\pm$ 0.13} & 0.37 \scriptsize{$\pm$ 0.13}\\
        \hline 
        Large & 0.46 \scriptsize{$\pm$ 0.18} & 0.64 \scriptsize{$\pm$ 0.28} & 0.11 \scriptsize{$\pm$ 0.03} & 0.14 \scriptsize{$\pm$ 0.07} & 0.57 \scriptsize{$\pm$ 0.21} & 0.78 \scriptsize{$\pm$ 0.35} \\
        \hline 
        Extra Large & 0.83 \scriptsize{$\pm$ 0.25} & 1.38 \scriptsize{$\pm$ 0.56} & 0.29 \scriptsize{$\pm$ 0.03} & 0.27 \scriptsize{$\pm$ 0.1} & 1.12 \scriptsize{$\pm$ 0.28} & 1.65 \scriptsize{$\pm$ 0.66}
    \end{tabular}
    \label{tab:vae-vs-vae-soap-time}
\end{table}

\clearpage
\section{Hyperparameters for the p-FNM} \label{appendix:fnm-hyperparams}

The hyperparameters of the p-FNM architecture are summarized in Table\ref{tab:appendix-fnm-hyperparams}. These hyperparameters were selected following an extensive series of preliminary experiments aimed at maximizing predictive performance.
Several architectural variants were investigated. In particular, different implementations of the V2F block were considered, including both linear mappings and multilayer perceptrons. Among the tested configurations, the linear V2F formulation consistently yielded the best performance. Additional experiments were conducted to evaluate the influence of the activation function, the number of retained Fourier modes in the spectral convolution layers, and the overall model capacity.
The effect of model size was also examined by varying the total number of trainable parameters. Although increasing the model capacity initially led to improved predictive performance, the gains rapidly diminished beyond a certain parameter count, indicating a saturation regime in which additional complexity provided only marginal benefits.
Furthermore, to enable a meaningful comparison with the TPM architecture, the number of trainable parameters in the p-FNM was chosen to be comparable to that of the TPM. Experiments with larger p-FNM models showed only negligible performance improvements, suggesting that the selected model capacity is sufficient to capture the dominant dynamics of the problem while maintaining computational efficiency.

\begin{table}[htp]
\centering
\caption{p-FNM model hyperparameters}
\begin{tabular}{l|r}
\hline
V2F block & \small{\verb|Linear(6, 1440) -> Truncated Fourier modes = 12x12|} \\
Lift operator $\mathcal{P}$ & \small{\verb|LinearChannelWise(6, 22)|} \\
FNO Layers count & \small{\verb|4|}  \\
FNO Layer $\mathcal{W}$ & \small{\verb|Conv2D(channels_in=22, channels_out=22, kernel_size=3, padding=1)|}  \\
FNO Layer SpectralConv & \small{\verb|SpectralConv2D(channels_in=22, channels_out=22, modes_w=12, modes_h=12)|}  \\
FNO Layer Activation & \small{\verb|ReLU|} \\
Projection operator $\mathcal{Q}$ & \small{\verb|LinearChannelWise(22, 128) -> LeakyReLU -> Linear(128, 1)|} \\
\hline
\end{tabular}
\label{tab:appendix-fnm-hyperparams}
\end{table}
\clearpage
\section{Detailed Results for the p-FNM Model} \label{appendix:detail-res}

\begin{table}[ht]
    \centering
    \caption{Detailed metrics for the p-FNM model across the training datasets. Metrics are reported as the mean $\pm$ 1.96 std for the test dataset across 30 training runs.}
    \begin{tabular}{c|l|c}
        Dataset & Metric & p-FNM \\
        \hline 
        \multirow[c]{8}{*}{\centering\rotatebox{90}{Small}} & Pressure MAE [kPa] & 1.21 \scriptsize{$\pm$ 0.095}\\
        & Pressure MAPE [\%] & 1.98 \scriptsize{$\pm$ 0.16}\\
        & Pressure MME [kPa] & 18.37 \scriptsize{$\pm$ 1.80} \\
        & Pressure $R^2$ &  0.9827 \scriptsize{$\pm$ 0.0026} \\
        & GAF norm $\mathrm{MAPE}_\rho$ [\%] & 32.13 \scriptsize{$\pm$ 7.30} \\
        & GAF norm $R^2_\rho$ & 0.57 \scriptsize{$\pm$ 0.21} \\
        & GAF phase $\mathrm{wRMSE}_\phi$ [rad] & 0.50 \scriptsize{$\pm$ 0.17} \\
        & Training time [GPU.h] & 0.04 \scriptsize{$\pm$ 0.01} \\
        \hline  
        \multirow[c]{8}{*}{\centering\rotatebox{90}{Medium}} & Pressure MAE [kPa] & 0.63 \scriptsize{$\pm$ 0.025}\\
        & Pressure MAPE [\%] & 1.02 \scriptsize{$\pm$ 0.039}\\
        & Pressure MME [kPa] & 10.27 \scriptsize{$\pm$ 0.52} \\
        & Pressure $R^2$ &  0.9952 \scriptsize{$\pm$ 4.27e-4} \\
        & GAF norm $\mathrm{MAPE}_\rho$ [\%] & 20.56 \scriptsize{$\pm$ 5.49} \\
        & GAF norm $R^2_\rho$ & 0.85 \scriptsize{$\pm$ 0.078} \\
        & GAF phase $\mathrm{wRMSE}_\phi$ [rad] & 0.38 \scriptsize{$\pm$ 0.12} \\
        & Training time [GPU.h] & 0.23 \scriptsize{$\pm$ 0.05} \\
        \hline 
        \multirow[c]{8}{*}{\centering\rotatebox{90}{Large}} & Pressure MAE [kPa] & 0.38 \scriptsize{$\pm$ 0.018}\\
        & Pressure MAPE [\%] & 0.63 \scriptsize{$\pm$ 0.030}\\
        & Pressure MME [kPa] & 6.25 \scriptsize{$\pm$ 0.38} \\
        & Pressure $R^2$ &  0.9986 \scriptsize{$\pm$ 1.33e-4} \\
        & GAF norm $\mathrm{MAPE}_\rho$ [\%] & 7.41 \scriptsize{$\pm$ 1.88} \\
        & GAF norm $R^2_\rho$ & 0.98 \scriptsize{$\pm$ 0.0084} \\
        & GAF phase $\mathrm{wRMSE}_\phi$ [rad] & 0.12 \scriptsize{$\pm$ 0.098} \\
        & Training time [GPU.h] & 0.60 \scriptsize{$\pm$ 0.07} \\
        \hline 
        \multirow[c]{8}{*}{\centering\rotatebox{90}{Extra Large}} & Pressure MAE [kPa] & 0.28 \scriptsize{$\pm$ 0.016}\\
        & Pressure MAPE [\%] & 0.46 \scriptsize{$\pm$ 0.027}\\
        & Pressure MME [kPa] & 5.67 \scriptsize{$\pm$ 0.36} \\
        & Pressure $R^2$ &  0.9992 \scriptsize{$\pm$ 8.84e-5} \\
        & GAF norm $\mathrm{MAPE}_\rho$ [\%] & 4.42 \scriptsize{$\pm$ 1.44} \\
        & GAF norm $R^2_\rho$ & 0.99 \scriptsize{$\pm$ 0.0050} \\
        & GAF phase $\mathrm{wRMSE}_\phi$ [rad] & 0.060 \scriptsize{$\pm$ 0.018} \\
        & Training time [GPU.h] & 1.21 \scriptsize{$\pm$ 0.08}
    \end{tabular}
    \label{tab:fnm-metrics-detailed}
\end{table}
\clearpage
\section{Pressure Field Predictions} \label{appendix:pressure-preds}
\begin{table}[!htp]
    \centering
    \caption{Example of predicted and reference pressure fields for a test-dataset simulation at a fixed time step. The red vertical line corresponds to the leading edge of the blade.}
    \vspace*{4mm}
    \begin{tabular}{c|c|c|c|c|c}
        \multicolumn{2}{c|}{Dataset} & Small & Medium & Large & Extra Large \\
        \hline
        \multirow{6}{*}{\rotatebox[origin=c]{90}{TPM}} & \rotatebox[origin=c]{90}{Prediction} &
        \raisebox{-0.5\height}{\shortstack[c]{\includegraphics[width=0.20\linewidth]{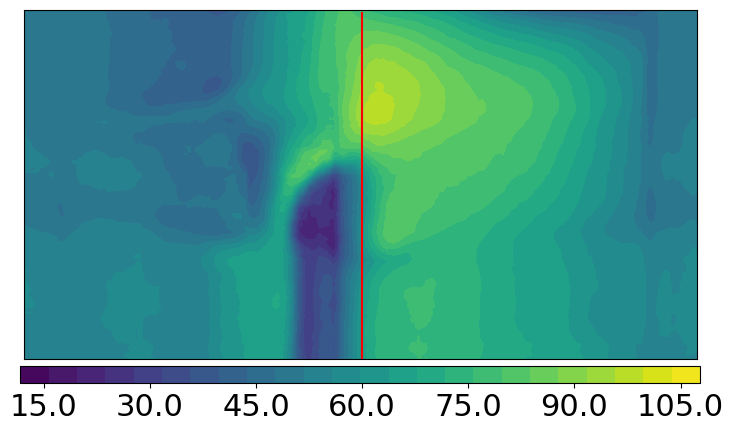}\\[-1mm]\tiny Pressure [kPa]\\[-1mm]}} &
        \raisebox{-0.5\height}{\shortstack[c]{\includegraphics[width=0.20\linewidth]{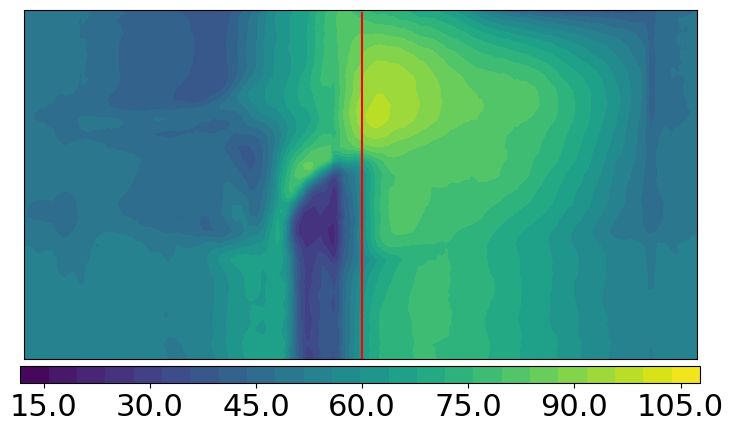}\\[-1mm]\tiny Pressure [kPa]}} &
        \raisebox{-0.5\height}{\shortstack[c]{\includegraphics[width=0.20\linewidth]{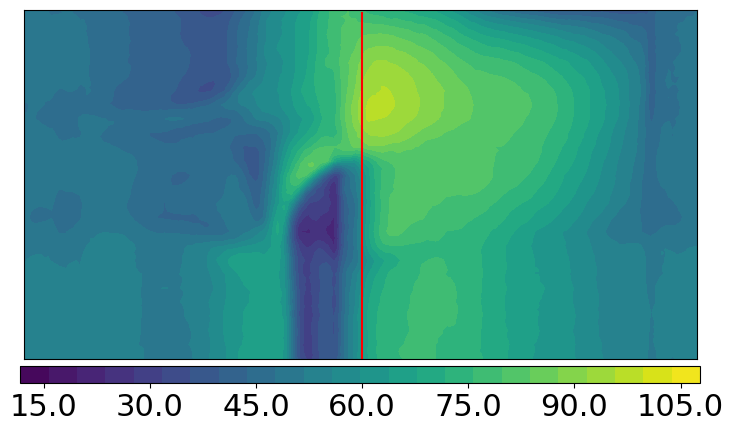}\\[-1mm]\tiny Pressure [kPa]}} &
        \raisebox{-0.5\height}{\shortstack[c]{\includegraphics[width=0.20\linewidth]{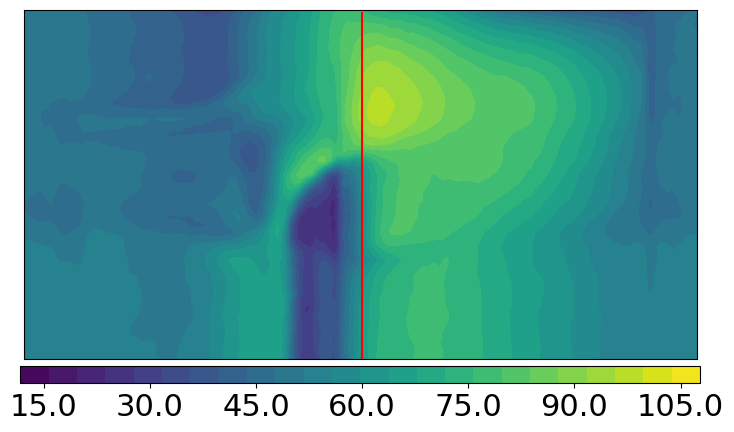}\\[-1mm]\tiny Pressure [kPa]}} \\
        \cline{2-6}
        & \rotatebox[origin=c]{90}{MAE} &
        \raisebox{-0.5\height}{\shortstack[c]{\includegraphics[width=0.20\linewidth]{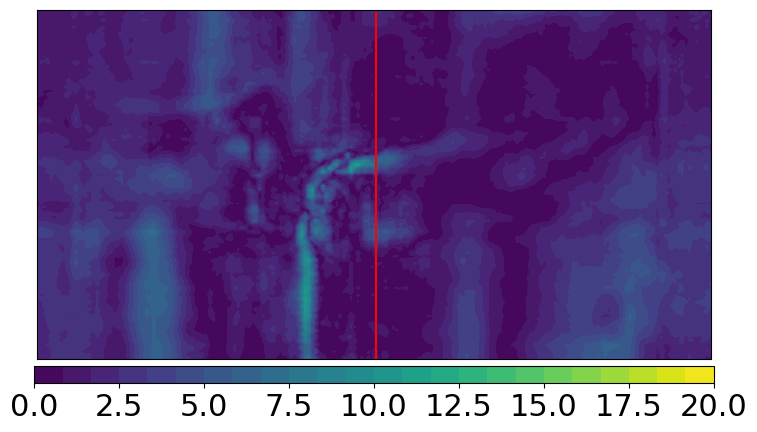}\\[-1mm]\tiny Pressure [kPa]}} &
        \raisebox{-0.5\height}{\shortstack[c]{\includegraphics[width=0.20\linewidth]{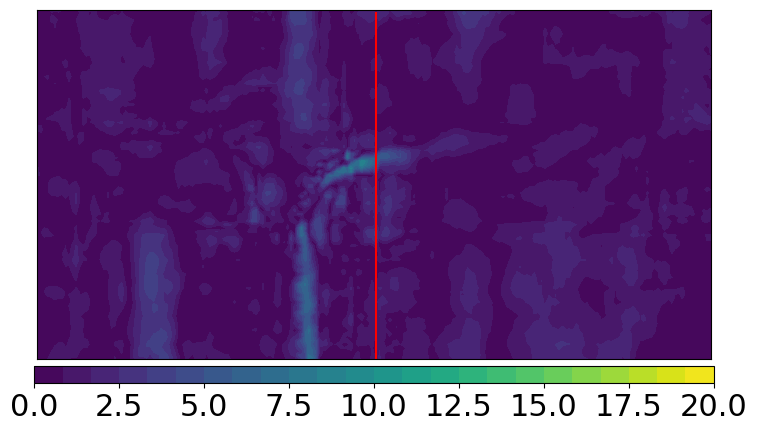}\\[-1mm]\tiny Pressure [kPa]}} &
        \raisebox{-0.5\height}{\shortstack[c]{\includegraphics[width=0.20\linewidth]{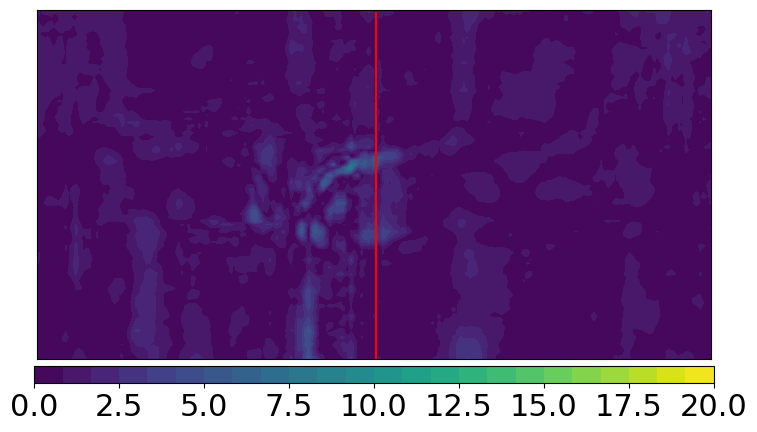}\\[-1mm]\tiny Pressure [kPa]}} &
        \raisebox{-0.5\height}{\shortstack[c]{\includegraphics[width=0.20\linewidth]{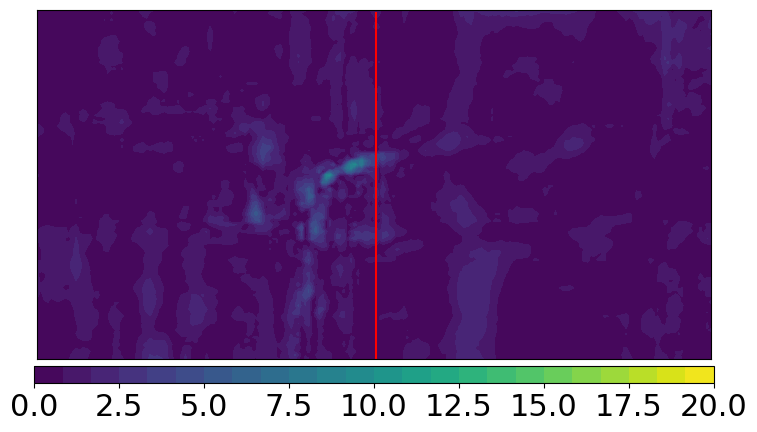}\\[-1mm]\tiny Pressure [kPa]}} \\
        \hline \hline
        \multirow{6}{*}{\rotatebox[origin=c]{90}{p-FNM}} & \rotatebox[origin=c]{90}{Prediction} &
        \raisebox{-0.5\height}{\shortstack[c]{\includegraphics[width=0.20\linewidth]{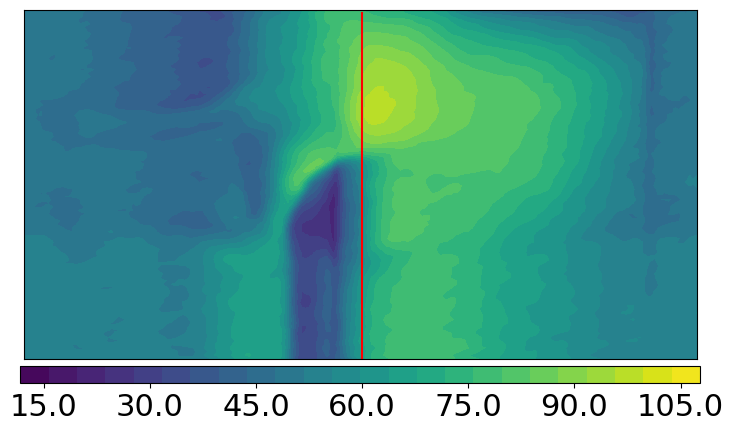}\\[-1mm]\tiny Pressure [kPa]}} &
        \raisebox{-0.5\height}{\shortstack[c]{\includegraphics[width=0.20\linewidth]{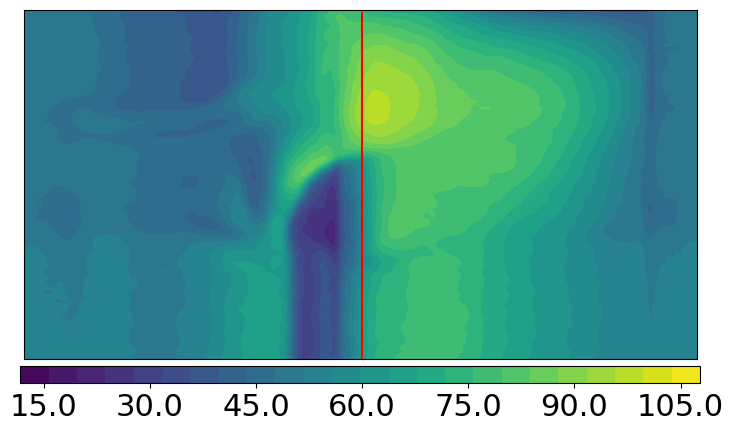}\\[-1mm]\tiny Pressure [kPa]}} &
        \raisebox{-0.5\height}{\shortstack[c]{\includegraphics[width=0.20\linewidth]{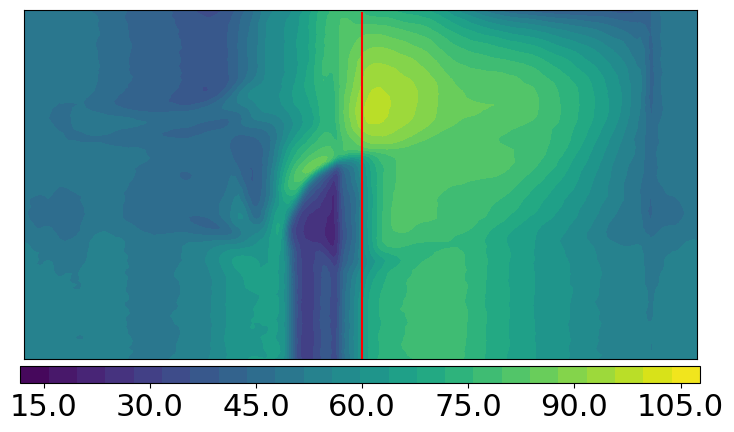}\\[-1mm]\tiny Pressure [kPa]}} &
        \raisebox{-0.5\height}{\shortstack[c]{\includegraphics[width=0.20\linewidth]{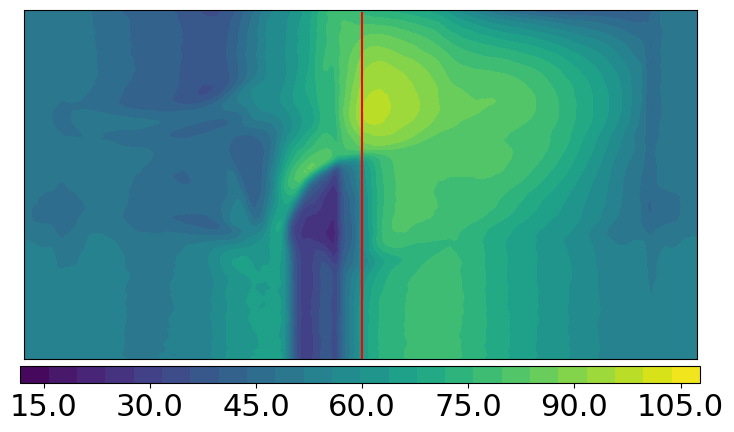}\\[-1mm]\tiny Pressure [kPa]}} \\
        \cline{2-6}
        & \rotatebox[origin=c]{90}{MAE} &
        \raisebox{-0.5\height}{\shortstack[c]{\includegraphics[width=0.20\linewidth]{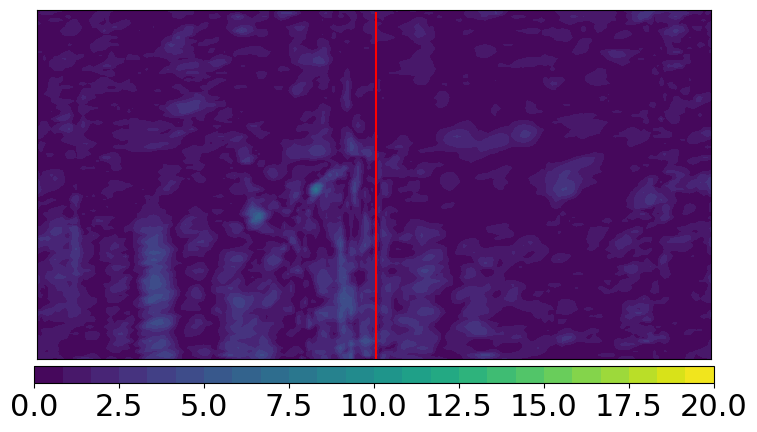}\\[-1mm]\tiny Pressure [kPa]}} &
        \raisebox{-0.5\height}{\shortstack[c]{\includegraphics[width=0.20\linewidth]{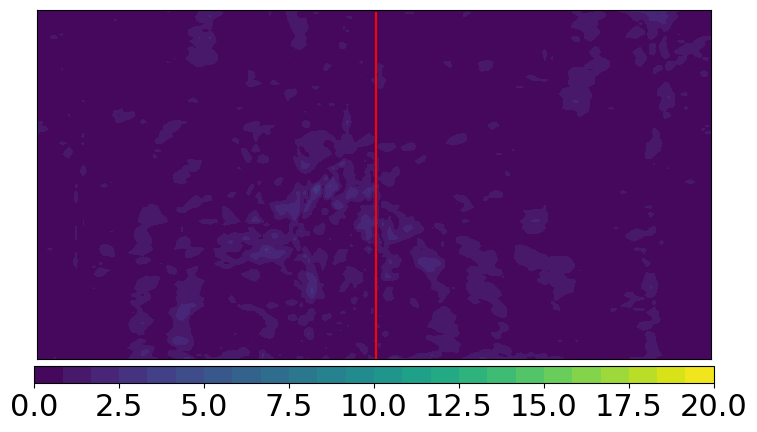}\\[-1mm]\tiny Pressure [kPa]}} &
        \raisebox{-0.5\height}{\shortstack[c]{\includegraphics[width=0.20\linewidth]{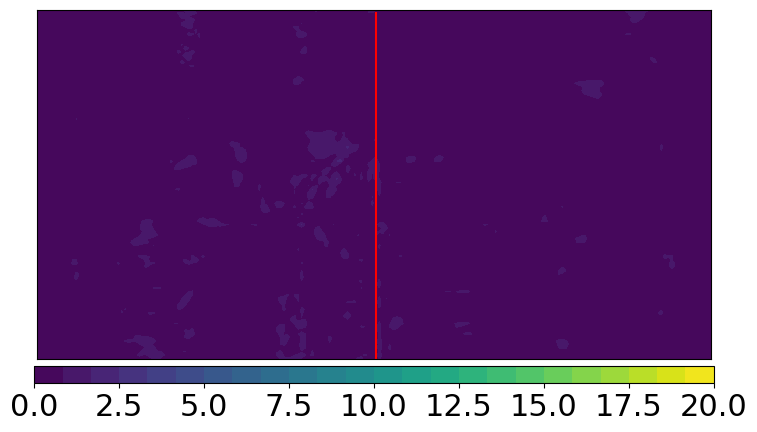}\\[-1mm]\tiny Pressure [kPa]}} &
        \raisebox{-0.5\height}{\shortstack[c]{\includegraphics[width=0.20\linewidth]{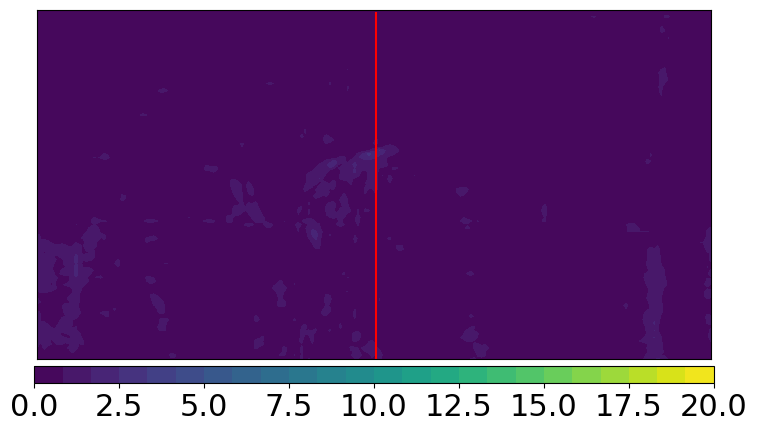}\\[-1mm]\tiny Pressure [kPa]}} \\
        \hline \hline
        \multicolumn{2}{c|}{\rotatebox[origin=c]{90}{Ground truth}} & 
        \multicolumn{4}{c}{\raisebox{-0.5\height}{\shortstack[c]{\includegraphics[width=0.30\linewidth]{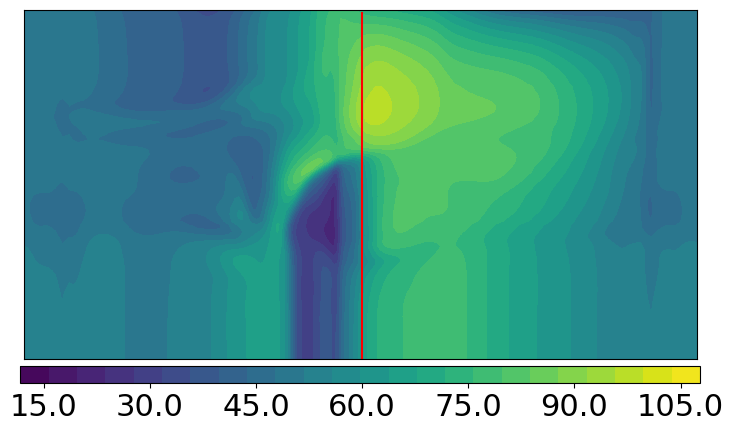}\\[-1mm]\tiny Pressure [kPa]}}} \\
        \hline
    \end{tabular}
    \label{fig:pressure-pred-sim0-t14}
\end{table}
\begin{table}[!htp]
    \centering
    \caption{Example of predicted and reference pressure fields for a test-dataset simulation at a fixed time step. The red vertical line corresponds to the leading edge of the blade.}
    \vspace*{4mm}
    \begin{tabular}{c|c|c|c|c|c}
        \multicolumn{2}{c|}{Dataset} & Small & Medium & Large & Extra Large \\
        \hline
        \multirow{6}{*}{\rotatebox[origin=c]{90}{TPM}} & \rotatebox[origin=c]{90}{Prediction} &
        \raisebox{-0.5\height}{\shortstack[c]{\includegraphics[width=0.20\linewidth]{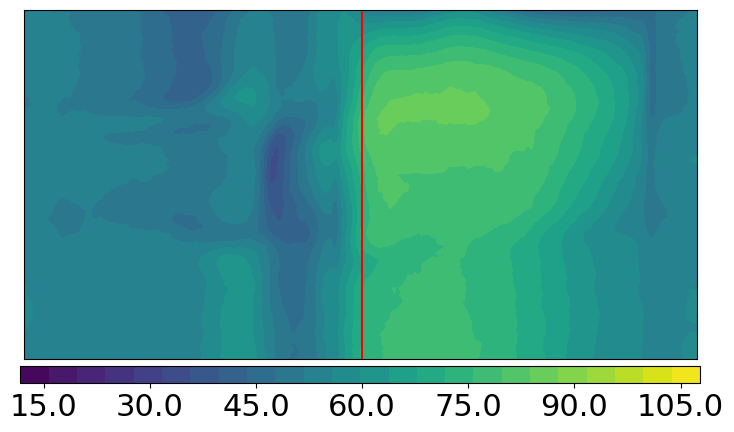}\\[-1mm]\tiny Pressure [kPa]}} &
        \raisebox{-0.5\height}{\shortstack[c]{\includegraphics[width=0.20\linewidth]{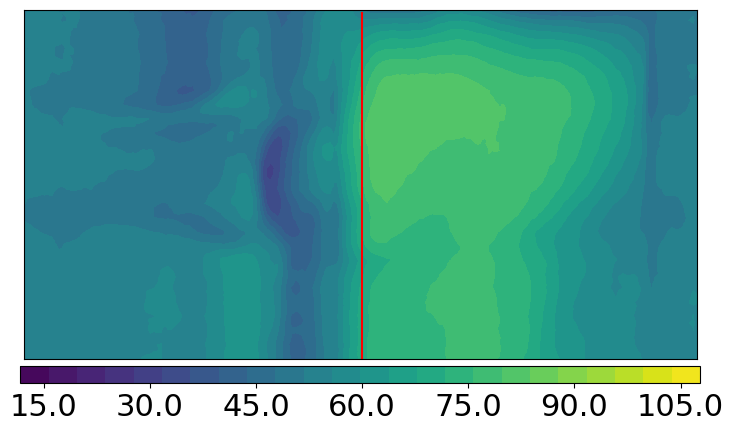}\\[-1mm]\tiny Pressure [kPa]}} &
        \raisebox{-0.5\height}{\shortstack[c]{\includegraphics[width=0.20\linewidth]{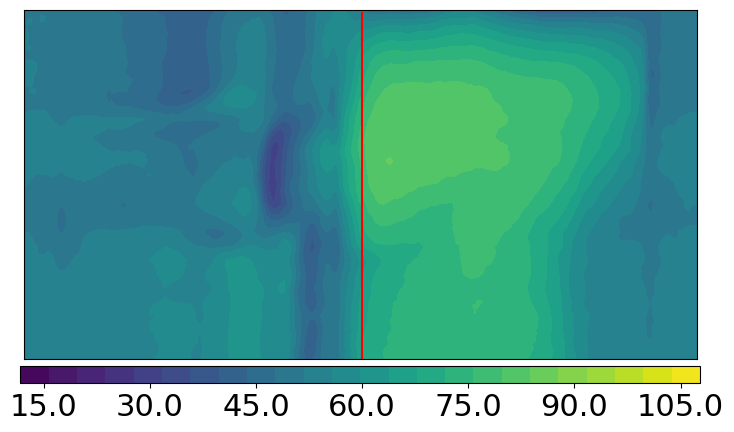}\\[-1mm]\tiny Pressure [kPa]}} &
        \raisebox{-0.5\height}{\shortstack[c]{\includegraphics[width=0.20\linewidth]{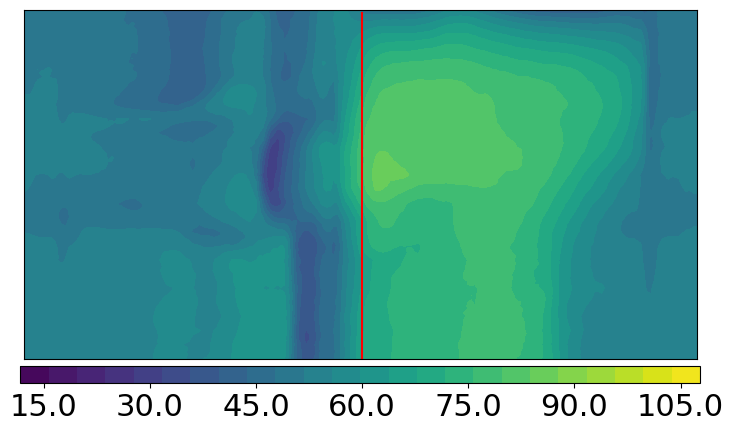}\\[-1mm]\tiny Pressure [kPa]}} \\
        \cline{2-6}
        & \rotatebox[origin=c]{90}{MAE} &
        \raisebox{-0.5\height}{\shortstack[c]{\includegraphics[width=0.20\linewidth]{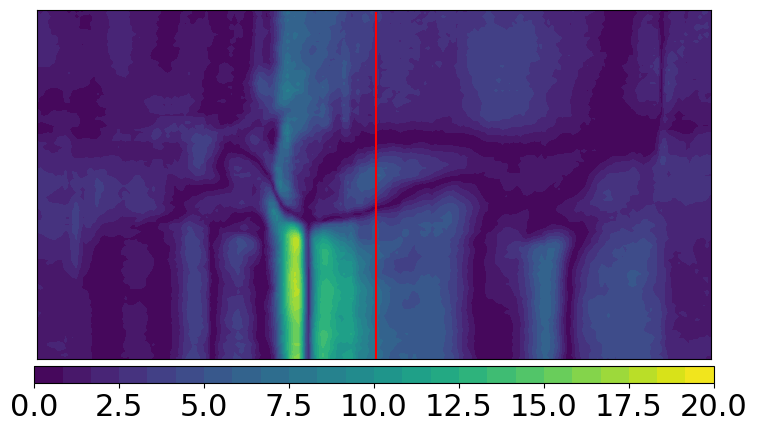}\\[-1mm]\tiny Pressure [kPa]}} &
        \raisebox{-0.5\height}{\shortstack[c]{\includegraphics[width=0.20\linewidth]{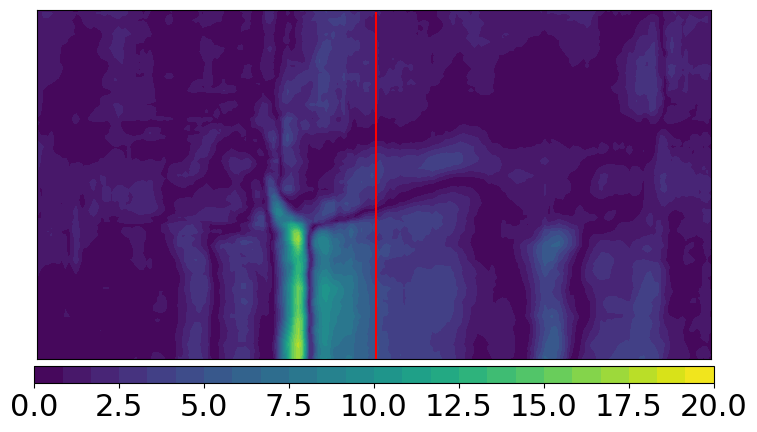}\\[-1mm]\tiny Pressure [kPa]}} &
        \raisebox{-0.5\height}{\shortstack[c]{\includegraphics[width=0.20\linewidth]{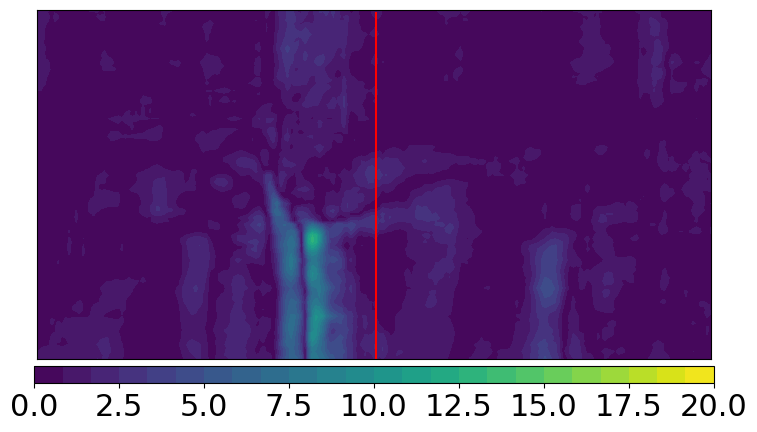}\\[-1mm]\tiny Pressure [kPa]}} &
        \raisebox{-0.5\height}{\shortstack[c]{\includegraphics[width=0.20\linewidth]{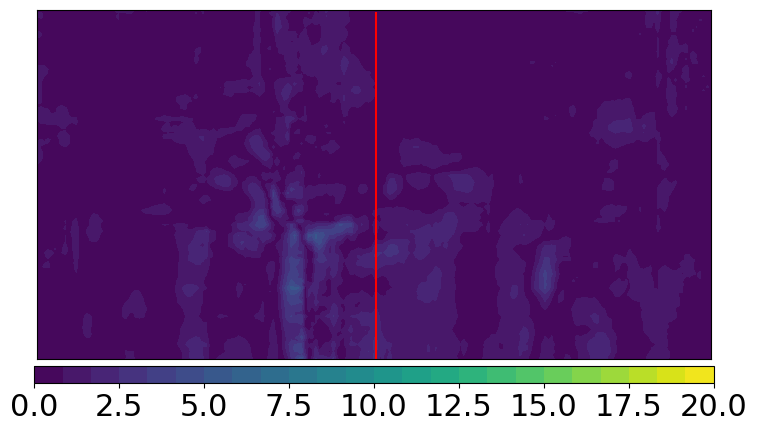}\\[-1mm]\tiny Pressure [kPa]}} \\
        \hline \hline
        \multirow{6}{*}{\rotatebox[origin=c]{90}{p-FNM}} & \rotatebox[origin=c]{90}{Prediction} &
        \raisebox{-0.5\height}{\shortstack[c]{\includegraphics[width=0.20\linewidth]{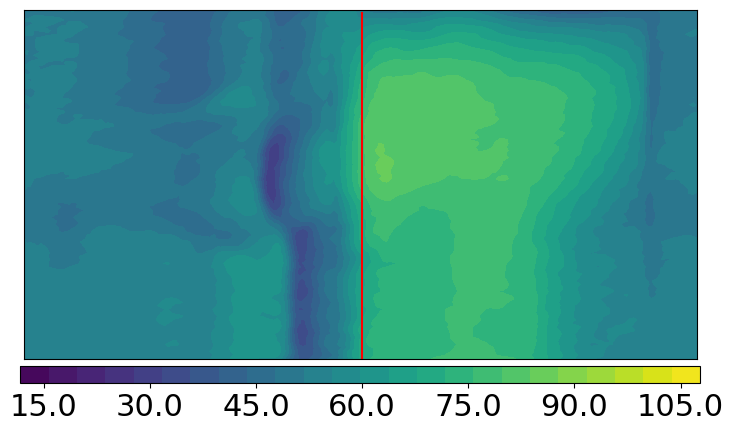}\\[-1mm]\tiny Pressure [kPa]}} &
        \raisebox{-0.5\height}{\shortstack[c]{\includegraphics[width=0.20\linewidth]{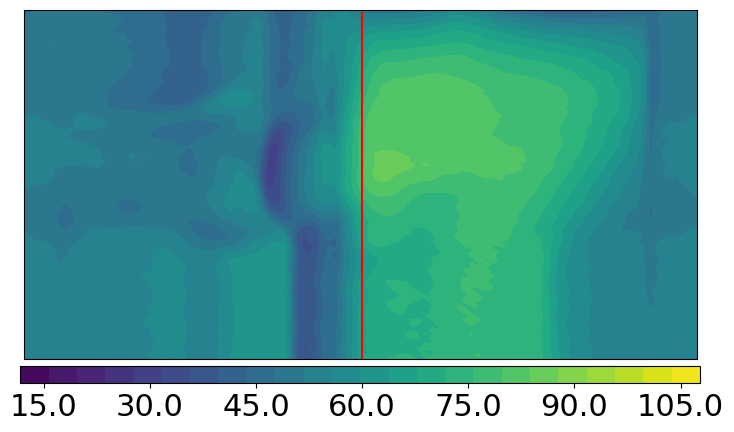}\\[-1mm]\tiny Pressure [kPa]}} &
        \raisebox{-0.5\height}{\shortstack[c]{\includegraphics[width=0.20\linewidth]{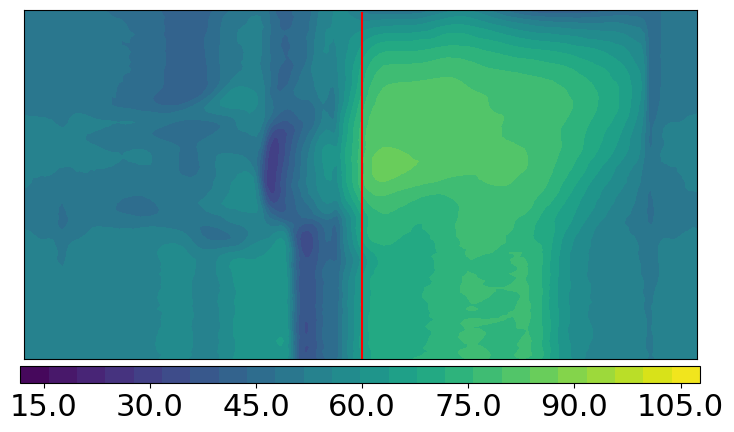}\\[-1mm]\tiny Pressure [kPa]}} &
        \raisebox{-0.5\height}{\shortstack[c]{\includegraphics[width=0.20\linewidth]{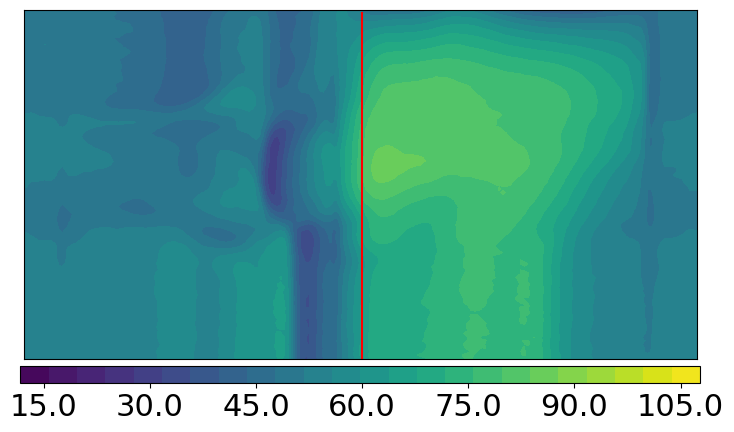}\\[-1mm]\tiny Pressure [kPa]}} \\
        \cline{2-6}
        & \rotatebox[origin=c]{90}{MAE} &
        \raisebox{-0.5\height}{\shortstack[c]{\includegraphics[width=0.20\linewidth]{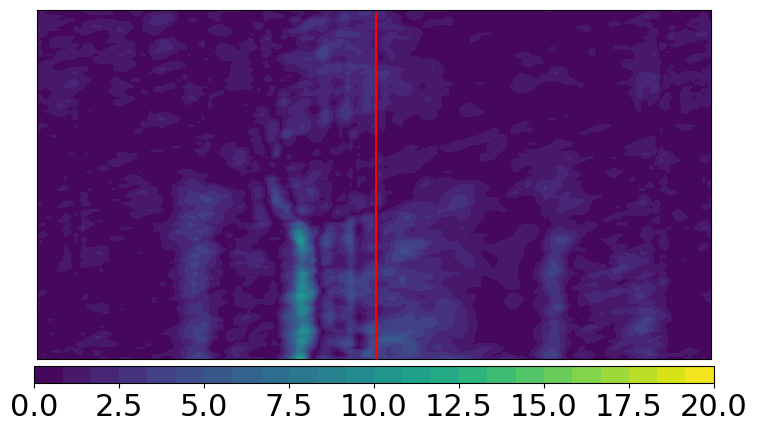}\\[-1mm]\tiny Pressure [kPa]}} &
        \raisebox{-0.5\height}{\shortstack[c]{\includegraphics[width=0.20\linewidth]{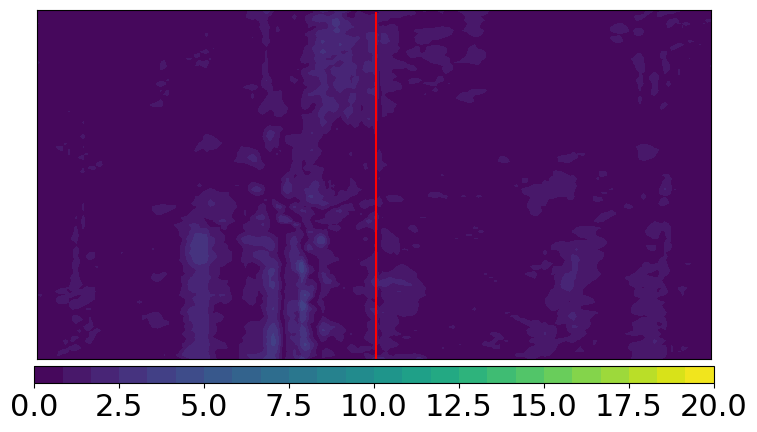}\\[-1mm]\tiny Pressure [kPa]}} &
        \raisebox{-0.5\height}{\shortstack[c]{\includegraphics[width=0.20\linewidth]{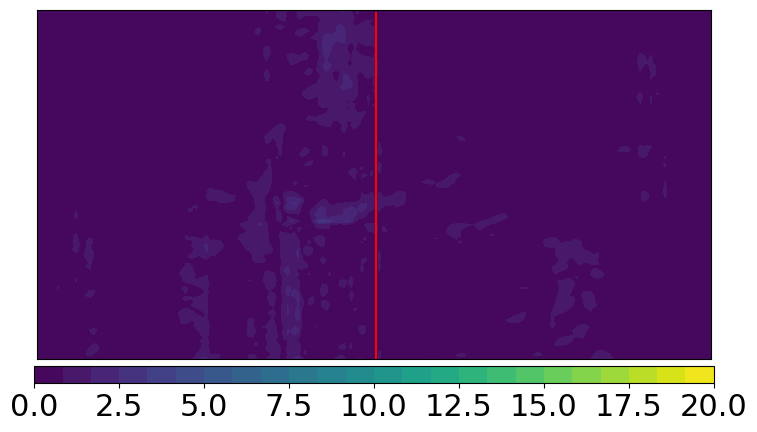}\\[-1mm]\tiny Pressure [kPa]}} &
        \raisebox{-0.5\height}{\shortstack[c]{\includegraphics[width=0.20\linewidth]{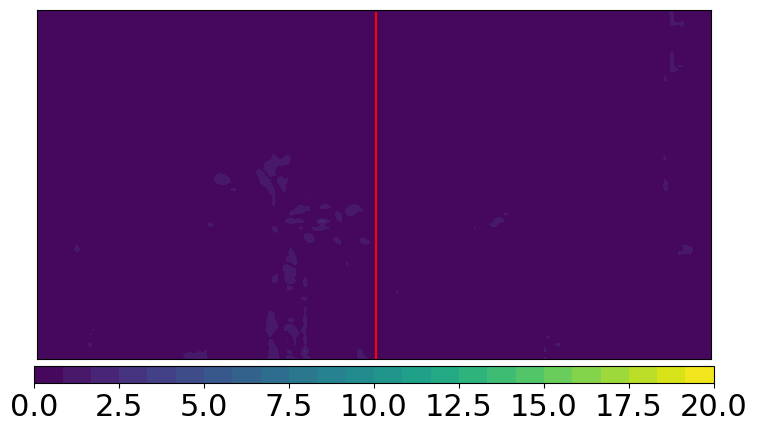}\\[-1mm]\tiny Pressure [kPa]}} \\
        \hline \hline
        \multicolumn{2}{c|}{\rotatebox[origin=c]{90}{Ground truth}} & 
        \multicolumn{4}{c}{\raisebox{-0.5\height}{\shortstack[c]{\includegraphics[width=0.30\linewidth]{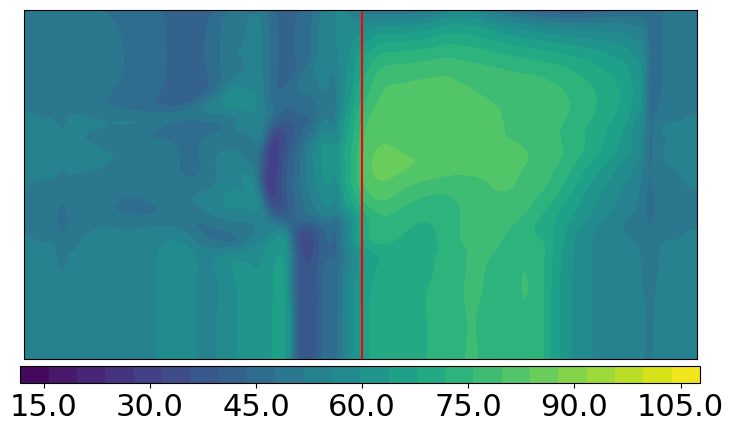}\\[-1mm]\tiny Pressure [kPa]}}} \\
        \hline
    \end{tabular}
    \label{fig:pressure-pred-sim1-t107}
\end{table}

\clearpage
\section{VAE Latent Space Analysis} \label{sec:appendix:vae-latent-analysis}


The VAE does not explicitly enforce any temporal structure, as it is trained only to reconstruct static pressure fields without direct information about temporal ordering. Nevertheless, due to the strong similarity between pressure fields at neighboring time steps, the VAE implicitly captures aspects of the underlying temporal dynamics. In particular, VAEs are known to organize similar samples into nearby latent representations \cite{higgins2017betavae, kingma2022autoencodingvariationalbayes}. More generally, the latent representation learned by a VAE is expected to preserve the intrinsic geometry of the data manifold, enabling smooth interpolation between similar flow states \cite{burgess2018understanding}.
To illustrate this behavior, all time steps from two simulations of the test set were encoded using a VAE trained on the large dataset. The resulting latent representations were subsequently projected into lower-dimensional spaces using t-SNE for visualization. Each latent representation consists of two components: the latent distribution means and the associated log-variances. The latent means were projected into a two-dimensional space, whereas the latent log-variances were projected into a three-dimensional space. The corresponding visualizations are reported in Table \ref{fig:vae-latent-analysis}.
A clear loop-like organization can be observed for both the latent means and the latent log-variances, indicating that temporal periodicity has been implicitly embedded within the latent-space structure. Furthermore, the smooth color gradients along these trajectories show that pressure fields associated with consecutive time steps remain close to one another in latent space.
These observations demonstrate that, although the VAE is trained exclusively on static pressure fields and receives no explicit temporal information, the learned latent representation nevertheless preserves an implicit notion of temporal continuity and periodicity.

A similar observation has recently been reported by Solera \textit{et al.} \cite{solera2024}. In their work, a $\beta$-VAE provides a compact latent representation of the flow, while a transformer learns its temporal evolution. The authors argue that increasing the $\beta$ parameter promotes a more disentangled and approximately decorrelated latent representation, drawing an analogy with the orthogonal modes produced by Proper Orthogonal Decomposition (POD). Although such a representation may improve interpretability, orthogonality is not a mathematical requirement for nonlinear reduced-order modeling \cite{otto2023learning}. Indeed, a nonlinear latent representation can accurately reconstruct and predict the dynamics of a system even when the latent coordinates are correlated. From the perspective of temporal prediction, a more fundamental property is that the latent manifold remains smooth and structured so that the underlying dynamics can be learned efficiently by the temporal model. The continuous trajectories observed in Fig.~\ref{fig:vae-latent-analysis} suggest that this property is naturally satisfied by the latent space learned in the present work, which helps explain why latent-space temporal predictors, such as the TPM, are able to recover the overall temporal evolution despite the absence of explicit temporal supervision during VAE training.
\begin{table}[!htp]
    \centering
    \caption{t-SNE projection of latent means and log-variances obtained by encoding all time steps of different test-dataset simulations using a VAE trained on the large dataset.}
    \vspace*{4mm}
    \begin{tabular}{c|c|c}
        & Latent means (t-SNE 2D) & Latent log-variances (t-SNE 3D) \\
        \hline
        \rotatebox[origin=c]{90}{Test simulation \#1} &
        \raisebox{-0.5\height}{\includegraphics[width=0.30\linewidth]{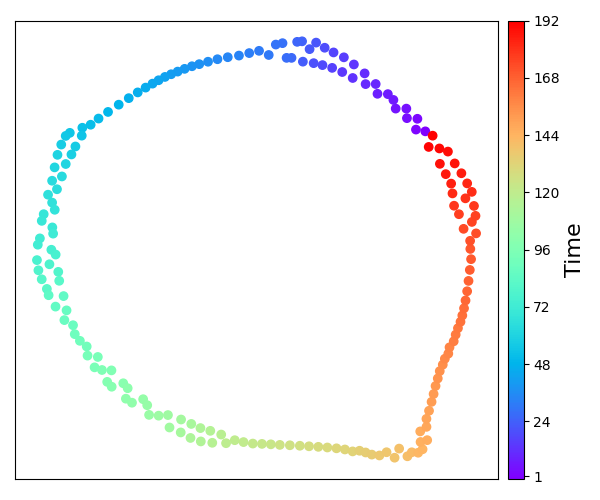}} &
        \raisebox{-0.5\height}{\includegraphics[width=0.30\linewidth]{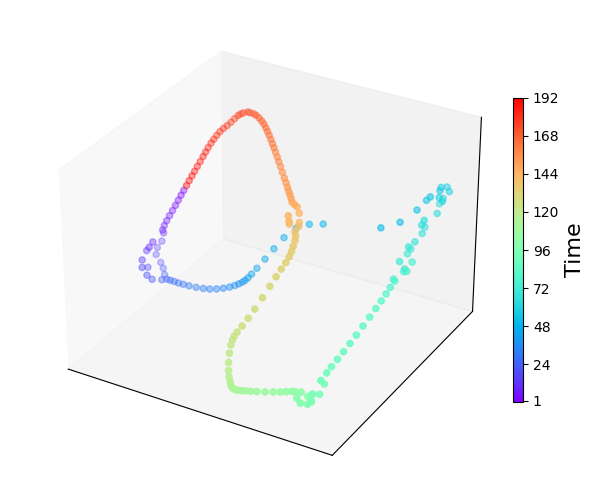}} \\
        \hline
        \rotatebox[origin=c]{90}{Test simulation \#2} &
        \raisebox{-0.5\height}{\includegraphics[width=0.30\linewidth]{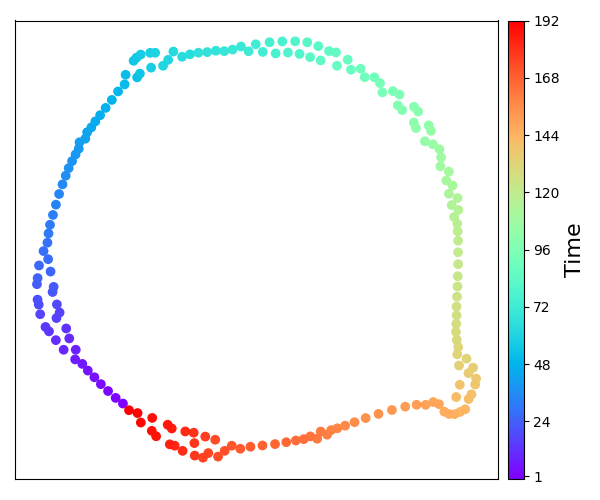}} &
        \raisebox{-0.5\height}{\includegraphics[width=0.30\linewidth]{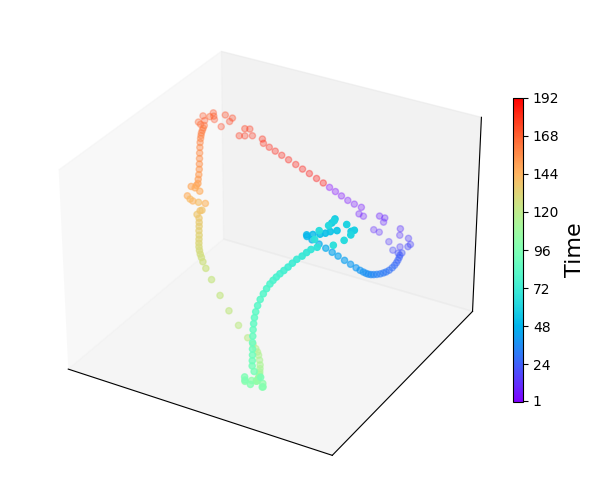}} \\
        \hline
    \end{tabular}
    \label{fig:vae-latent-analysis}
\end{table}

\end{document}